\pdfoutput=1
\documentclass[epjc3,twocolumn,a4paper,fleqn]{svjour3} 
\usepackage[fleqn]{amsmath}
\usepackage{graphicx}
\usepackage{epstopdf}
\usepackage{fix-cm}
\usepackage{txfonts}
\usepackage{lipsum}
\usepackage{mathtools}
\usepackage{cuted}
\usepackage{multirow}
\usepackage{comment}
\usepackage{empheq}
\usepackage[titletoc,title]{appendix}

\usepackage{listings}
\usepackage{lineno}
\usepackage{mdwlist}
\usepackage{pennames}
\usepackage[british]{babel}
\usepackage{color}
\usepackage{subfigure}
\usepackage{siunitx}
\usepackage{import}
\usepackage{amssymb}
\usepackage{epsfig}
\usepackage{cite}
\usepackage{hyperref}
\hypersetup{
colorlinks=true,
linkcolor=blue,
citecolor=blue,
urlcolor=blue,
}

\DeclareSIUnit \clight  {\textit{c}}
\RequirePackage[normalem]{ulem} 
\RequirePackage{color}\definecolor{RED}{rgb}{1,0,0}\definecolor{BLUE}{rgb}{0,0,1} 
\RequirePackage[normalem]{ulem} 
\RequirePackage{color}\definecolor{RED}{rgb}{1,0,0}\definecolor{BLUE}{rgb}{0,0,1} 

\DeclareGraphicsExtensions{.eps,.pdf,.jpeg,.jpg,.png}

\newcommand{\bea}{\begin{eqnarray}}
\newcommand{\eea}{\end{eqnarray}}
\newcommand{\be}{\begin{equation}}
\newcommand{\ee}{\end{equation}}
\newcommand{\fref}[1]{Fig.~\ref{#1}}
\newcommand{\Fref}[1]{Figure~\ref{#1}}
\newcommand{\sref}[1]{Sect.~\ref{#1}}
\newcommand{\Sref}[1]{Section~\ref{#1}}
\newcommand{\tref}[1]{Table~\ref{#1}}
\newcommand{\eref}[1]{Eq.~(\ref{#1})}
\newcommand{\erefs}[1]{Eqs.~(\ref{#1})}

\newcommand*{\muonp}          {\ifmmode\mathrm{\muup^+}\else$\mathrm{\muup^+}$\fi}
\newcommand*{\muon}           {\ifmmode\mathrm{\muup}\else$\mathrm{\muup}$\fi}
\newcommand*{\tauon}          {\ifmmode\mathrm{\tauup}\else$\mathrm{\tauup}$\fi}

\newcommand*{\egamma}         {E_{\mathrm{\gammaup}}}
\newcommand*{\photon}         {\ifmmode{\gammaup}\else${\gammaup}$\fi}
\newcommand*{\positron}       {\ifmmode{\mathrm{e}^+}\else${\mathrm{e}^+}$\fi}
\newcommand*{\electron}       {\ifmmode{\mathrm{e}}\else${\mathrm{e}}$\fi}
\newcommand*{\epositron}      {{E_\mathrm{e^+}}}
\newcommand*{\ppositron}      {{p_\mathrm{e^+}}}
\newcommand*{\tpositron}      {{t_\mathrm{e^+}}}

\newcommand*{\tegamma}        {{t_{\mathrm{e^+ \gammaup}}}}
\newcommand{\tg}{{\ifmmode t_{\gammaup_1\mathrm{e}^+}\else$t_{\gammaup_1\mathrm{e}^+}$\fi}}
\newcommand*{\tgg}{{\ifmmode t_{\gammaup\gammaup}\else$t_{\gammaup\gammaup}$\fi}}
\newcommand*{\Thetaegamma}    {{\Theta_{\mathrm{e}^+ \gammaup}}}

\newcommand*{\thetaegamma}    {{\theta_{\mathrm{e}^+ \gammaup}}}
\newcommand*{\phiegamma}      {{\phi_{\mathrm{e}^+ \gammaup}}}
\newcommand*{\thetae}         {{\theta_\mathrm{e^+}}}
\newcommand*{\phie}           {{\phi_\mathrm{e^+}}}

\newcommand{\meg}{\ifmmode{\muup \to e \gammaup}\else$\mathrm{\muup \to e \gammaup}$\fi}
\newcommand{\megp}{\ifmmode{\muup^+ \to \mathrm{e}^+ \gammaup}\else$\mathrm{\muup^+ \to e^+ \gammaup}$\fi}
\newcommand{\michel}{\ifmmode{\muup^+ \to e^+ \nuup\bar{\nuup}}\else$\mathrm{\muup^+ \to e^+ \nuup\bar{\nuup}}$\fi}
\newcommand{\radiative}{\ifmmode{\muup^+ \to e^+\nuup\bar{\nuup}\gammaup} \else$\mathrm{\muup^+ \to e^+ \nuup\bar{\nuup}\gammaup}$\fi}
\newcommand{\conv}{\ifmmode{\muup^- \to e^-}\else$\mathrm{\muup^- \to e^-}$\fi}
\newcommand{\convN}{\ifmmode{\muup^-N \to e^-N}\else$\mathrm{\muup^-N \to e^-N}$\fi}
\newcommand{\mute}{\ifmmode{\muup \to 3e}\else $\mathrm{\muup \to 3e}$\fi}
\newcommand{\mutec}{\ifmmode{\muup^+ \to e^+e^+e^-}\else $\mathrm{\muup^+ \to e^+e^+e^-}$\fi}
\newcommand{\aif}{\ifmmode\mathrm{e}^+ \mathrm{e}^- \to \gammaup\gammaup \else$\mathrm{e}^+ \mathrm{e}^- \to \gammaup \gammaup$\fi}
\newcommand{\teg}{\ifmmode{\tauup \to e \gammaup} \else$\mathrm{\tauup \to e \gammaup}$\fi}
\newcommand{\tmg}{\ifmmode{\tauup \to \gammaup} \else$\mathrm{\tauup \to \muup \gammaup}$\fi}
\newcommand{\tmueg}{\ifmmode{\mathrm\tauup \to \ell \gammaup}\else$\mathrm{\tauup \to \ell \gammaup}$\fi}
\newcommand{\tautl}{\ifmmode{\mathrm\tauup \to 3\ell} \else$\mathrm\tauup \to 3\ell$\fi}

\newcommand*{\nphe} {N_{\mathrm{phe},i}}
\newcommand*{\npho} {N_{\mathrm{pho},i}}

\newcommand*{\xpos}          {x_\mathrm{e^+}}
\newcommand*{\ypos}          {y_\mathrm{e^+}}
\newcommand*{\zpos}          {z_\mathrm{e^+}}

\newcommand*{\ugamma}         {u_{\gammaup}}
\newcommand*{\vgamma}         {v_{\gammaup}}
\newcommand*{\wgamma}         {w_{\gammaup}}

\newcommand*{\epositronerr}     {\sigma_\epositron}
\newcommand*{\thetaeerr}        {\sigma_{\theta_\mathrm{e^+}}}
\newcommand*{\phieerr}          {\sigma_{\phi_\mathrm{e^+}}}

\newcommand*{\yposerr}          {\sigma_{y_{\mathrm{e^+}}}}
\newcommand*{\zposerr}          {\sigma_{z_{\mathrm{e^+}}}}

\newcommand*{\mathtentative}{}
\def\mathtentative#1#{\mathcoloraux{#1}}
\newcommand*{\mathcoloraux}[3]{%
  \protect\leavevmode
  \begingroup
    \color#1{#2}#3%
  \endgroup
}

\journalname{Eur. Phys. J. C} 

\begin{document}


\title{Operation and performance of the MEG~II detector}

\author{The MEG~II collaboration}
\newcommand*{\INFNPi}{INFN Sezione di Pisa$^{a}$; Dipartimento di Fisica$^{b}$ dell'Universit\`a, Largo B.~Pontecorvo~3, 56127 Pisa, Italy}
\newcommand*{\INFNGe}{INFN Sezione di Genova$^{a}$; Dipartimento di Fisica$^{b}$ dell'Universit\`a, Via Dodecaneso 33, 16146 Genova, Italy}
\newcommand*{\INFNPv}{INFN Sezione di Pavia$^{a}$; Dipartimento di Fisica$^{b}$ dell'Universit\`a, Via Bassi 6, 27100 Pavia, Italy}
\newcommand*{\INFNRm}{INFN Sezione di Roma$^{a}$; Dipartimento di Fisica$^{b}$ dell'Universit\`a ``Sapienza'', Piazzale A.~Moro, 00185 Roma, Italy}
\newcommand*{\INFNNa}{INFN Sezione di Napoli, Via Cintia, 80126 Napoli, Italy}
\newcommand*{\INFNLe}{INFN Sezione di Lecce$^{a}$; Dipartimento di Matematica e Fisica$^{b}$ dell'Universit\`a del Salento, Via per Arnesano, 73100 Lecce, Italy}
\newcommand*{\ICEPP} {ICEPP, The University of Tokyo, 7-3-1 Hongo, Bunkyo-ku, Tokyo 113-0033, Japan }
\newcommand*{\Kobe} {Kobe University, 1-1 Rokkodai-cho, Nada-ku, Kobe, Hyogo 657-8501, Japan}
\newcommand*{\UCI}   {University of California, Irvine, CA 92697, USA}
\newcommand*{\KEK}   {KEK, High Energy Accelerator Research Organization, 1-1 Oho, Tsukuba, Ibaraki 305-0801, Japan}
\newcommand*{\PSI}   {Paul Scherrer Institut PSI, 5232 Villigen, Switzerland}
\newcommand*{\Waseda}{Research Institute for Science and Engineering, Waseda~University, 3-4-1 Okubo, Shinjuku-ku, Tokyo 169-8555, Japan}
\newcommand*{\BINP}  {Budker Institute of Nuclear Physics of Siberian Branch of Russian Academy of Sciences, 630090 Novosibirsk, Russia}
\newcommand*{\JINR}  {Joint Institute for Nuclear Research, 141980 Dubna, Russia}
\newcommand*{\ETHZ}  {Institute for Particle Physics and Astrophysics, ETH Z\" urich, 
Otto-Stern-Weg 5, 8093 Z\" urich, Switzerland}
\newcommand*{\NOVS}  {Novosibirsk State University, 630090 Novosibirsk, Russia}
\newcommand*{\NOVST} {Novosibirsk State Technical University, 630092 Novosibirsk, Russia}
\newcommand*{\ScuolaPi}{Scuola Normale Superiore, Piazza dei Cavalieri 7, 56126 Pisa, Italy}
\newcommand*{\INFNLNF}{\textit{Present Address}: INFN, Laboratori Nazionali di Frascati, Via 
E. Fermi, 40-00044 Frascati, Rome, Italy}
\newcommand*{\Liverpool}{Oliver Lodge Laboratory, University of Liverpool, Liverpool, L69 7ZE, United Kingdom}

\date{Received: date / Accepted: date}

\author{
The MEG~II collaboration\\\\
        K.~Afanaciev~\thanksref{addr12} \and
        A.~M.~Baldini~\thanksref{addr1}$^{a}$ \and
        S.~Ban~\thanksref{addr10} \and
        V.~Baranov~\thanksref{addr12}\thanksref{e5}  \and
        H.~Benmansour~\thanksref{addr1}$^{ab}$ \and
        M.~Biasotti~\thanksref{addr3}$^{a}$ \and
        G.~Boca~\thanksref{addr4}$^{ab}$ \and         P.~W.~Cattaneo~\thanksref{addr4}$^{a}$\thanksref{e1} \and
        G.~Cavoto~\thanksref{addr5}$^{ab}$ \and
        F.~Cei~\thanksref{addr1}$^{ab}$ \and
        M.~Chiappini~\thanksref{addr1}$^{ab}$ \and
        G.~Chiarello~\thanksref{addr6}$^{a}$\thanksref{e2} \and
        A.~Corvaglia~\thanksref{addr6}$^{a}$ \and
        F.~Cuna~\thanksref{addr6}$^{ab}$\thanksref{e3} \and
        G.~Dal~Maso\thanksref{addr2,addr16} \and
        A.~De~Bari~\thanksref{addr4}$^{a}$ \and
        M.~De~Gerone~\thanksref{addr3}$^{a}$ \and
        L.~Ferrari~Barusso~\thanksref{addr3}$^{ab}$ \and
        M.~Francesconi~\thanksref{addr17} \and 
        L.~Galli~\thanksref{addr1}$^{a}$ \and
        G.~Gallucci~\thanksref{addr3}$^{a}$ \and
        F.~Gatti~\thanksref{addr3}$^{ab}$ \and
        L.~Gerritzen~\thanksref{addr10}  \and
        F.~Grancagnolo~\thanksref{addr6}$^{a}$ \and
        E.~G.~Grandoni~\thanksref{addr1}$^{ab}$ \and 
        M.~Grassi~\thanksref{addr1}$^{a}$ \and 
        D.~N.~Grigoriev~\thanksref{addr7,addr8,addr9} \and
        M.~Hildebrandt~\thanksref{addr2} \and
        K.~Ieki~\thanksref{addr10}  \and
        F.~Ignatov~\thanksref{addr15} \and
        F.~Ikeda~\thanksref{addr10}  \and
        T.~Iwamoto~\thanksref{addr10}  \and
        S.~Karpov~\thanksref{addr7,addr9} \and
        P.-R.~Kettle~\thanksref{addr2} \and
        N.~Khomutov~\thanksref{addr12} \and
        S.~Kobayashi~\thanksref{addr10}  \and
        A.~Kolesnikov~\thanksref{addr12}  \and
        N.~Kravchuk~\thanksref{addr12}  \and
        V.~Krylov~\thanksref{addr12} \and
        N.~Kuchinskiy~\thanksref{addr12}  \and
        W.~Kyle~\thanksref{addr11} \and
        T.~Libeiro~\thanksref{addr11} \and  
        V.~Malyshev~\thanksref{addr12}  \and
        A.~Matsushita~\thanksref{addr10}  \and
        M.~Meucci~\thanksref{addr5}$^{ab}$ \and   
        S.~Mihara~\thanksref{addr13}  \and
        W.~Molzon~\thanksref{addr11} \and
        Toshinori~Mori~\thanksref{addr10}  \and
        F.~Morsani~\thanksref{addr1}$^{a}$  \and
        M.~Nakao~\thanksref{addr10} \and 
        D.~Nicol\`o~\thanksref{addr1}$^{ab}$ \and
        H.~Nishiguchi~\thanksref{addr13}  \and
        A.~Ochi~\thanksref{addr14}  \and
        S.~Ogawa~\thanksref{addr10}  \and
        R.~Onda~\thanksref{addr10}  \and
        W.~Ootani~\thanksref{addr10}  \and
        A.~Oya~\thanksref{addr10} \and
        D.~Palo~\thanksref{addr11} \and
        M.~Panareo~\thanksref{addr6}$^{ab}$ \and
        A.~Papa~\thanksref{addr1}$^{ab}$\thanksref{addr2} \and
        V.~Pettinacci~\thanksref{addr5}$^{a}$ \and
        A.~Popov~\thanksref{addr7,addr9} \and
        F.~Raffaelli~\thanksref{addr1}$^{a}$ \and
        F.~Renga~\thanksref{addr5}$^{a}$ \and
        S.~Ritt~\thanksref{addr2} \and
        M.~Rossella~\thanksref{addr4}$^{a}$ \and
        A.~Rozhdestvensky~\thanksref{addr12}  \and
        P.~Schwendimann~\thanksref{addr2} \and
        K.~Shimada~\thanksref{addr10} \and
        G.~Signorelli~\thanksref{addr1}$^{a}$ \and
        A.~Stoykov~\thanksref{addr2} \and
        M.~Takahashi~\thanksref{addr14}  \and
        G.F.~Tassielli~\thanksref{addr6}$^{ab}$\thanksref{e4} \and
        K.~Toyoda~\thanksref{addr10} \and
        Y.~Uchiyama~\thanksref{addr10,addr14} \and
        M.~Usami~\thanksref{addr10} \and
        A.~Venturini~\thanksref{addr1}$^{ab}$ \and        B.~Vitali~\thanksref{addr1}$^{a,}$\thanksref{addr5}$^{b}$ \and
        C.~Voena~\thanksref{addr5}$^{ab}$ \and   
        K.~Yamamoto~\thanksref{addr10}  \and
        K.~Yanai~\thanksref{addr10} \and
        T.~Yonemoto~\thanksref{addr10}  \and
        K.~Yoshida~\thanksref{addr10} \and
        Yu.V.~Yudin~\thanksref{addr7,addr9} 
}

\institute{\JINR   \label{addr12}
           \and
             \INFNPi \label{addr1}
           \and
             \ICEPP \label{addr10}
           \and
             \INFNGe \label{addr3}
           \and
             \INFNPv \label{addr4}
           \and
             \INFNRm \label{addr5}
           \and
             \INFNLe \label{addr6} 
           \and
             \PSI \label{addr2}
            \and
             \ETHZ    \label{addr16}
            \and
             \INFNNa \label{addr17}
           \and
             \BINP   \label{addr7}
           \and
             \NOVST  \label{addr8}
           \and
             \NOVS   \label{addr9}
            \and
             \Liverpool  \label{addr15}
           \and
             \UCI    \label{addr11}
           \and
             \KEK    \label{addr13}
           \and
             \Kobe    \label{addr14}
}

\thankstext[*]{e1}{Corresponding author: paolo.cattaneo@pv.infn.it} 
\thankstext[**]{e2}{Presently at Department of Engineering, University of Palermo, Viale delle Scienze,
Building 9, 90128 Palermo, Italy} 
\thankstext[***]{e3}{Presently at INFN Sezione di Bari, Via Giovanni Amendola, 173, 70126, Bari, Italy} 
\thankstext[****]{e4}{Presently at Dipartimento di Medicina e Chirurgia, Università LUM “Giuseppe
Degennaro”, 70010, Casamassima, Bari, Italy} 

\thankstext[$\dagger $]{e5}{Deceased} 

\maketitle 
 
\begin{abstract}
The MEG~II experiment, located at the Paul Scherrer Institut (PSI) in Switzerland, is the successor to the MEG experiment, which completed data taking in 2013.
MEG~II started fully operational data taking in 2021, with the goal of improving the sensitivity of the \megp\ decay down to \num{\sim 6.e-14} almost an order of magnitude better than the current limit.
In this paper, we describe the operation and performance of the experiment 
and give a new estimate of its sensitivity versus data acquisition time.
\end{abstract}


\tableofcontents 


\section{Introduction}
\label{sec:introduction}
The MEG~II detector is the upgrade, with significant improvements and additions, to the MEG detector \cite{megdet}, which by 2013 had recorded data establishing 
the best limit to date for the SM violating decay \megp\ \cite{baldini_2016}. 
The goal of the upgraded experiment is to improve the sensitivity of this decay by about one order of magnitude.
The physical rationale for this upgrade and the design criteria for the MEG~II subdetectors are outlined in
\cite{baldini_2018}. 

The signal is a $\photon$-ray and a positron back-to-back with energy $\egamma\sim\epositron\sim m_{\muup}c^2/2=\SI{52.83}{\MeV}$ emanating at the same time
from a common vertex. 
The main background results from the accidental time coincidence of high-momentum positrons from Michel decay \michel\ and high-energy $\photon$-rays from radiative muon decay (RMD) \radiative, positron Bremsstrahlung and positron annihilation in flight \aif. An additional smaller background is RMDs with high-momentum positrons and high-energy $\photon$-rays.

\begin{figure*}[!htb]
\centering
  \includegraphics[width=1\textwidth,angle=0] {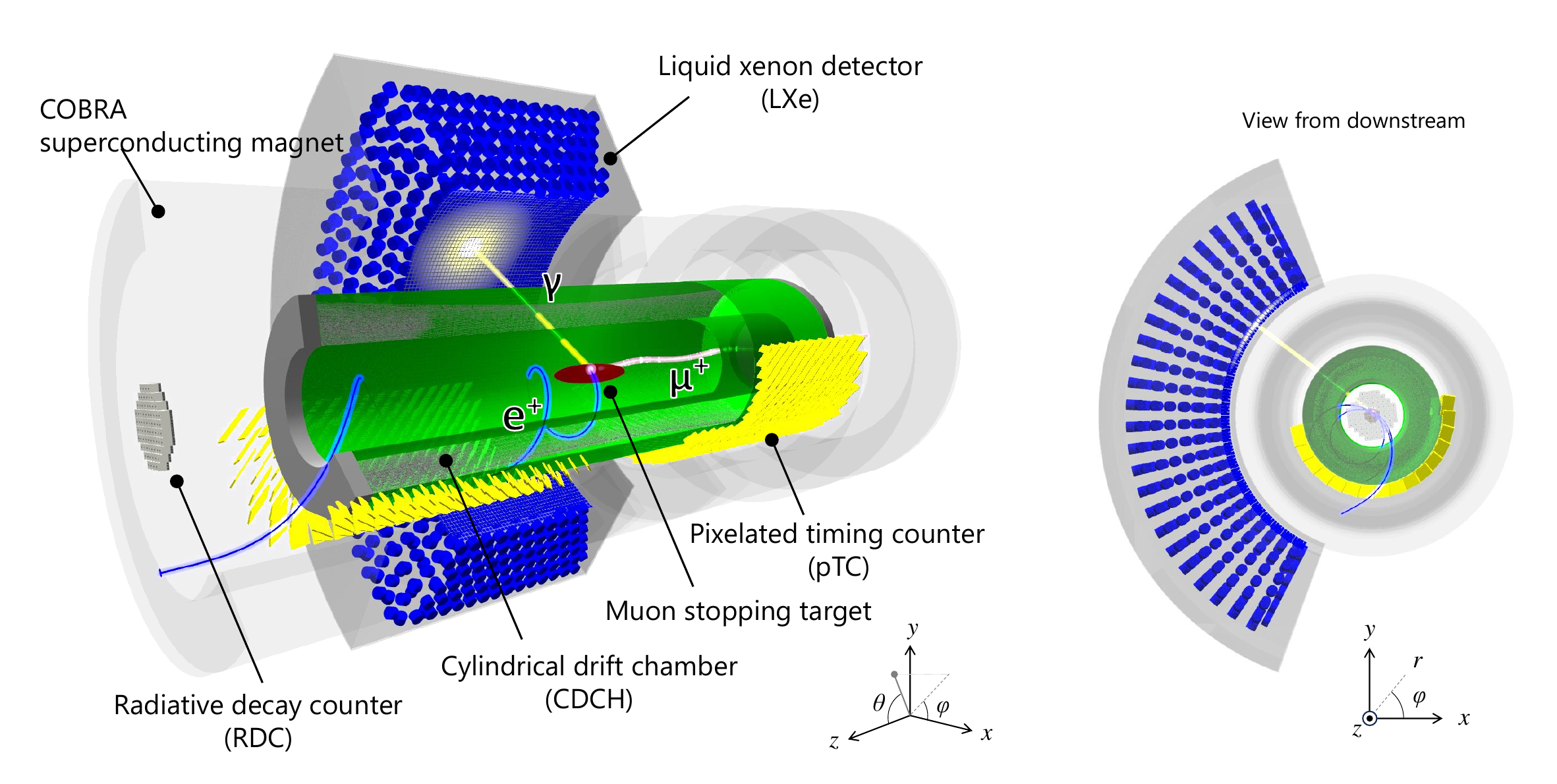}
  \caption{A sketch of the MEG~II detector with a simulated \megp\ event.}
 \label{introduction:meg2det}
\end{figure*}

A schematic view of the detector highlighting the main components can be found in \fref{introduction:meg2det}.
The detector is located at the $\piup$E5 beamline in the high-intensity proton accelerator facility at the Paul Scherrer Institut (PSI) in Switzerland, allowing the world's most intense continuous positive muon beam to be stopped in a thin target and to measure the decay products.
The positron spectrometer relies on the COnstant Bending RAdius (COBRA) superconducting 
magnet generating a gradient magnetic field ranging from \SI{1.27}{\tesla} in the centre to 
\SI{0.50}{\tesla} at either end of the magnet cryostat to measure positron momenta by the Cylindrical Drift CHamber (CDCH). 
Positron time is measured by the pixelated Timing Counter (pTC), and the Radiative Decay Counter (RDC) helps reject background.
Finally, the Liquid Xenon (LXe) detector uses on the scintillation process to measure the energy, position and timing of the incident \photon-ray.

We use a cylindrical coordinate system $(r,\phi, z)$ with the origin at the centre of COBRA. 
The $z$-axis is aligned along the COBRA axis in the direction of the incident muon beam. The 
azimuthal angle $\phi = 0$ is opposite the centre of the LXe detector, and corresponds to the $x$-axis of the associated Cartesian coordinate system, 
while the $y$-axis points upward.
Positrons follow trajectories with decreasing $\phi$ coordinate. 
The polar angle $\theta$ with respect to the $z$-axis is also used. The region with $z < 0$ is called upstream, while the region with $z > 0$ is called downstream.

The geometrical acceptance of the experiment is defined by the size of the LXe fiducial volume, which is approximately equal to 
$\phi_\photon \in \left( \frac{2}{3} \pi, \frac{4}{3} \pi\right)$ and
$|\cos \theta_\photon | < 0.35$, giving an overall acceptance of \SI{\sim 11}{\percent}. The efficiencies given below refer to this geometrical acceptance.

After the years 2017 to 2020 during which the engineering runs required to commission the detector component were performed, data acquisition with the detector fully operational began in 2021 with the \megp\ trigger activated. 

In this paper, we give only a brief summary of the designs of the subdetectors, describing only the differences between the design and the actual realisation, and we describe the operation and performance of the detector and compare the measured resolutions with the design \cite{baldini_2018}. 
Finally, the expected sensitivity for the coming years is presented.
\section{Beam}
\label{sec:beam}

Pions and muons are produced by impinging  protons, accelerated up to \SI{590}{\MeV} by 
the PSI ring cyclotron, onto the production target TgE at an angle of \SI{8}{\degree} with respect to the beam direction. 
This configuration, in operation since the second half of 2020, has proved to be less sensitive to variation of the 
secondary beam intensities as a function of the proton beam position.
For a comprehensive and historical description of the PSI facility, see \cite{PSICyclotron,10.21468/SciPostPhysProc.5.002}.

\begin{figure*}[tb]
		\centering
 		\includegraphics[width = 1\textwidth]{
        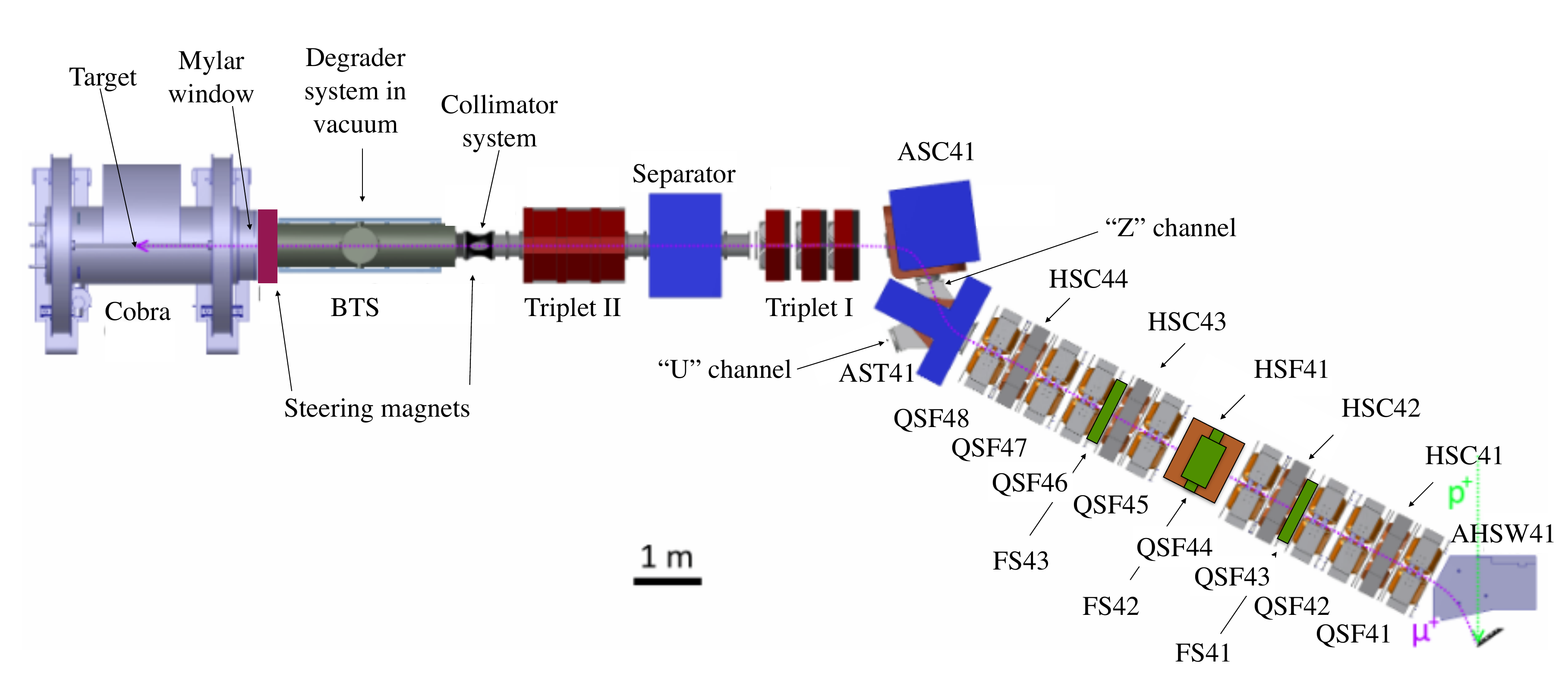
        }
		\caption{$\piup$E5 beamline section connected to Z-channel. The elements up to Triplet II are shared between the MEG~II and the Mu3e \cite{Mu3ePhaseI} experiments.}
		\label{f:piE5DS}
\end{figure*}

The $\piup$E5 area is served by a low-energy secondary beamline capable of delivering pions and muons in the \SIrange[range-units = single]{10}{120}{\mega\electronvolt}/
${\mathrm c}$ momentum range. 
It has a \SI{165}{\degree} angle of view with respect to the proton beam on TgE. 
\Fref{f:piE5DS} shows the beamline layout from TgE up the MEG II detector. 
The AHSW41 dipole, which is part of the proton beamline, captures the pions and muons in the backward direction and defines the momentum accepted by the $\piup$E5 beamline. 
%
%
The collected charged particles are deflected by \SI{47.5}{\degree} and coupled into a straight section consisting of quadrupoles (QSF4*) and sextupoles (HSC4*). 
Along this section three slit systems (FS41--42--43) are used to reduce the beam intensity delivered to the experimental area and \textcolor{black}{(FS41--43)}
to cut the momentum distribution.
This is possible because the dipoles introduce dispersion into the lattice and because the beams are not monochromatic 
and there is a correlation between the horizontal transverse coordinate of the beam and the momentum distribution.

In the MEG~II experiment, where surface \muonp\ (\SI{28}{\MeV}/$\mathrm{c}$) are selected and stopped in the muon stopping target 
(see \sref{sec:target}), FS41 is used to reduce the beam intensity which also reduces the 
width of the momentum distribution and results in less straggling at the target and a higher stopping efficiency.

After QSF48, the dipole magnet AST41 produces a symmetric deflection of the beam, again at \SI{47.5}{\degree}, to either the Z-channel or the U-channel. When the Z-channel mode is selected, the beam is transmitted through a second dipole ASC41, which generates a \SI{75}{\degree} deflection and then delivers the beam into the experimental area. 

 The beam is coupled into a Wien filter via Triplet I, a quadrupole triplet, 
 to separate the muon beam from the main contaminants: pions and positrons. The separation is achieved vertically at the downstream collimator system. 
 The task of Triplet I is to shape the beam so that it has a horizontal waist, while being vertically parallel to enhance the separation.

 The beam is then focused by Triplet II 
 at a collimator, dumping 
 the contaminating beams. For beam tuning campaigns in $\piup$E5, the first measurement point is immediately downstream of the collimator.

Triplet II couples the beam into the Beam Transport Solenoid (BTS). There, a \SI{300}{\micro\meter} thick Mylar$^\textsuperscript{\textregistered}$ moderator is positioned at the focus to minimise the effects of multiple scattering.
%
A \SI{190}{\micro\meter} thick Mylar window separates the vacuum in the
beamline from the helium atmospheric volume containing the muon-stopping target at the centre of the COBRA magnet.

The field shape of the COBRA magnet 
was optimised to reduce the dependence of the bending radius on the emission angle of charged particles coming from the centre of the experiment and to avoid multiple turn trajectories through the tracking system, which severely affected the MEGA experiment \cite{PhysRevD.65.112002}. 
A detailed description of the final part of the beamline, the BTS, COBRA and the magnetic field map, its calculation and measurement, can be found in \cite{megdet}.
%

\begin{figure}[tb]
		\centering
		\includegraphics[width = 1\columnwidth]{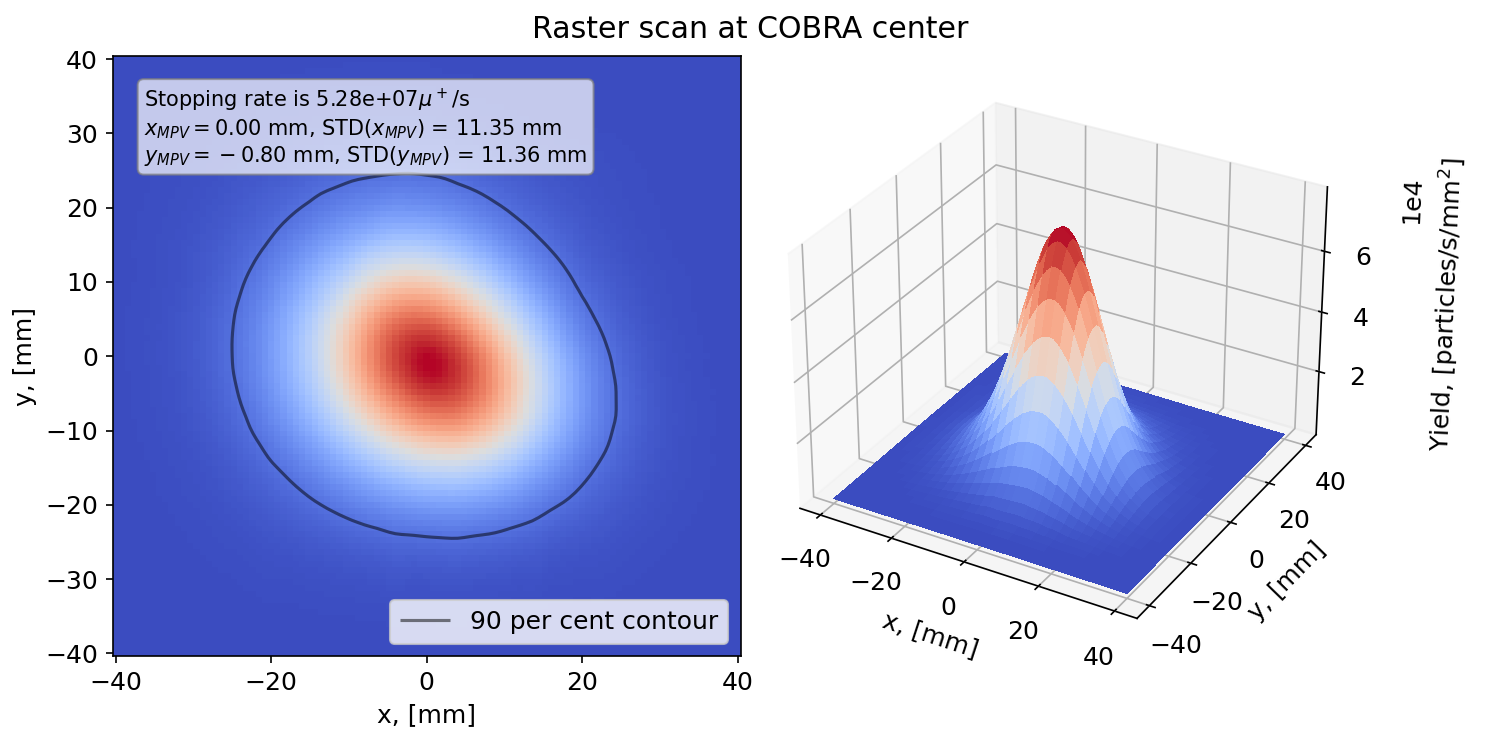}
		\caption{Beam profile at COBRA centre for a stopped muon rate $R_{\muon} = \SI{5.3e7}{\per\second}$ at $I_\mathrm{p}= \SI{2.2}{\milli\ampere}$.
        }
		\label{f:rasterscanCobra}
\end{figure}

Beam tuning at different intensities was performed to meet different requirements, including specific calibration data sets at low beam intensity (down to $10^6$ particles/s).
\Fref{f:rasterscanCobra} shows a typical beam profile measured at the centre of the COBRA magnet, where the stopping target is located. The beam is centred at $x_{\rm b} = \SI{0\pm 0.5}{\mm}$, $y_{\rm b} = \SI{-0.8 \pm 0.5}{\mm}$ with a standard deviation of the two coordinates equal to $\sigma_{x} = \SI{11.35\pm 0.5}{\mm}$ and $\sigma_{y} = \SI{11.36\pm 0.5}{\mm}$. In this figure, the slits were tuned to obtain a stopped muon rate $R_{\muon} = \SI{5.3e7}{\per\second}$
at the primary proton beam current $I_\mathrm{p} = \SI{2.2}{\milli\ampere}$\footnote{This is the nominal value of the facility. In 2021--2022, the actual current was lower; see \tref{table:Beam}.}. 
Beam settings with similar profiles were achieved in the range of stopped muon rates $R_{\muon}=\SIrange[range-phrase=-,range-units=single,range-exponents=combine-bracket]{2e7}{5e7}{\per\second}$.

During the 2021--2022 physics runs, different $R_{\muon}$
were used, starting with a lower value, to study the detector stability and rate capability, to tune the data acquisition and trigger systems, and to optimise the reconstruction algorithms, as explained in the next sections.
\tref{table:Beam} summarises the used $R_{\muon}$, normalised to the typical $I_\mathrm{p}$ for each year. The listed numbers include the MEG~II target-stopping efficiency estimated by simulation to be \SI{89}{\percent}. A \SI{5}{\percent} systematic uncertainty is attached to each measurement of $R_{\muon}$, due to uncertainties on the beam measurement setup and variations of the proton beam position on TgE.

\begin{table}[]
\caption{Stopped muon rates $R_{\muon}$ for the runs 2021--2022 normalised to the typical $I_\mathrm{p}$ for each year.}
\begin{center}
\begin{tabular}{ l l l l l } 
 \hline
 Year & Typical $I_\mathrm{p}$  (\unit{\mA}) & \multicolumn{3}{l}{$R_{\muon}$ ($10^7$~\unit{\per\second})}\\
 \hline
 2021 & 1.965 & 3.12 & 4.01 & 5.21 \\ 
 2022 & 1.765 & 2.80 & 4.07 & 5.02 \\  
 \hline
\end{tabular}
\label{table:Beam}
\end{center}
\end{table}

The measurement reported here has been performed using an avalanche photodiode detector (SPL4419 Hamamatsu). 
It has a \SI{1.5}{\mm} active area radius, with \SI{130}{\micro\meter} depletion layer. 
The depletion layer is the sensitive part of the detector, 
allowing for a clear separation between the energy deposit by the muons (stopping particles) and positrons (minimum ionising particles) mainly from stopped muon decay. 
A scan on the plane perpendicular to the incident beam direction is performed in steps of \SI{2}{\mm}. 
Details about the beam monitoring detectors used and developed for the MEG~II experiment can be found in \cite{baldini_2018}.

\section{Target}
\label{sec:target}
The role of the target is to stop the \muonp\ beam over a limited axial 
region at the centre of the COBRA magnet satisfying a number of contradictory requirements.
The target must intercept the largest possible fraction of the beam, yet with a minimal amount of material to reduce the interaction of particles from muon decay. 
The target parameters are the results of extensive calculations and simulations as well as of the experience gained in MEG.
Most relevant, the position and planarity of the target must be precisely known to limit the systematic errors in \muonp\ decay vertex position.

\subsection{Concept and design}

The design of the MEG~II target has been optimised, on the basis of 
measurements and simulations. 
Different combinations of material/target thickness/degrader thickness 
have been considered \cite{berg_2017}.
A scintillator material (BC400) was selected  for the target, 
which has the advantage, over non-scintillating materials, 
to allow non-destructive beam intensity and profile measurements 
with dedicated equipment. 
After some experimental investigation, the current effort is to exploit the 
feasibility of this technique within the strict constraints dictated by the MEG~II experiment.

In order to identify a \megp\ event, it is necessary to measure the 
angles of the \positron\ trajectory ($\phie,\thetae$) at the point where the \muonp\ decays 
by back-propagating the trajectory measured by the 
spectrometer up to the target surface. 
The MEG~II spectrometer provides a precision of \SI{\sim7}{\milli\radian}.
A precise knowledge of the target position is 
then required: with a radius of curvature \SI{\sim13}{\cm} for the 
\positron\ trajectory in \megp\ events, a displacement of the target by 
\SI{500}{\micro\meter} along its normal direction implies a 
systematic deviation of \SI{\sim4}{\milli\radian} in $\phie$ for 
$\phie = 0$, and a larger effect for non-zero $\phie$. 

Moreover, deformations of the target planarity, which were observed during the MEG data taking, 
produce a similar effect. The uncertainty on the target position
and deformation was the dominant systematic error in the MEG result
\cite{baldini_2016}, causing a \SI{5}{\percent} variation of the upper limit
on the branching ratio while other contributions were below \SI{1}{\percent}.

\subsection{The target mechanics}

The MEG~II target is an elliptical foil (length of \SI{270}{\milli\meter} and 
height of \SI{66}{\milli\meter})
with \SI{174\pm 20}{\micro\meter} average thickness (the error is the maximum deviation not the Gaussian $\sigma$).
The direction normal to the target foil lies on 
the horizontal plane $(x,z)$ and forms an angle of 
\SI{75.0 \pm 0.1}{\degree} with respect to the beam axis ($z$-axis). 
The nominal position and the error (maximum not Gaussian) are evaluated on the basis of the mounting procedure
and fixation to the adaptation mechanism, but are not used in the analysis. The target inclination as well as its position are estimated with the methods described in the following.

\begin{figure}[tbp]
\begin{center}
\includegraphics[width=1\columnwidth]{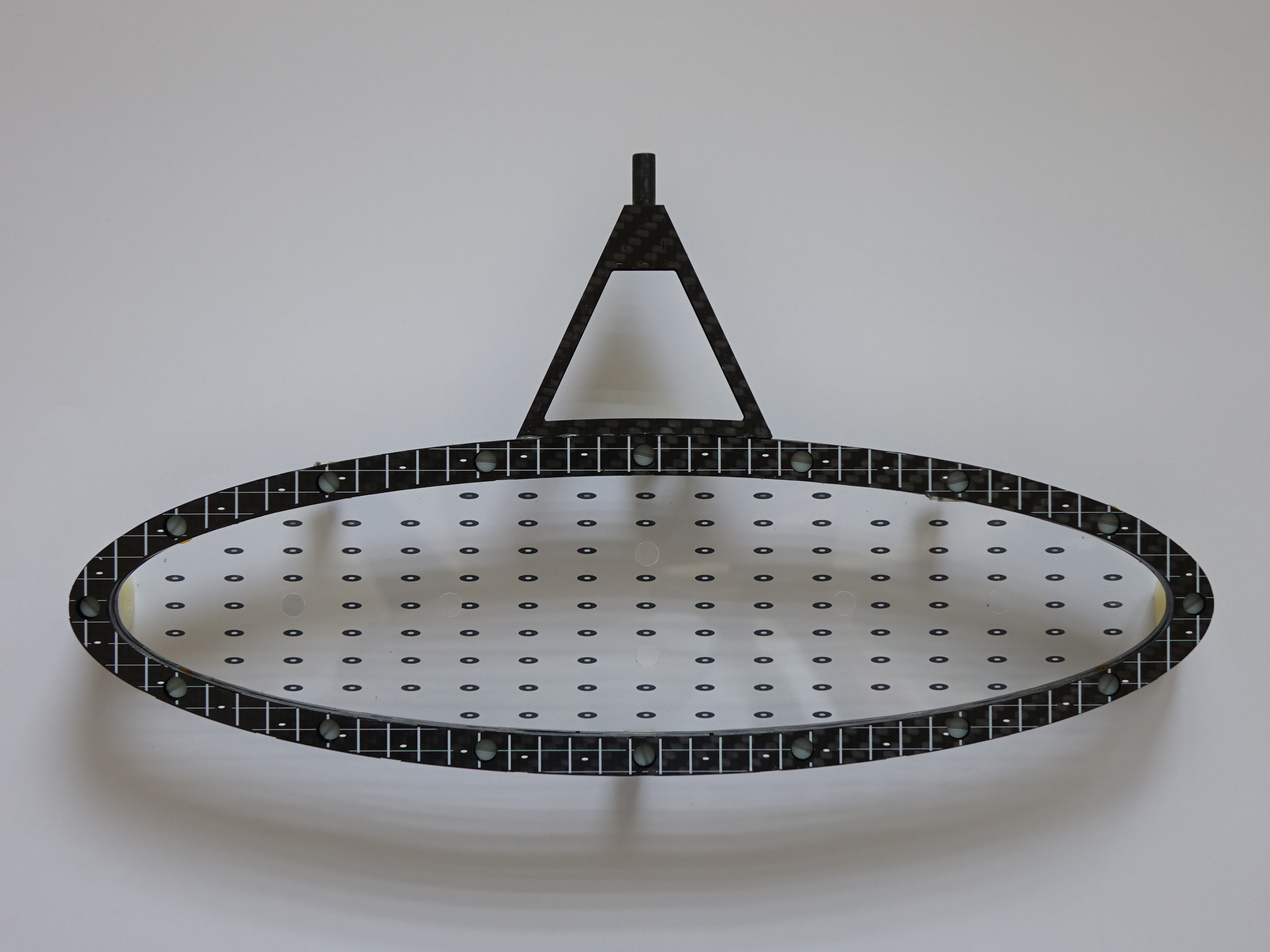}
\caption{The MEG~II target with the dot pattern on the foil and on the frame. The six holes are located along the ellipse axes.}
\label{fig:pattern}
\end{center}
\end{figure}

The target foil is supported by two hollow carbon fibre 
frames. A pattern of white dots, superimposed on a black background, is printed 
on both the frame and the foil. The dots are elliptical with a height and a 
width of \SI{0.51}{\milli\meter} and \SI{1.52}{\milli\meter} on the target and 
\SI{0.42}{\milli\meter} and \SI{1.27}{\milli\meter} on the frame such that the 
dots appear circular when imaged at an oblique angle with respect to the target’s surface.
{\color{black}
Six holes are bored into the target, that are (barely) visible along the ellipse axis 
on \fref{fig:pattern}: four along the major axis, two along the minor one located symmetrically with respect to the centre.
Starting from the dot at the center, they are located at the place of the (missing) third dots along the short axis, at the place of the (missing) seventh points and between the fourth and fifth dots along the long axis.

The holes and the dots are instrumental in measuring the relative alignment between the target and the tracking detectors as discussed in \fref{sec:bfield-cdch}.}

\subsection{The photo camera system for position measurement}

\textcolor{black}{
}

\textcolor{black}{The target position at the beginning of a MEG II data-taking run is precisely determined, with improved
accuracy with respect to MEG, thanks to reflectors installed on the target frame for a optical survey using a laser.}

\textcolor{black}{
Target position monitoring over long data taking periods was also possible by 
reconstructing the position of several fiducial holes made in
the target itself. A map of the reconstructed muon decay vertices on the 
target clearly showed the position of such holes. If the target position 
assumed in the trajectory reconstruction procedure is not exact, the holes 
artificially appear at different positions for different \positron\ angles. 
This method allowed one to reconstruct deviations of the target position from 
the nominal one. It was also effective to identify and correct the 
deformation of the target planarity. On the other hand, it required a large 
amount of data, so that it could only be
used to monitor the average target position over a few months of data taking, 
while the target was frequently moved far from its working position (at least 
every week) to perform the calibration of the LXe  detector through
a pneumatic system, which did not ensure micrometric repeatability or 
reproducibility of the target positioning.
}

\textcolor{black}{
The improved resolutions of the MEG~II positron spectrometer imposed the 
development of an additional method for more frequent monitoring of the target 
relative position over the data taking period to ensure that the systematic errors due to target position and deformation remain limited. 
The method is based on a photogrammetric survey of a pattern of dots printed on the target itself.
}

The dots are imaged with two digital CMOS photo cameras placed outside the beam halo, hosted in two independent supports.
Two LEDs are placed on one of these support
to provide illumination during the acquisition of the target pictures.
The supports are fixed to the target insertion system  at a distance of \SI{\sim 1000}{\milli\meter} from the centre of the MEG~II reference system, in correspondence to the CDCH (see \sref{sec:dch}) end-plate. The transverse distance from the $z$-axis is \SI{\sim 120}{\milli\meter}, at $\theta=\ang{6.3}$. 

One camera (IDS, mod.\ UI-3282SE)\footnote{www.1stvision.com/cameras/models/IDS-Imaging/UI-3282SE-M/C} has a Sony IMX264 sensor having 
\numproduct{2456 x 2054} pixels each of \SI{3.5}{\micro\meter} size, for a total sensor size of \qtyproduct[product-units = bracket-power]{8.473 x 7.086}{\milli\meter} and a TUSS optical system, mod. LVK7518, with a focal length of \SI{75}{\milli\meter}  and a maximum aperture of $f/1.8$.
The USB3 protocol is used for communications with front-end computers, since it has been proven to be immune to the magnetic field \cite{Cavoto:2020etw}.

The second camera (RVT-1001700S)\footnote{specinstcameras.com/wp-content/uploads/2021/07/RVT100-Brochure.pdf
} is a radiation tolerant camera from Spectral Instruments equipped with \num{4e6} pixels each of \SI{5.5}{\micro\meter} size, and a \SI{50}{\milli\meter} lens. 
This camera connects directly to a server communicating via fibre optic \cite{Palo:2019gzw}.

\subsection{The photogrammetric method}
\label{subsec:photogram}

The photogrammetric method is based on frequent and regular measurements (pictures) \textcolor{black}{taken during the run}.
The photo cameras image the pattern of dots,
and the position of dots on the picture can be determined with standard image processing algorithms. 
If the target moves between two successive camera shoots, or if it deforms over time, the position of these patterns in the pictures will change. 
The displacement of the target can be then determined with respect to a reference position, 
measured at the time of the optical survey. 

Two approaches are used to determine the target position, \textcolor{black}{orientation} and deformation, 
starting from the \textcolor{black}{differents} sets of pictures taken from the two photo cameras. 
The target position and orientation are described by the coordinates of the target centre in the MEG~II reference system \textcolor{black}{and by the Euler angles, respectively}.
The deformation is taken into account differently by the two \textcolor{black}{approaches}.

In the first approach, a $\chi^2$ goodness of fit is performed, 
in which  the $\chi^2$ is computed from the measured and the expected dot positions, 
where the latter depend on the target position and deformation and on the parameters of the optical system. 
The deformation is parameterised with Zernike polynomials; 
target position and deformation are floating parameters in the fit, 
while the parameters of the optical system are measured  from pictures taken just after the 
\textcolor{black}{optical} survey, when the target parameters are known. 
There are seven optical parameters: the position of the optical centre (three parameters), 
the independent components of the unit vector of the optical axis (two parameters), 
the orientation of the sensor around the optical axis (one parameter), 
and the distance of the sensor from the centre of the optical system (one parameter). 

\begin{figure}[tbp]
\begin{center}
\includegraphics[width=1\columnwidth]{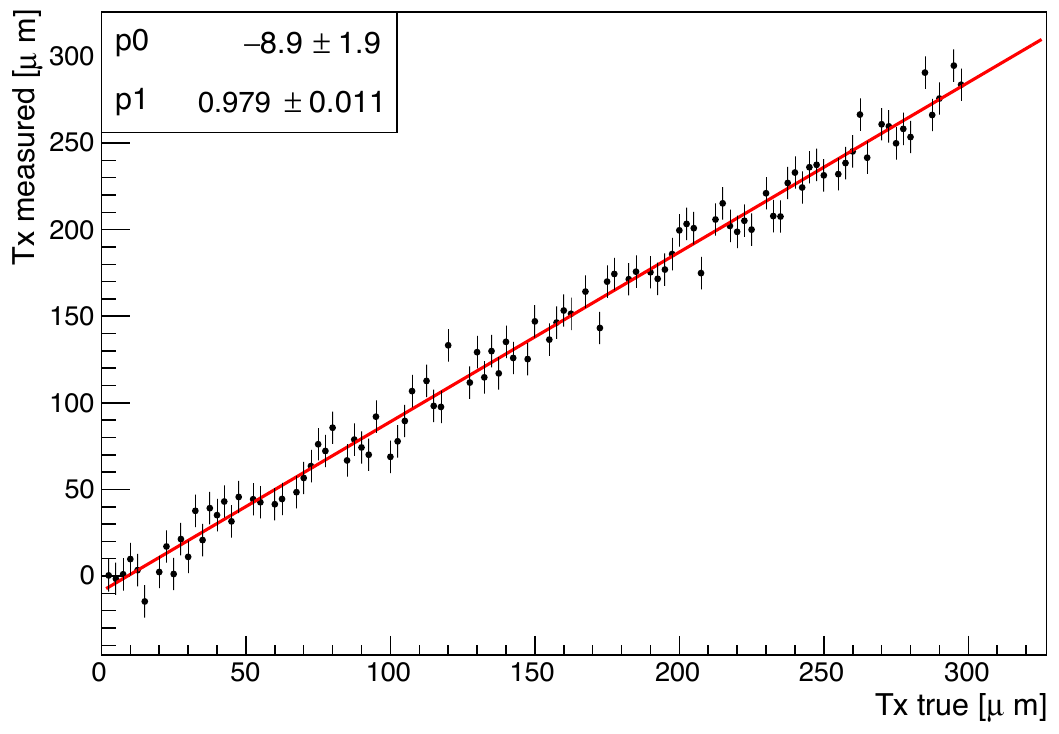}
\caption{Difference between the measured shift of the 
$x$-coordinate of the target centre obtained with the 
photogrammetric method as a function of the true shift, 
for a position scan in the $x$-direction.}
\label{fig:accuracy}
\end{center}
\end{figure}

The method has been validated on a bench-top test, by installing the photo camera, 
a LED, and a target mock-up on an optical table in a configuration similar to that in the experiment. 
The target was mounted on a linear stage with \SI{2.5}{\micro\meter} position accuracy. 
\Fref{fig:accuracy} shows the difference between the measured shift in the $x$-coordinate of the target centre, 
obtained with the photogrammetric method, as a function of the true shift, 
for a position scan performed along the $x$-direction (normal to the target plane). 
A linear fit has been performed to the distribution.
The resulting uncertainty in the $x$-coordinate of the target centre $T_x$ is $\sigma_{T_x} = \SI{12}{\micro\meter}$. 
Given that the direction transverse to the target plane in
the experiment is almost coincident with the $x$-axis, 
we can conclude that we fully satisfy the precision requirements \textcolor{black} {\SI{<500}{\micro\meter}}. 
The angular coefficient \textcolor{black}{p1 is consistent with 1} 
and the \textcolor{black}{difference of the intercept p0 with 0 is negligible compared with the required precision}. 

The second approach minimises a $\chi^{2}$ at the camera's image plane between the measured, imaged dot coordinates $\left(X'_\mathrm{CCD},Y'_\mathrm{CCD}\right)$ and estimated 3D dot coordinates in the camera coordinate system $\left(\vec{X}_\mathrm{CAM}\right)$ for each image. 
Here, $X_\mathrm{CAM},Y_\mathrm{CAM}$ are parallel with the camera's CCD and $Z_\mathrm{CAM}$ is parallel with the camera's optical axis. 
Images are taken every \SI{\sim 10}{\minute} to track the position, orientation, and shape of the target in the camera coordinate system. 
The analysis relies on a single position and orientation of the target with respect to the CDCH to map the coordinates in the camera coordinate system to the standard MEG~II coordinate system. 
This is taken either from an optical survey or the hole analysis. Here, the relative 3D dot coordinates in a local target coordinate system were taken from a CT-scan of the target. 
This CT-scan contains $O(10~\mathrm{M})$ data points including the 3D target deformation (\SI{<1}{\mm}). 
Given the measured distances between the dots on the foil and the camera's optical focal length, the technique is self-calibrating. 

We rely on the following optical equations to project the 3D coordinates in the camera coordinate system onto the image plane where $f$ is the camera's focal length :
\begin{align}
X_\mathrm{CCD} = \frac{X_\mathrm{CAM} \cdot f} {Z_\mathrm{CAM} - f}, \hspace{0.5cm} Y_\mathrm{CCD} = \frac{Y_\mathrm{CAM} \cdot f} {Z_\mathrm{CAM} - f}.
\end{align}
The $\chi^{2}$ contains a rigid body transformation $R\left(\vec{X}_\mathrm{T},\vec{\theta}_\mathrm{T}\right)$ with a translation and three Euler angles:
\begin{align}
\chi^{2} = \sum_{i}^{N}{ 
 \left[R\left(\vec{X}_\mathrm{T},\vec{\theta}_\mathrm{T}\right) \cdot X_{i, \mathrm{CCD}} - X'_{i, \mathrm{CCD}}\right]^{2}} +\\ \nonumber 
\left[R\left(\vec{X}_\mathrm{T},\vec{\theta}_\mathrm{T}\right) \cdot Y_{i, \mathrm{CCD}} - Y'_{i, \mathrm{CCD}}\right]^{2}.
\end{align}
\begin{figure}[tb]
\centering
  \includegraphics[width=1\columnwidth,angle=0] {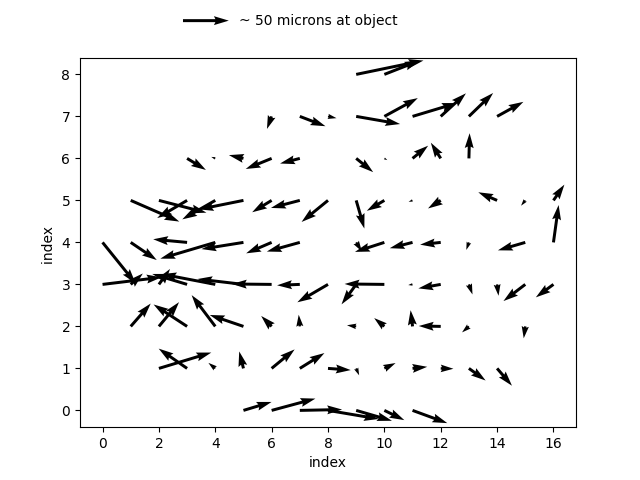}
  \caption{2D residuals at the camera image plane scaled to the object plane assuming the 3D CT-scan shape. The axes are the dot indices. }
 \label{target:TargetUCIResiduals}
\end{figure}

The residuals dot-by-dot at the image plane scaled to the object are shown in \fref{target:TargetUCIResiduals}. Without any additional shape parameters, the fit results in  residuals 
\SI{<50}{\micro\meter} at the object. These are due to additional deformations between the time of the CT-scan and that of the image, errors in the measured dot coordinates on the image plane, and errors in the projection equations. The residuals can be further suppressed by including elliptical Bessel function parameters (i.e. ``drumhead modes") into the fit, but the rigid body transformation was sufficient for the full 2021 data set given the 3D CT-scan. 

The two methods have been compared on pictures taken during the 2021 run, and found to yield consistent results within \SI{100}{\micro\meter} in the direction normal to the target plane, which is the most sensitive to the angular resolutions.


\subsection{The target hole method}
The hole analysis is performed after correcting the temporal variation of the target position traced by the photogrammetric method to determine the target position relative to the CDCH. The method and the results are described in \sref{sec:target-cdch}.


\section{Cylindrical drift chamber}
\label{sec:dch}
The positron momentum and position vectors are measured with CDCH. 
In this section we review the main advantages of CDCH over the MEG drift chamber system, its construction, commissioning, reconstruction, alignment and 
performance. 
\subsection{Concept and design}
\label{sec:dchconc}
The MEG drift chamber system consisted of sixteen individual radially aligned chambers. With this detector configuration, relevant efficiency loss and resolution degradation were caused by the material of the chamber's mechanical support, electronic cards and cables, which were frequently crossed by the positron trajectories. 
The CDCH, extensively discussed in \cite{baldini_2018, Chiappini:2023egy}, overcomes these limitations by replacing the segmented structure of the MEG chambers with a single volume one.

\begin{figure*}[tb]
\centering
  \includegraphics[width=1\textwidth,angle=0] {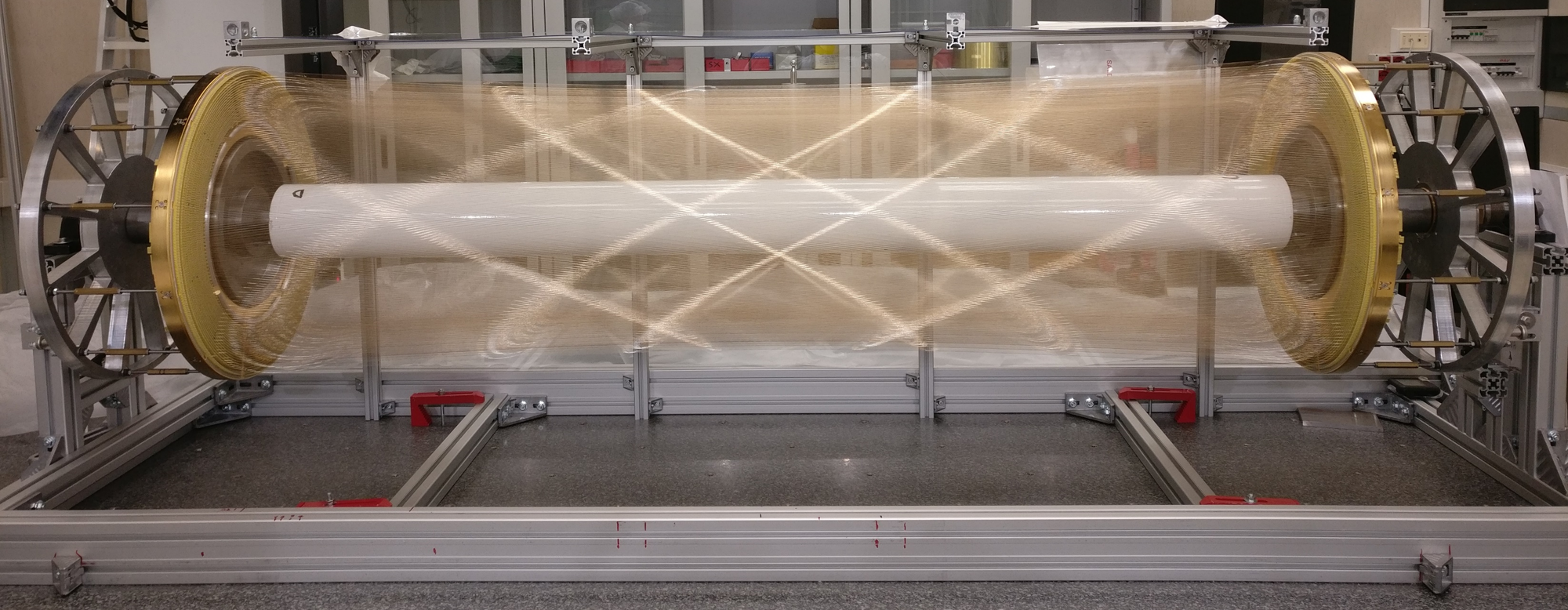}
  \caption{Picture of the open CDCH equipped with all the wires.}
 \label{dch:chamber}
\end{figure*}

The CDCH, \textcolor{black}{shown in~\fref{dch:chamber}}, is a \SI{1.93}{\meter} long low-mass cylindrical volume, filled with a helium--isobutane gas mixture (the exact composition is specified in \sref{sec:dchgeo}) and equipped with nine concentric layers of \num{192} gold-plated tungsten sense wires each, \num{1728} in total, arranged in a stereo configuration with two views and \num{\sim 10000} silver-plated aluminium cathode and guard wires. The sense wires collect the signals from the drift electrons, while the cathode and guard wires form nearly square drift cells and define the electric field within and at the boundaries of the sensitive volume; the cell dimensions range from \SIrange{5.8}{7.5}{\mm} at center and from \SIrange{6.7}{8.7}{\mm} at the end-plates.  The sense wires within the experimental acceptance, about 1200, are read out by the data acquisition system.

\begin{figure}[tb]
\centering
  \includegraphics[width=1.\columnwidth] {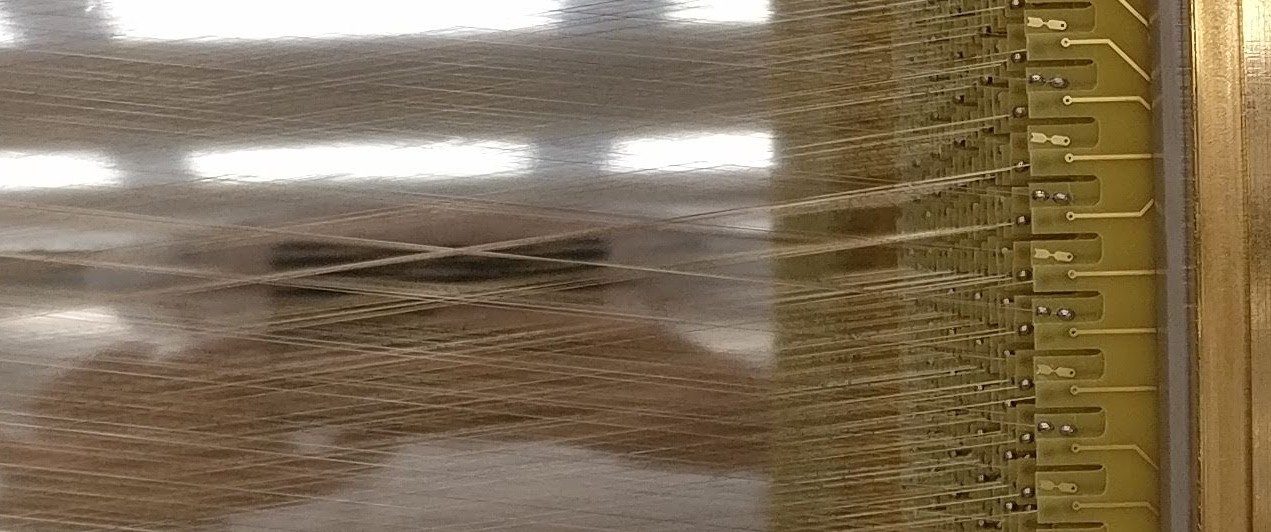}
  \caption{\textcolor{black}{Zoom on the stereo wire geometry, radial perspective.}}
 \label{dch:CDCHWireZoom.jpg}
\end{figure}

\textcolor{black}{A zoom on the endplate region to magnify the wire stereo geometry is shown in~\fref{dch:CDCHWireZoom.jpg}.}

%
\subsection{Construction}
\label{sec:dchcons}
Construction of the CDCH, which began in late 2016 and was completed in spring 2018, was performed modularly by soldering wires in groups of sixteen onto printed circuit boards (PCBs), 
which were then mounted between radial spokes at the ends of the chamber.  The mechanical structure of the detector achieves its final form by integrating a carbon fibre component securely fastened to both end-plates, effectively enclosing the sensitive volume.

The chamber was delivered to PSI for the commissioning phase in summer 2018 and  integrated into the MEG~II detector in winter 2018. 
After installation, the CDCH was operated at PSI with engineering runs in $2019$ and $2020$ and with physics data taking runs in 2021 and 2022.

\subsubsection{DAQ and services}

High voltage (HV) is supplied with a commercial system by ISEG Spezialelektronik GmbH, made of mod. EHS F230 and EHS F430 boards in a WIENER crate with MPOD controller, for a total of 144 channels. HV channels are split between two cables by custom-built distribution boards, and eight wires are powered through each cable.

Ionisation signals are read at both ends of the chamber wires by \num{216} front-end (FE) boards with a bandwidth of \SI{\sim 400}{\mega\hertz}, which are connected to the wire PCBs.
The signals are digitised  at \SI{1.2}{GSPS} by the WaveDREAM boards of the integrated trigger and DAQ system WaveDAQ, described in \sref{sec:rtdaq}. 
The low voltage (LV, 5~V) is supplied to the FE boards via a dedicated distribution system through the WaveDREAM  boards. 
As the FE boards have a total power consumption of \SI{\approx 350}{\watt}, they have to be cooled with water and glycol by a cooling system 
built into the board holders and purged with dry air to avoid condensation of water vapour.

The gas is supplied by a dedicated gas system~\cite{Baldini:2018ing} that mixes helium with isobutane and oxygen in the required proportions. Isopropyl alcohol is added by passing a fraction of the helium flow through a thermostated, alcohol-filled bubbler.

\subsection{Final geometry and operation}
\label{sec:dchgeo}
The final geometry of the CDCH differs in some details from the original design. 
The chamber was designed to have ten layers of sense wires, but during the wiring phase there were delays in establishing the environmental 
conditions and procedures needed to eliminate the risk of wire breakage. 
Then, the outermost layer was not installed to fit into the schedule of the CDCH operation. 
We used Monte Carlo (MC) simulations to verify that the degradation of tracking efficiency and performance due to the missing layer is \SI{<1}{\percent}. 

First, using Garfield++ simulations \cite{garfield++}, the HV operating point was set in the range \SIrange[range-units = single]{1400}{1480}{\volt}, 
with the innermost layer at the highest voltage, to achieve a gas gain of $\sim$\num{5e5}.
This configuration gives a reasonable sensitivity to individual ionisation clusters, as confirmed by the distance of closest approach (DOCA) measurements presented in \sref{sec:dchhitrec}.
The HV can be set individually for each cell at \SI{10}{\volt} steps to account for difference in size of cell due to their radial position within the chamber. 
The mechanical tension of the wires, which is required to ensure electrostatic stability and to avoid short circuits between the anode and cathode wires, 
was also calculated using Garfield++ simulations. Several tests were carried out, both on a full-size prototype and on the CDCH itself. 
The required tension value was determined by stretching the wires by +\SI{5.2}{\mm} with 
respect to the nominal length (\SI{65}{\percent} of the the elastic limit), based on calculations. 

The gas mixture was also optimised in $2019$ and $2020$ to avoid corona discharges and current spikes and to restore normal operation after a sustained short circuit between an anode and a cathode wire.
The original He--isobutane gas mixture $\left(90:10\right)$ was modified by the addition of oxygen and isopropyl alcohol (\SI{1.5}{\percent}). The addition of oxygen is a particularly delicate operation
due to its electronegativity, 
as it can capture drift electrons. The oxygen level was initially set to \SI{2}{\percent} to dampen anomalous currents of \SI{\sim400}{\micro\ampere} and then gradually reduced to a stable value of \SI{0.5}{\percent} after normal current levels were restored. 

\begin{figure}[tb]
\centering
  \includegraphics[width=1\columnwidth,angle=0] {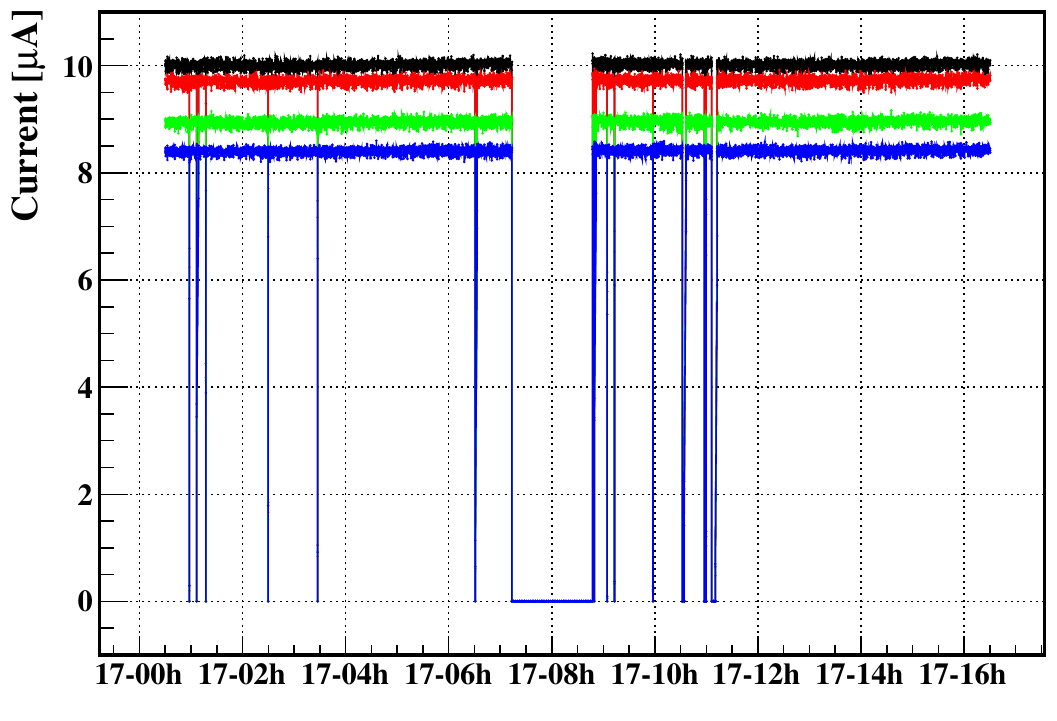}
  \caption{Typical currents drawn by six CDCH HV channels when the \muonp\ beam is on. Currents fall abruptly to zero when the beam goes off.}
 \label{dch:current}
\end{figure}

In \fref{dch:current} we show typical currents on six HV channels during normal MEG~II beam operation; each HV channel powers 16 sense wires.

\subsubsection{Issues and problems}
\label{sec:dchproblems}
The main problem encountered during the construction and commissioning of CDCH was the breakage of \num{107} cathode wires. 
The wire breaks were investigated in detail by microscopic inspections, chromatography, SEM/EDX analyses and immersion tests in water, 
and were proved to be due to galvanic corrosion of the aluminium core, caused by air humidity penetrating through small cracks in the silver coating. 
A phenomenological model \cite{wireanalysis} was developed to predict the number of breaks
as a function of the time the wire is exposed to the ambient moisture and the elongation of the wire.
The problem was solved by keeping the CDCH in a controlled dry atmosphere during maintenance and operation. 
When these precautions were taken, the rate of wire breakage dropped to 0 so far.

The impact of \num{\sim 100} missing cathodes (\SI{< 1}{\percent} of the total) was assessed by using MC simulations and found to be negligible. Nevertheless, the presence of broken wire fragments within the CDCH is dangerous as they can cause short circuits affecting several sectors of the chamber. All fragments were carefully removed with a dedicated tool, but one of them caused the strong short circuit cited in \sref{sec:dchgeo}, which increased the chamber current up to \SI{\sim 400}{\micro\ampere}. 
A dedicated optimisation procedure for the gas mixture was necessary to restore normal current values. 

\begin{figure}[tb]
\centering
  \includegraphics[width=1\columnwidth,angle=0] {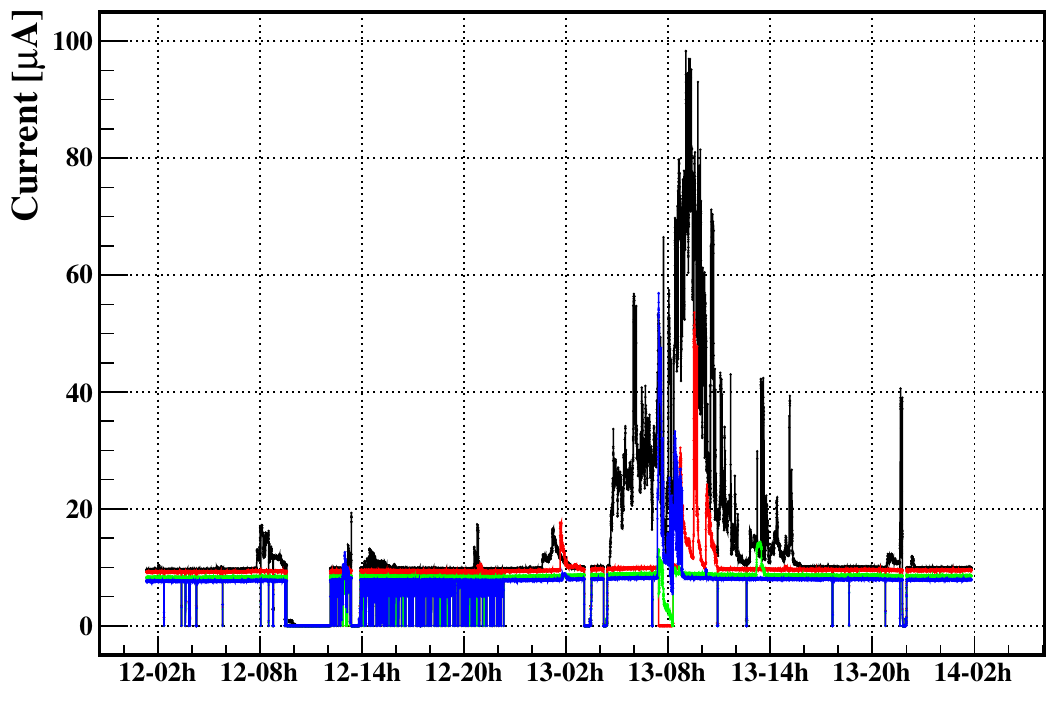}
  \caption{Currents in four sectors of layer 1. Spikes are due to shortage of isopropyl alcohol.}
 \label{dch:anomaly}
\end{figure}

During the standard data taking it is very important to keep the isopropyl alcohol level stable, to avoid current spikes, by continuously bubbling it within the chamber; an example of the current spikes occurring in case of a shortage of isopropyl alcohol is shown in \fref{dch:anomaly}. 
An additional problem is the noise caused by occasional short circuits on FE boards. Because the FE boards are densely mounted on the back of the chamber end plates, 
small movements of a board can cause short circuits between the capacitors on the board and the aluminium support structure of the board itself; 
such short circuits cause oscillations in the amplifier circuits that can affect the behaviour of large parts of the chamber.
These oscillations can only be stopped by switching off the LV of the boards, but since the LV is distributed in groups of eight channels, 
a single noisy channel will cause eight channels to fail. This problem was solved by inserting suitable insulating plastic blocks
to ensure the proper separation between the board elements and the metallic parts of the support structure.  
\subsection{Reconstruction algorithms}
\label{sec:dchhitrec}
The purpose of the CDCH analysis is to identify and reconstruct the positron tracks 
using the information on the sense wires provided by the ionisation clusters. 
The CDCH operates in a high-rate environment; the hit rate per cell can be \SI{>1}{\mega\hertz} 
at the innermost wires for $R_{\muon} = \SI{5e7}{\per\second}$, 
corresponding to a cell occupancy of \SI{25}{\percent} in the time window of maximum drift time.
Moreover, the geometrical characteristics, with only nine layers and small stereo angles, makes the track finding complicated.

A more detailed description on the reconstruction algorithms is available elsewhere \cite{CDCHPerfor}.

\subsubsection{Waveform processing and hit reconstruction}
\begin{figure}[tb]
\centering
  \includegraphics[width=1\columnwidth,angle=0] {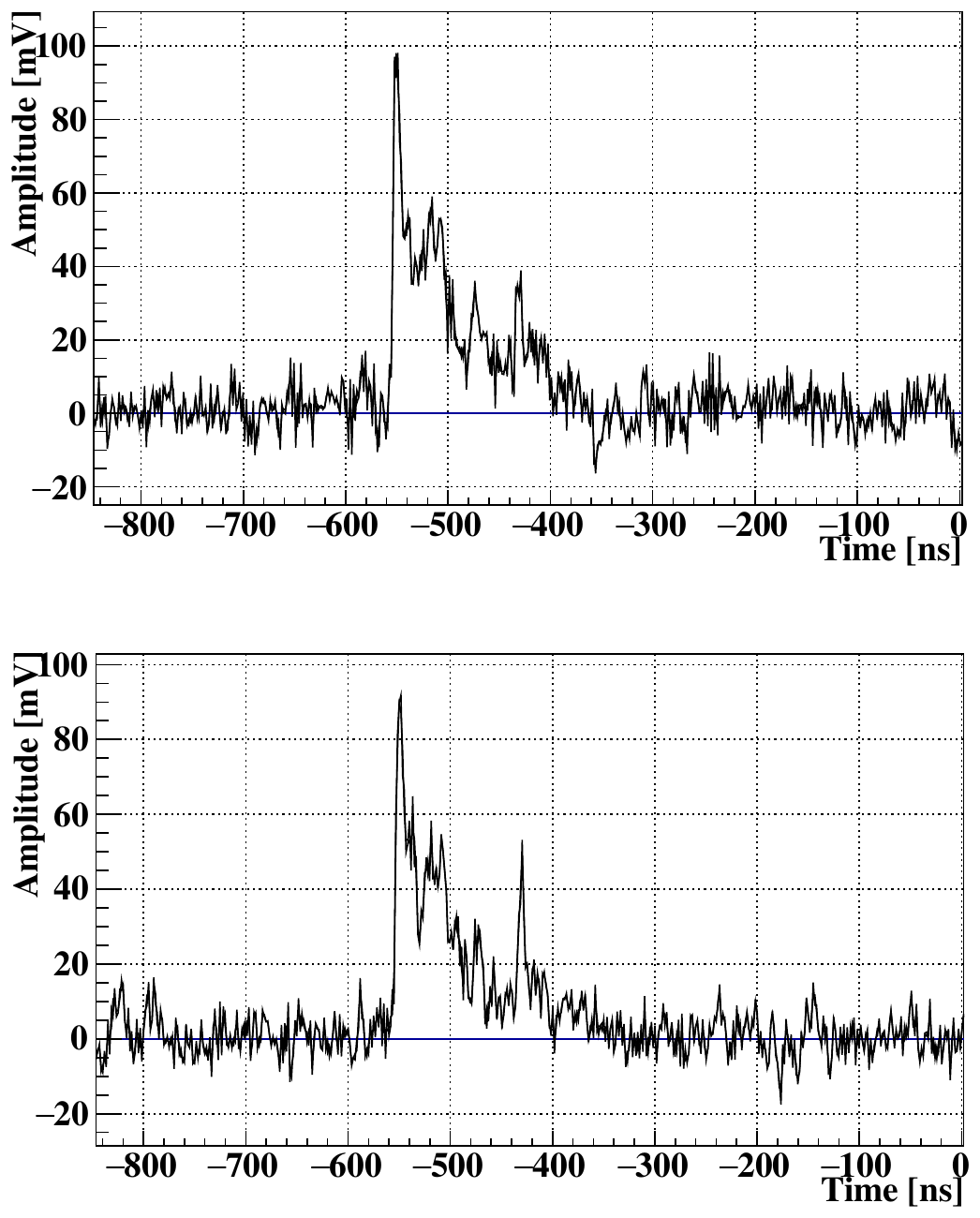}
 \caption{Example of typical waveforms in a drift cell crossed by a \positron\ track after the noise reduction applied. The top plot is read by the upstream electronics, the bottom by the downstream.}
 \label{dch:waveform}
\end{figure}

The first step is to identify the signals induced by drift electrons
in the waveforms of the cells traversed by positrons; 
such signals are called \lq\lq hits\rq\rq. \Fref{dch:waveform} shows typical waveforms due to a hit, which consists of several temporally separated pulses from different ionisation clusters that are stretched by the slow drift time of the electrons. For good hit reconstruction efficient signal-to-noise discrimination and pile-up identification are required.
Two waveform processing algorithms have been developed to detect hits. 

A coherent low frequency noise was observed mainly over adjacent 16 wires. 
Therefore, the first algorithm starts with coherent noise subtraction using the averaged waveform for adjacent channels excluding the region with signal pulses.
This reduces the noise level (FWHM) from \SI{23}{\mV} at \SI{13}{\mV}.
Significant incoherent high frequency noise is observed for \SI{>200}{\mega\hertz}, while the signal power in this frequency range is negligible.
Therefore, we apply high-frequency cut-off at \SI{225}{\mega\hertz} using a discrete Fourier transform technique.
These noise reduction algorithms have been optimised by maximising the number of hits per track and the tracking efficiency, and minimising the chi-square of track fitting.

Hit detection is based on a fixed voltage threshold for two adjacent sampling points and a fixed threshold for the integration over \SI{20}{\ns} from the two initial points.
After the hit is detected, the thresholds are lowered to look for a low-amplitude cluster pulse before the detected pulse.

The second method uses a deep-learning algorithm based on a convolutional neural network (CNN). 
The network model accepts waveforms from eight neighbouring cells as input to learn the pattern of the coherent noise as well as that of the signal
and outputs the probability of the first cluster arrival time of a hit at each sampling point.
It was trained with samples of simulated waveform data, where hits are added randomly at the expected rate for $R_{\muon} = \SI{3e7}{\per\second}$, 
overlaid with real noise data taken without beam.

Combining the results by the two methods results in a higher hit efficiency but also a higher fake hit rate than the first method. 
To make the best use of the results of the two methods, the following reconstruction
is repeated twice, with only the hits found with the first method and with the hits 
found with the two methods after combining them. The results are combined after 
the reconstruction is completed.
If the reconstruction is successful with both methods, the higher quality tracks are selected.
This approach improves the final tracking efficiency (see \sref{sec:dchperformeff}), 
by a factor of \num{1.26} compared to applying only the first method, but at the cost of 
4 times the computational time for the track finding process described in \sref{sec:track_rec} due to the higher number of hits.

{\color{black}
The difference in the arrival times ($\alpha$) and the ratio of the amplitudes ($\beta$)
between the signals measured on the two ends of a wire 
provides information about the $z$-coordinate of the hit along the wire. 
Therefore, once a hit is detected in at least one of the two waveforms from a wire, these values are computed by minimising the following chi-square function,
\begin{align}
\chi^2(\alpha,\beta) = \sum_{k} 
 \frac{ \left[ v_i(t_k) - \beta\cdot v_j(t_k - \alpha) \right]^2}{ \sigma_i^2 + \sigma_j^2} ,
 \label{eq:cross-fit}
\end{align}
where $i$ and $j$ are indices for the wire ends (0: upstream end, 1: downstream end, and $i\neq j$), $v_{i}(t_k)$ is the waveform voltage at the $k$-th sampling point on the end where the hit is detected  while 
$v_{j}(t_k - \alpha)$ is the voltage at the time $(t_k - \alpha)$ on the other end of the wire,\footnote{The voltage between points is calculated with a linear interpolation.} and
  $\sigma_{i(j)}$ is the RMS noise on the waveform.
The index  $k$ runs for the points in the range $[\SI{-20}{\ns}, \SI{10}{\ns}]$ around the detected signal timing.} 
The resolution of the $z$-coordinate is several centimetres. 
Although this is moderate, it helps to ensure that the track finding process is efficient and robust against pile-up. The $z$-coordinate 
resolution is improved by exploiting the stereo configuration of the wires in the tracking stage. 

Another piece of information provided by the digitised waveforms is the arrival time of the ionisation clusters on the sense wires. 
Since multiple clusters are frequently generated on one waveform, the time of the first cluster must be identified to correctly reconstruct the drift circle.
This is measured from the summed waveform of the two ends after adjusting the relative timing of the two.
The arrival time of cluster is the sum of the common track time ($T_{0}$) and the drift time of the cluster, 
where $T_0$ is measured by pTC measurement (see \sref{sec:pTC}). 
The drift time of the first cluster is converted to the DOCA using the time-distance relationship (TXY tables) described in \sref{sec:track_rec}.
 


\subsubsection{Track finding and fitting}
\label{sec:track_rec}
After identification and reconstruction, the hits are fed into a pattern recognition algorithm (track finder), followed by a track-fitter algorithm, both based on Kalman filters. 
The former combines hits belonging to the same positron track into a track candidate with a preliminary estimate of the positron's kinematics.
The latter reconstructs the complete trajectory of the positron and provides the best estimate of the kinematics at the target:
the positron energy $\epositron$, the emission angles 
($\thetae,\phie$) and the coordinates of the intersection of the track with the target, $(\xpos, \ypos, \zpos)$.

The track-finding algorithm is based on a track following method that starts from hit pairs in outer layers, where occupancy is lower.
All compatible combinations of two pairs in different layers form a set of track seeds.
Each seed is propagated backward to the adjacent layers, with checking the consistency between the track and hits and updating the track parameters using the Kalman filter algorithm, until the innermost layer, and then propagated forward with attempting to find additional compatible hits to form full single turn track candidates. A track candidate is required to have at least seven hits.

The track fitter uses an extension of the Kalman filter, namely the deterministic annealing filter (DAF) \cite{daf}
with iterative weighting and annealing process, implemented in the GENFIT package \cite{genfit1,genfit2} with a proper treatment of the material effect. 
The left/right ambiguity is also resolved by DAF. 
The fitter first fits individual track candidates from the track finder and then merge the fitted segments to form full multi-turn tracks inside CDCH.
Then, the tracks are propagated forward to the pTC and backward to the target.
Once the track matches a pTC cluster, the track is re-fitted with updated DOCA using the best estimated $T_{0}$, with a correction of the time of flight from each hit to the pTC, and the best method of DOCA estimation discussed bellow. 
During this re-fitting, missing hits that the track finder was unable to associate to the track are searched for. 
Frequently, hits in the final half turn are missed by the track finder but can be added in this process, resulting in improved momentum resolution.

The deviation of the target from planarity is modelled by a triangle mesh in the fitter. The track is first propagated to a virtual plane, a few mm in front of the real target, and then to the nearest triangle of the mesh, where the positron's kinematics at the emission is reconstructed.
The target crossing point  ($\xpos,\ypos,\zpos$) is used as a candidate of muon decay point to correlate the track direction with the first interaction point of the \photon-ray observed in the LXe  detector. 
The length of the trajectory from the target to the matched pTC counter is converted to the time of flight and subtracted from the pTC hit time to determine the positron emission time at the target $\tpositron$.


During the tracking process, the DOCA of each hit is iteratively refined. 
The DOCA value is estimated at the beginning using angle-averaged TXY tables
extracted using Garfield++ simulations, which determine the drift velocity and lines taking 
into account the electric and magnetic fields, the ionisation pattern, the diffusion process, etc.
The Garfield++ simulations are performed in two-dimensions in $(r, \phi)$ sampling at slices at fixed $z$-coordinates for each layer and then are interpolated
to take into account the longitudinal change in cell size and shape as well as the effect of the magnetic field. 
Once the track is reconstructed, the DOCA is recalculated taking into account the cell crossing angle of the track.  
This DOCA estimation is biassed by the low cluster density; the small number of clusters within the cell overestimates the hit distance from the wire. 

Another DOCA estimate less prone to intrinsic biases is obtained by neural network approaches \cite{DocaUCI}.  
The networks take as input various hit properties (wire and plane number, charge, timing, hit coordinates, track angle, $T_0$ etc.) and waveforms (for the CNN model) and train on  the fitted track DOCA (made using the Garfield++ TXY tables) as an estimator of the true DOCA to create a ``data-driven'' TXY table. 
This TXY optimally accounts for ionisation statistics biases, removes errors from the simulated TXY, and uses information from all ionisation clusters. 

\begin{figure}[tb]
\centering
  \includegraphics[width=1\columnwidth,angle=0] {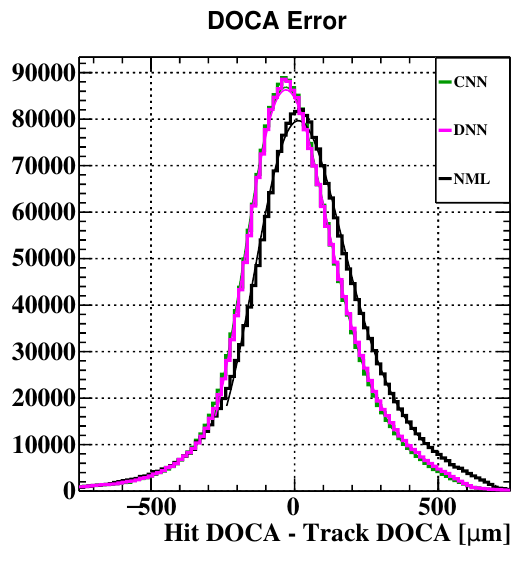}
  \caption{The distribution of the DOCA residuals (hit DOCA $-$ track DOCA) estimated with the conventional (NML) and the two (DNN and CNN) neural network based approaches. The curves are fitted double Gaussian functions
  with a core sigma of \SI{114}{\micro\metre} (\SI{119}{\micro\metre}) and a tail sigma of
  \SI{236}{\micro\metre} (\SI{259}{\micro\metre}) with a core fraction of \num{0.59} (\num{0.70}) for the neural network (conventional) approaches.}
 \label{dch:docaresiduals}
\end{figure}

The distribution of DOCA residuals for the conventional and two neural network based approaches, a dense neural network (DNN) and a CNN, are compared in \fref{dch:docaresiduals}.  
The best result is obtained with the CNN that processes the waveforms from all ionisation clusters.
The main improvement from the neural network is the suppression of the positive right tail presumably from the suppression of the ionisation statistics bias.
This improves the positron kinematics by \SI{\sim10}{\percent} as shown in \sref{sec:dchperformreso}.
Therefore, the DOCA is finally updated with the CNN method in the re-fit process.
\subsection{Alignment}
\subsubsection{Wire alignment}
\label{sec:dchcalib}
The global position of CDCH was measured in both $2021$ and $2022$ with an optical survey. The information on the relative wire-by-wire alignment was extracted from several measurements during the construction of the chamber.
However, after the complete reconstruction of the track, the distributions of the difference between the nominal position of the wires and the wire position calculated by the tracking algorithm using the measured DOCA show systematic deviations of the order of \SI{\sim 100}{\micro\meter}, which worsen the tracking resolutions. The deviations were resolved by implementing a track-based alignment.
%

Two types of tracks can be used to improve relative alignment: Michel positrons and cosmic rays.
The former have the advantage of being collected and reconstructed during normal data taking, 
but also the disadvantage of requiring a more complex reconstruction and being more affected by possible uncertainties in the knowledge of the magnetic field. 
The latter have the advantage of being straight tracks, but the disadvantage 
of requiring dedicated data taking periods and a different coverage of the tracking volume. 
Currently, alignment to cosmic rays is being investigated;
in the following we only present the alignment based on Michel positrons.  

The alignment procedure is an iterative adjustment of the wire coordinates 
driven by the mean residual. The residual $r$ is fitted, wire by wire, as a 
function of the longitudinal position along the wire ($z$-coordinate) with a parabolic shape:
\begin{align}
r \left( z \right) = p_{0} + p_{1} z + 
p_{2} \left[ \left( \frac{2z}{L} \right)^{2} - 1 \right].
\label{eq:dchalign}
\end{align}
The parameter $p_{0}$ corresponds to a global wire displacement, the linear 
term results from the inclination of the wire to the chamber axis and the 
quadratic term takes into account the wire sagitta, due to the electrostatic 
and gravitational forces acting on the wire. 
$p_{2}$ is the (absolute) maximum value of the sagitta, which can reach \SI{\sim 100}{\micro\meter}. 

The number of events required for high quality alignment of a single wire 
depends strongly on the position of the wire in the chamber due to the trigger criteria to select \megp\ candidate events. 
%
Furthermore, the alignment algorithm is almost insensitive to alignment errors 
when the track direction is parallel to the misalignment vector. 
This is because the DOCA estimate 
does not change when the wire position is shifted along the track direction. 
Therefore, chamber sectors crossed by a large variety of angles are aligned more efficiently than those crossed with a small scatter of angles.


The alignment procedure was based on \num{17} million hits, with an average \num{40} hits per track.
We required \num{>5000} hits for each wire to be aligned; 152 wires were  excluded from this alignment and the survey results are used.
They are at the edges of each layer or with electronic problems.
With larger data statistics, the alignment for the edge region can be improved. 
One way to increase the statistics for the region 
is to use tracks outside the time window that do not cross the pTC
but cross more frequently the CDCH regions outside or at the boundaries of the 
MEG~II geometric and trigger acceptance. These tracks 
are collected accidentally when another event that meets the trigger conditions opens the WaveDREAM time window.  

\begin{figure}[tb]
\centering
  \includegraphics[width=1\columnwidth,angle=0] {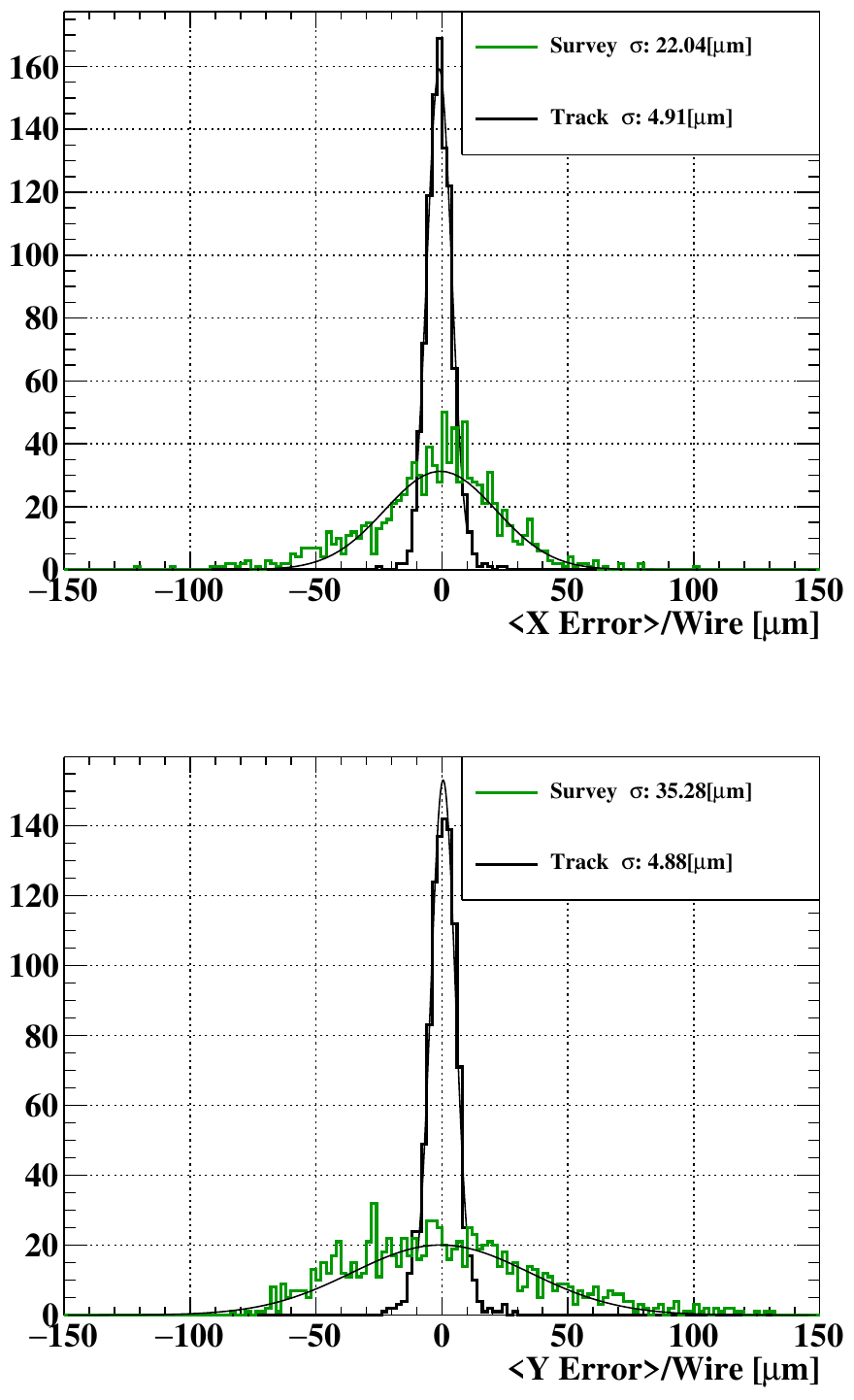}
  \caption{The mean residuals in the $x$- (top plot) and in the $y$-coordinates (bottom plot). The black curves are obtained after \num{14} alignment iterations, the green curves using the alignment based on the survey. The curves are the results of Gaussian fits.
  }
 \label{dch:xalign}
\end{figure}

 \Fref{dch:xalign} shows the final mean residuals in the $x$- (top plot) and $y$-coordinates (bottom plot)  compared to the same residuals obtained using the survey-based alignment.
The spreads of the Gaussian cores
of the two distributions are $\sigma^{\rm{sur}}_x=\SI{22}{\micro\meter}$ and $\sigma_y^{\rm{sur}}=\SI{35}{\micro\meter}$ when the survey-based alignment is used and decrease to $\sigma_{x,y}^{\rm{trk}}<\SI{5}{\micro\meter}$ for both coordinates after the alignment procedure.

The fit parameters $p_{0}$ (global displacement) and $p_{2}$ (sagitta) are 
strongly correlated, as expected. The remaining sagitta is 
$\sigma_{p_2}\sim\SI{13}{\micro\meter}$ and the wires with the largest sagittas are concentrated in the peripheral chamber sectors, crossed by fewer tracks than central sectors. 
The distributions of the displacements of the wire centres with respect to those 
in the survey have $\sigma_x \sim \SI{100}{\micro\meter}$, $\sigma_y \sim \SI{80}{\micro\meter}$ and $\sigma_z \sim \SI{30}{\micro\meter}$. 
The distributions of the differences of the wire angles $\Theta$ and 
$\Phi$ with respect to the survey have $\sigma_\Theta\sim\SI{0.1}{\milli\radian}$ 
and $\sigma_\Phi\sim\SI{1.1}{\milli\radian}$, respectively.
For the $\Phi$ angle the mean value is \SI{1.05}{\milli\radian}, 
which means that the alignment requires a global azimuthal rotation, while the required polar rotation is minimal, i.e. \SI{<0.1}{\milli\radian}.

The remaining error in wire-alignment is estimated to be between \SI{2}{\micro\meter} at the centre and \SI{15}{\micro\meter} at the end plates. 
Some correlations between the wire centre translation or the wire rotation and 
the wire number layer-by-layer have been observed and are currently being investigated.

\begin{figure}[tb]
\centering
  \includegraphics[width=1\columnwidth,angle=0] {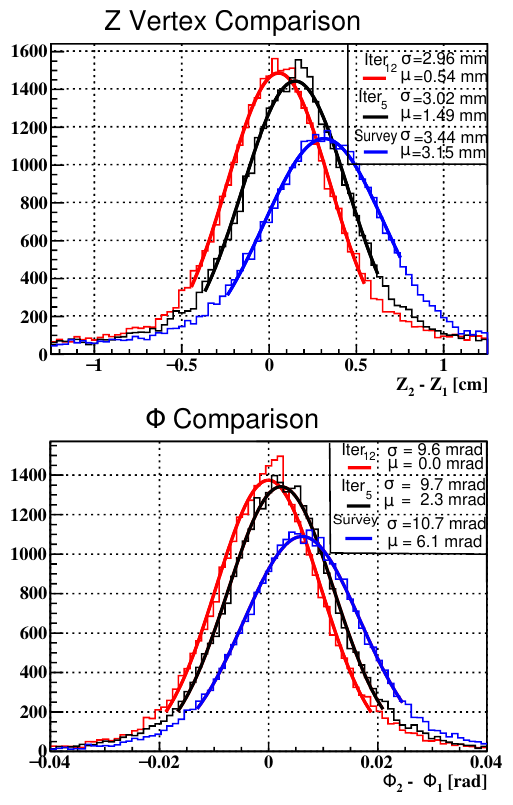}
  \caption{The double-turn analysis results for $\zpos$ and $\phie$. The blue curves are obtained 
  using the survey-based alignment, the black and the red curves using the 
  alignment algorithm results after $5$ and $12$ steps, respectively.}
 \label{dch:DTalign}
\end{figure}

The effects of the alignment procedure on the resolutions of the kinematic variables  
are evaluated using the double-turn method; see \sref{sec:dchperformreso} for the details. 
The distributions of $\zpos$ and $\phie$ are shown in \fref{dch:DTalign}.
%
The advantage of the alignment procedure is highlighted by the strong reduction of the systematic biases and by the narrowing of the distributions. 

\subsubsection{Relative alignment between magnetic field and CDCH}
\label{sec:bfield-cdch}
Due to the gradient magnetic field, misalignment between the CDCH and the magnet 
results in a non-uniformity of the energy scale with respect to the positron emission angle.
In the track reconstruction, we use a 3D map of the magnetic field calculated using a finite element method based on the measured coil dimensions and taking into account the thermal shrinkage of the coil.
The angular dependence is minimised by shifting the calculated magnetic field by $(\SI{100}{\micro\meter}, \SI{700}{\micro\meter}, \SI{300}{\micro\meter})$
from the nominal position, with an 
estimated alignment accuracy of \SIrange[range-phrase=--,range-units = single]{100}{200}{\micro\meter}. 
\begin{figure}[tbp]
  \centering
  \includegraphics[width=\linewidth]{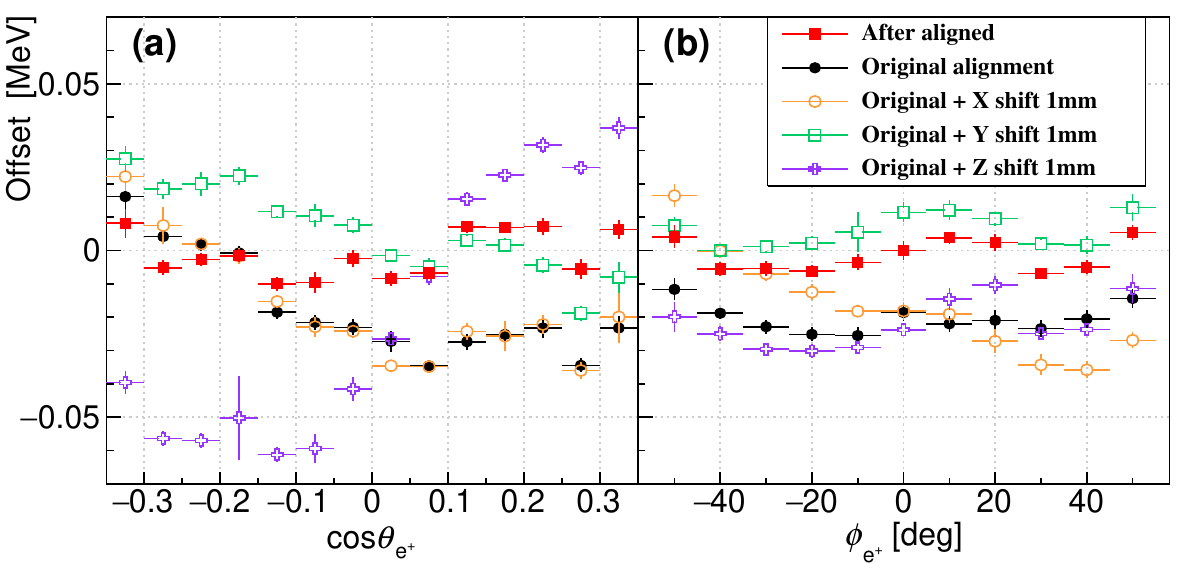}
  \caption{The angular dependence of the positron energy scale
  versus the angular kinematic variables before and after alignment.
  The offset on the $y$-axis is the difference of the measured value with the expected value of the Michel edge. 
  The three superimposed plots show the effects of shifting the magnetic field by \SI{1}{\mm} in $x$, $y$, and $z$. }
  \label{fig:BFieldAlignment}
\end{figure}
After the shift, there is no more bias as shown in
\fref{fig:BFieldAlignment}, and the scatter
of the order of \SI{\sim 10}{\keV} is negligible compared to the energy resolution.

\textcolor{black}{The misalignment in the positive $x$ direction results in a decreasing energy scale versus $\phie$ and $\cos\thetae$.
The misalignment in the positive $y$ direction results also in an decreasing energy scale versus $\cos\thetae$.
The misalignment in the positive $z$ direction results in an increasing energy scale versus $\cos\thetae$. }

\subsubsection{Relative alignment between target and CDCH}
\label{sec:target-cdch}
\begin{figure}[tbp]
  \centering
  \includegraphics[width=\linewidth]{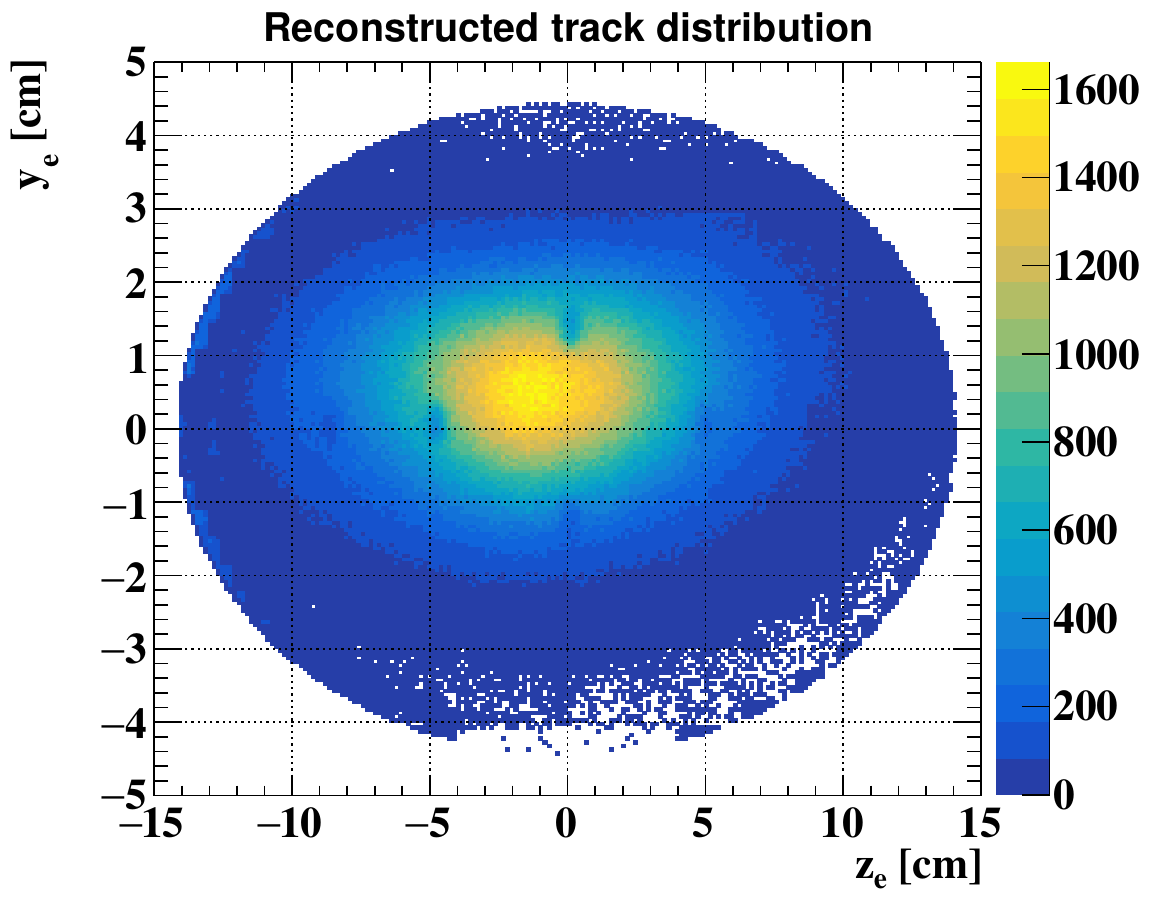}
  \caption{ Distribution of the reconstructed positron origin ($\ypos$,$\zpos$) on the target.}
  \label{fig:PositronTGTDist2D}
\end{figure}

The precise alignment of the muon stopping target with respect to the spectrometer is crucial in determining the positron emission angle and position as discussed in detail in \sref{sec:target}.
We use the holes on the target to align the target with respect to the CDCH. 
The six holes are visible \textcolor{black}{(as shortage of events)} in the reconstructed track distribution in \fref{fig:PositronTGTDist2D}, from which the $y$- and $z$-coordinates of the holes (thus those of the target) can be easily estimated.
\textcolor{black}{With reference to \fref{fig:pattern} the hole centres are located approximately at \SI{\pm1}{\cm} along the short axis and 
 \SI{\pm5}{\cm} and \SI{\pm8.5}{\cm} along the long axis.}

\begin{figure}[tbp]
  \centering
  \includegraphics[width=\linewidth]{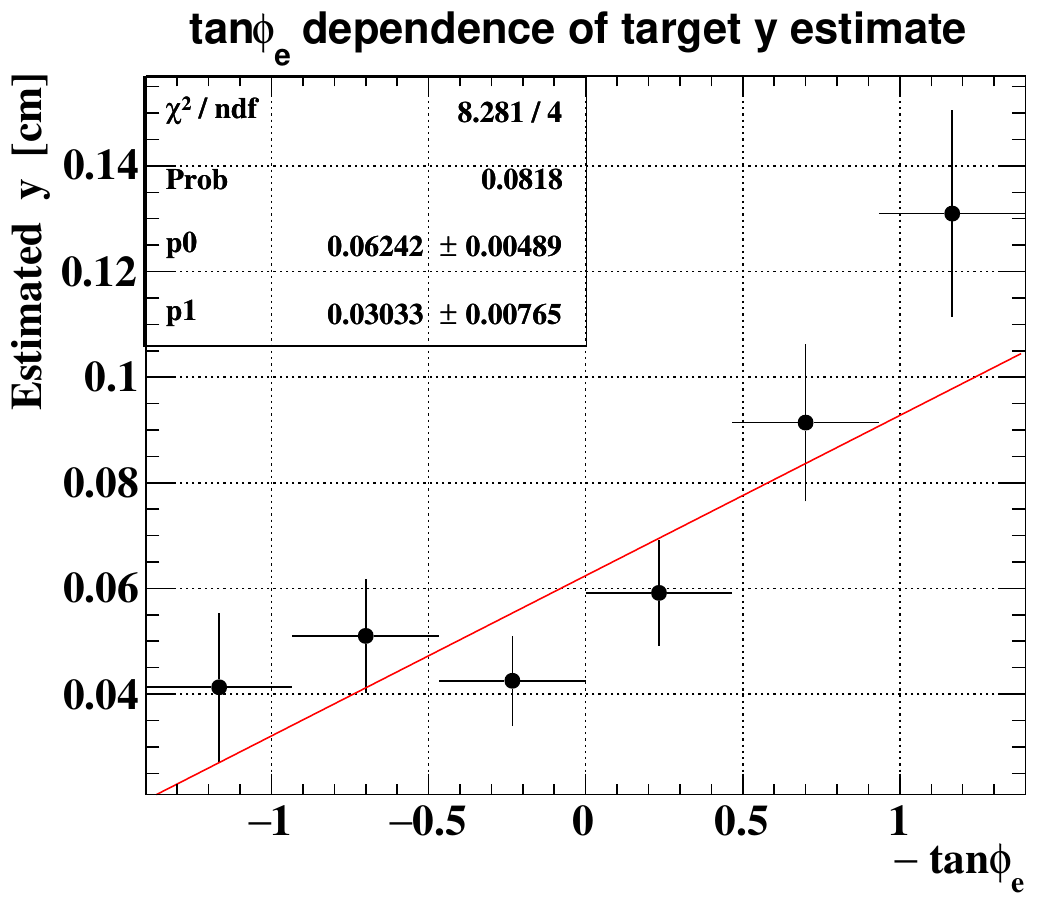}
  \caption{Estimated hole $y$ position as a function of $\phie$. The slope parameter in the fit gives the residual $x$ misalignment while the offset parameter gives the $y$ offset of the target.}
  \label{fig:TGTXYAlignment}
\end{figure}

The $x$-alignment exploits the dependence of the estimated $y$-position on $\phie$, 
as shown in \fref{fig:TGTXYAlignment}.
Combined with the photogrammetric method using the photo-camera system, which corrects for temporal variation in the target positions, the hole analysis yields the hole-by-hole residual misalignment.

The precision of the hole-by-hole position estimation is limited both by the statistical uncertainty of the positron tracks and by the systematic uncertainties of the method. 
The statistical uncertainty is \SIrange[range-phrase=--,range-units = single]{100}{200}{\micro\meter} in each coordinate.
All the systematic uncertainties originate from the non-uniform vertex position distribution due to the beam profile, which biases the $(y,z)$ distribution produced by the holes to be closer to the beam centre.
This effect is corrected in the hole position estimation, but the uncertainties in the correction factor dominates the hole-by-hole systematic uncertainties of 
$(\SI{50}{\micro \meter}, \SI{100}{\micro \meter}, \SI{200}{\micro \meter})$.

The global target geometry, both the rotational and translational parameters, is finally fitted to the residuals of the holes, yielding 
a translation of $(\SI{73}{\micro\metre}, \SI{799}{\micro\metre}, \SI{437}{\micro\meter})$ and a rotation of \SI{10}{mrad} mainly along the minor axis of the target from the results of the photogrammetric method. 
The accuracy, including systematic uncertainties, is \SI{100}{\micro\meter} for the translation in each axis and \SI{1.4}{mrad} (\SI{6}{mrad}) for the rotation along the minor (major) axis.

\subsubsection{Relative alignment between LXe detector and CDCH}
\label{sec:dchxcali}
The alignment procedure described in \sref{sec:dchcalib} is insensitive to a global misalignment between CDCH and other subdetectors. 
Among them, the LXe detector is especially important since it provides information about the \photon-ray direction.

Events that result with hits in both the CDCH and the LXe detector are used to compare information from both.
The appropriate event category is the cosmic rays,
which can release energy in the LXe detector and then (or vice versa, depending on the trajectory) 
pass through the CDCH, producing hits on multiple wires. The cosmic ray events are collected 
with a specific triggering scheme and with the magnetic field turned off, 
to obtain straight tracks that are easier to reconstruct and fit. 
The idea is to select cosmic ray tracks that cross the LXe detector inner face 
almost perpendicularly and to determine the coordinates of this crossing 
point using the LXe detector and the CDCH information independently.

On the CDCH side, a method based on the Legendre transform \cite{fruhwirth2021pattern,Primor_2007} is used for finding and fitting the cosmic ray tracks. The fitted tracks are then extrapolated to the inner face of the LXe detector. 

On the LXe detector side, the usual position reconstruction algorithm (see \sref{subsubsec:LXec-position}) is applied. 
The selection of cosmic ray tracks almost perpendicular to the detector entrance 
face is motivated by the fact that the continuous stream of energy release by 
such cosmic rays is seen from the photosensors on the inner face as a point-like 
energy deposit, whose position is more reliably reconstructed. 

\begin{figure}[tb]
\centering
  \includegraphics[width=1\columnwidth,angle=0] {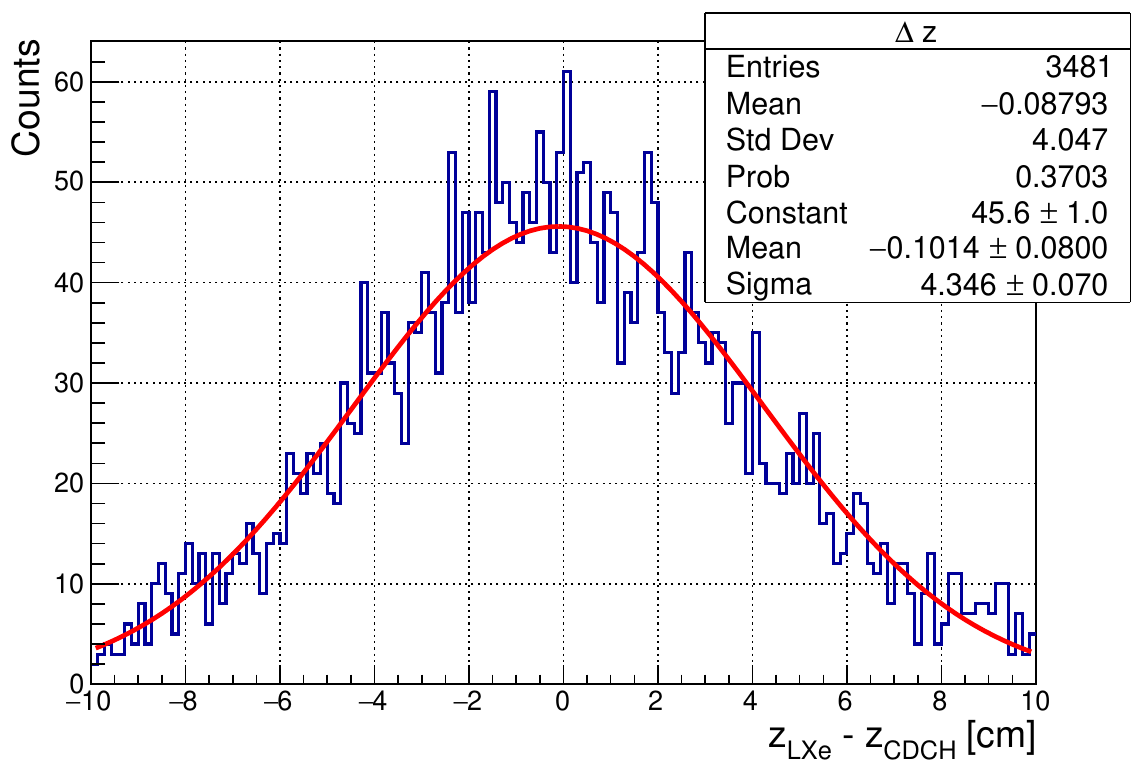}
  \caption{Difference in the reconstructed $z$-coordinate on the entrance face of 
  the LXe detector for cosmic rays performed by using the LXe detector and 
  CDCH information independently. The red curve is a Gaussian fit.}  
\label{fig:dchxecali}
\end{figure}

\Fref{fig:dchxecali} shows the LXe--CDCH difference in the reconstructed 
$z$-coordinate $\mathrm{\Delta} z = z_\mathrm{LXe} - z_\mathrm{CDCH}$.
The distribution has a Gaussian shape, centred at $\mathrm{\Delta} z = \SI{ -1.0 \pm 0.8}{\milli\meter}$.
This result shows that a small global shift can be applied to better align the two detectors even if it is compatible with zero.

We do not observe any significant deviation in the difference in $\phi$ within its uncertainty, which is more subject to systematic errors in the measurement because of non-uniform distribution of the cosmic rays with respect to $\phi$. 

The relative alignment between the LXe detector and the CDCH, combined with the relative alignment between the CDCH, the magnetic field and the target, is sufficient to determine the relative positron--photon angle, which is independent of the absolute position of the detectors.

For practical reasons, we need to define an absolute alignment. Reference marks on the COBRA cryostat are used to define the absolute MEG~II reference frame. The  position of the CDCH in this reference frame is determined by an optical survey with an estimated resolution of a few hundred micrometres, while the absolute position of the LXe detector is measured with uncertainties of \SI{\sim 600}{\micro\metre} as discussed in \sref{subsubsec:LXec-position}.
On this basis, we decided to use the CDCH position as a reference and apply global shifts to the LXe detector, the magnetic field and the target, according to the results of the data analysis for the relative alignments. 
The shifts of the magnetic field and target, reported in \sref{sec:bfield-cdch} and \sref{sec:target-cdch}, are quite similar in size and direction, which could indicate a bias in the CDCH survey. The shift in the $z$ direction is also confirmed by the relative alignment between the CDCH and the LXe detector, but with a large uncertainty.
For these reasons, the shift to be applied to the LXe detector was determined by combining all available information on a possible bias in the CDCH survey. We obtained $(\SI{90}{\micro\metre},\SI{800}{\micro\metre}, \SI{630}{\micro\metre})$, with a systematic uncertainty of \SI{1}{\mm} assigned in all directions.

%
%
\subsection{Performance}
\label{sec:dchperform}
In this section, we discuss the performance of the CDCH in terms 
of resolution and tracking efficiency. A more extensive discussion is given in a dedicated paper \cite{CDCHPerfor}.
We recall that the main motivation
for building the CDCH was the 
unsatisfactory performance of the MEG drift chamber system in terms of angular and 
energy resolutions and tracking efficiency due to its segmented structure. 
\subsubsection{Resolutions}
\label{sec:dchperformreso}

One technique for evaluating the resolutions of the kinematic variables is the \lq\lq double-turn\rq\rq~method, which was already used in the MEG experiment\cite{baldini_2016}. In the MEG~II experiment, \SI{\sim 15}{\percent} of the positron tracks cross the chamber volume five times, passing through \numproduct{9 x 5} sense wire layers.
In these tracks, one can identify two separate track segments (\lq\lq turns\rq\rq), the first of which corresponds to two chamber crossings 
and the second to three chamber crossings. 
Both track segments are independently fitted and propagated to a plane parallel to the target between the two turns and the distributions of the differences between the kinematic variables reconstructed by the two segments are compared. 

\begin{figure}[tb]
\centering
  \includegraphics[width=1\columnwidth,angle=0] {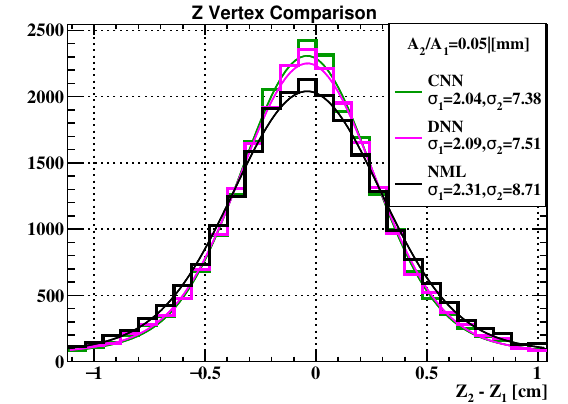}
  \caption{The distribution of the double-turn difference for $\zpos$ obtained using the conventional (NML) and neural network (DNN and CNN) approaches to DOCA reconstruction fit with a double Gaussian.
  \textcolor{black}{$A_{1,2}$ are the amplitudes of the two Gaussians, the ratio is fixed to 0.05; $\sigma_{1,2}$ are their standard deviations in \unit{\mm}. }} 
 \label{dch:doubleturn}
\end{figure}

\Fref{dch:doubleturn} shows the double-turn distributions for $\zpos$; the shapes of the $\ypos$, $\phie$ and $\thetae$ distributions are similar. 
The three distributions of \fref{dch:doubleturn} are based on the conventional and the neural network based DOCA reconstructions described in \sref{sec:track_rec}. The neural network approaches improve the width of the distribution by \SIrange[range-phrase=--,range-units = single]{1}{11}{\percent}.

These distributions do not directly represent the effective resolutions of the CDCH, but a combination in quadrature of the resolutions of the first and second turn.
The following corrections are required to convert the double-turn results into reliable estimates of the real resolutions.
One is to correct for the bias in the double-turn evaluation \textcolor{black}{due to difference in magnetic filed and properties between the two turns and with Michel positrons}. Another is to correct for the difference between Michel positrons and the signal positrons \textcolor{black}{due to the different numbers of hits}.
Both correction factors were evaluated via MC simulations. 
Good similarity was obtained between the double-turn results on data and those on the Michel MC simulation, showing that we have a solid  knowledge of the response of the CDCH. 
In addition, we can correct for the correlation effect between the variables for the detection of \megp\ events as discussed in \Sref{sec:sensitivity}.
The correlation parameters are also evaluated with the double turn analysis.
The effective resolutions after the corrections are summarized in \tref{dch:resolutions}.

\begin{table}[tb]
    \caption{Effective resolutions (core $\sigma$s) for experimental data obtained by combining the double-turn results on data with the MC correction factors. \textcolor{black}{The design values are in \tref{sensitivity:perf}.}}
    \centering
  \begin{minipage}{1\linewidth}
   \renewcommand{\thefootnote}{\alph{footnote})}	
   \renewcommand{\thempfootnote}{\alph{mpfootnote})}	
\centering
    \begin{tabular}{c c c c } 
    \hline 
      $\yposerr \left( \unit{mm} \right)$ & $\zposerr \left( \unit{mm} \right)$ & $\phieerr \left( \unit{\milli\radian} \right)$ & $\thetaeerr \left( \unit{\milli\radian} \right)$\\
    \hline
      $0.74$ & $2.0$ & $4.1$ & $7.2$ \\
    \hline
    \end{tabular}
\end{minipage}
    \label{dch:resolutions} 
\end{table}

\begin{figure}[tb]
\centering
  \includegraphics[width=1\columnwidth,angle=0] {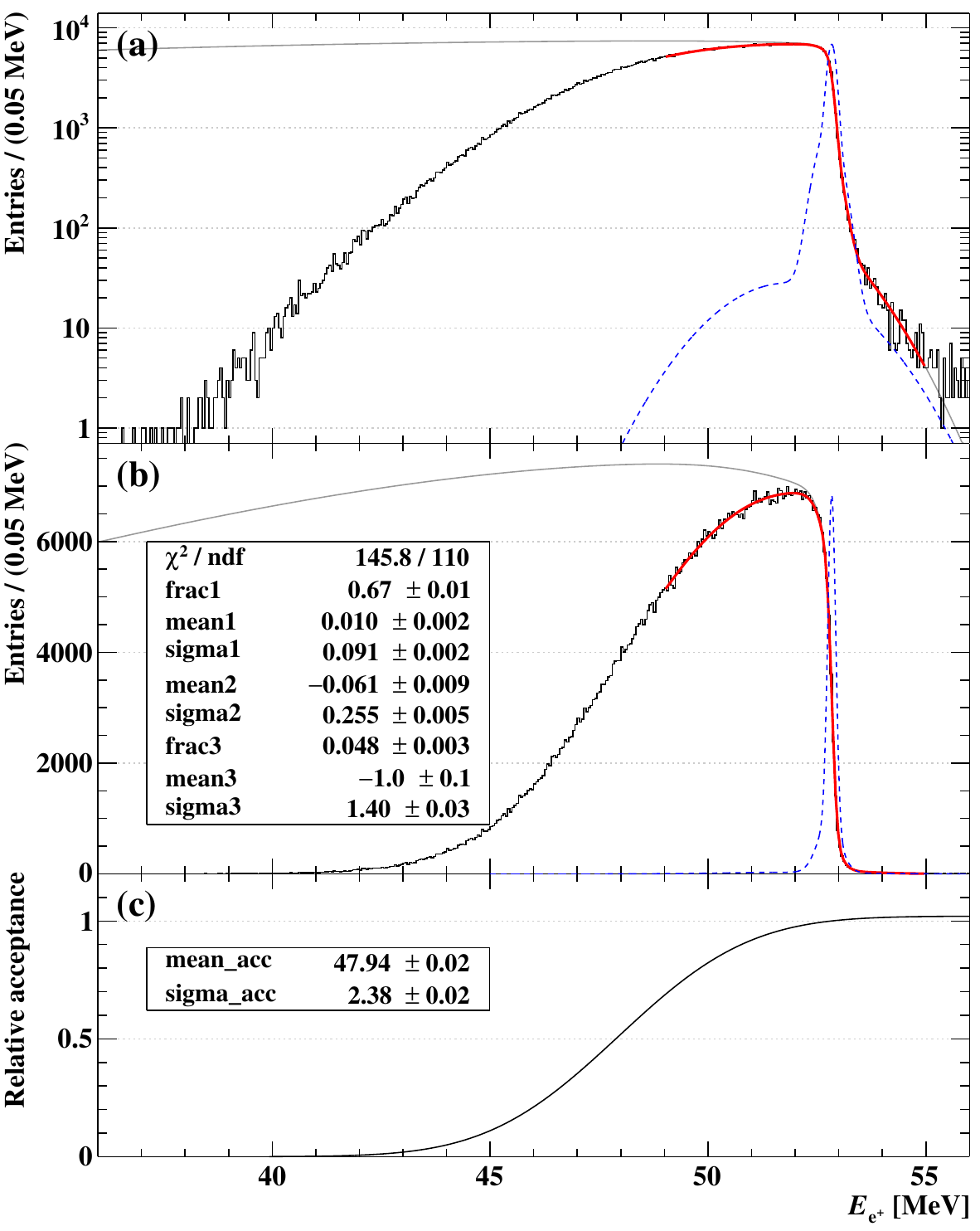}
  \caption{Fit of the Michel positron spectrum in logarithmic (a) and linear (b) scales. The black histogram is the experimental distribution, the blue curve is the sum of three Gaussian functions describing the resolution around the signal region and the red curve is the fitted function, obtained \textcolor{black}{from the theoretical spectrum multiplied by the acceptance curve shown in (c) and then convoluted with the resolution function.}  (c) The  acceptance curve of the spectrometer modelled with an error function.}
 \label{dch:michel}
\end{figure}

The energy resolution for experimental data is measured by fitting the theoretical Michel spectrum multiplied by an efficiency function that takes into account the high-energy selection of the spectrum by the spectrometer acceptance \textcolor{black}{and then convoluted with the} resolution function formed by the sum of three Gaussian functions to the measured Michel positron spectrum. The result is shown in \fref{dch:michel} in both logarithmic (a) and linear (b) scales.  
The $\sigma$ of the core Gaussian function, which accounts for \SI{\sim 67}{\percent} of the integral of the resolution curve, is $\epositronerr=
\SI{91}{\keV}$, better by \SI{40}{\keV} of the value quoted in the MEG~II proposal. The corresponding value for the MEG experiment was $\epositronerr=\SI{320}{\keV}$.
\subsubsection{Efficiency}
\label{sec:dchperformeff}
\begin{figure}[tb]
\centering
  \includegraphics[width=1\columnwidth,angle=0] {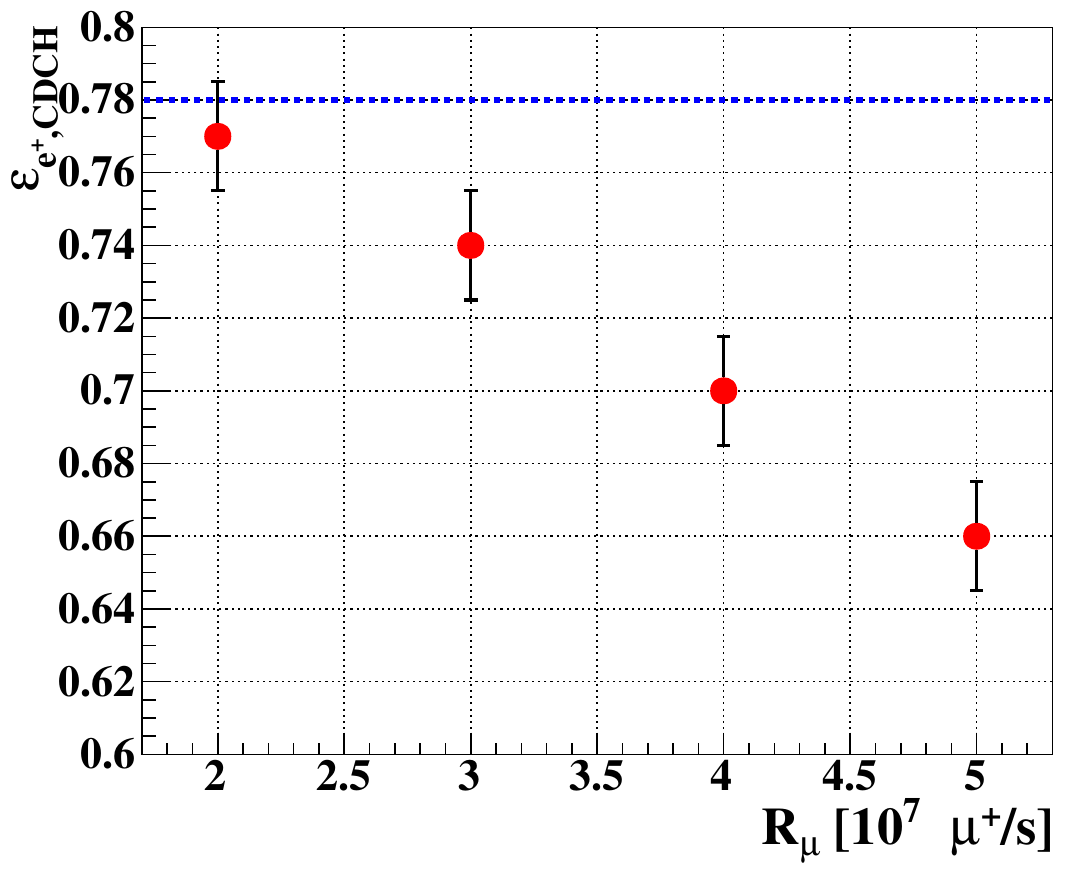}
    \caption{CDCH tracking efficiency as a function of $R_{\muon}$ for signal positrons. The blue dotted line is the design value.}  
\label{dch:cdcheff}
\end{figure}

\Fref{dch:cdcheff} shows the CDCH tracking efficiency for signal positrons $\varepsilon_{\positron, {\rm CDCH}}$ versus $R_{\muon}$. 
The efficiency is defined as the ratio of the number of reconstructed positrons in the signal energy region 
to the number of emitted positrons in the direction opposite to the LXe acceptance region and detected by the pTC. Since the efficiency depends on the positron energy, it is measured for energies just below the signal energy and its value is extrapolated to the signal energy.
The sample used for the calculation is taken through a minimum bias trigger that requires only one hit on the pTC.
It is expected that the efficiency decreases with $R_{\muon}$, because as $R_{\muon}$ increases, the probability of pile-up  also increases, making the track-finder algorithm 
less effective at identifying hits belonging to individual tracks. Nevertheless, the reduction in efficiency from the lowest value of $R_{\muon} = \SI{2e7}{\per\second}$ to the highest $R_{\muon} = \SI{5e7}{\per\second}$ is moderate \SI{< 15}{\percent}.
The blue dotted line represents the design value, which is almost reached at the smallest value of $R_{\muon}$ and is not far from being reached also at the highest. As a reference we measure $\varepsilon_{\positron, {\rm CDCH}} \sim \SI{74}{\percent}$ at $R_{\muon} = \SI{3e7}{\per\second}$.

The CDCH is highly transparent towards pTC and allows a high signal detection efficiency for pTC, $\varepsilon_{\positron,\mathrm{pTC}} = \SI{91(2)}{\percent}$ (discussed in \sref{sec:ptc_performance}), which is twice higher than with the MEG drift chamber. 
The MEG experiment used $R_{\muon} = \SI{3e7}{\per\second}$, with a positron efficiency of $\varepsilon_\positron \sim \SI{30}{\percent}$. At the same $R_{\muon}$, the positron efficiency is $\varepsilon_{\positron} = \varepsilon_{\positron,\mathrm{pTC}} \times \varepsilon_{\positron, {\rm CDCH}} \sim \SI{67}{\percent}$.
%
%

%
\subsection{Back-up chamber}
\label{sec:dchbackup}
The detailed analysis of the problems of wire breakage stimulated a long R\&D effort to explore 
possible alternatives to the silver-coated aluminium wires, which were found to be 
fragile and easily damaged by ambient humidity, with the aim of designing and 
building a new chamber (CDCH2) without the above problems.
Several possible types of wire were explored, with different coating materials (gold and nickel), 
without coating etc. They were studied from multiple aspects: sensitivity to corrosion 
processes, mechanical strength, soldering technique on PCBs, etc. We developed a system for measuring wire tension  \cite{tensionsystem}, based on the resonant frequency method, 
as well as with wiring and assembly stations \cite{chiarello_wiring}. 
The final choice was a wire made of pure aluminium with a diameter of \SI{50}{\micro\meter}, which is almost 
insensitive to corrosion and can be efficiently and solidly fixed on PCBs by a combination 
of soldering and gluing with suitable chemical products.
The CDCH2 is scheduled for completion by the end of $2023$
and delivery to PSI in spring 2024. At that time, the collaboration will 
decide whether and when to replace the current chamber.

\section{Pixelated timing counter}
\label{sec:pTC}

Measuring the time coincidence between a \positron\ 
and a \photon-ray with the highest resolution is crucial for reconstructing \megp\ like events in a high $R_{\muon}$ environment. 
The pTC was developed to measure the \positron\ impact time, 
from which the emission time at the target plane 
$\tpositron$ can be derived by correcting for the track length reconstructed by CDCH. 
It also plays an active role in the trigger algorithms by 
providing information for event selection based 
on time coincidence and directional matching with the $\photon$-ray measurement with the LXe detector.

\subsection{Concept and design}

\begin{figure*}[tbp]
\centering
\includegraphics[width=0.9\textwidth]{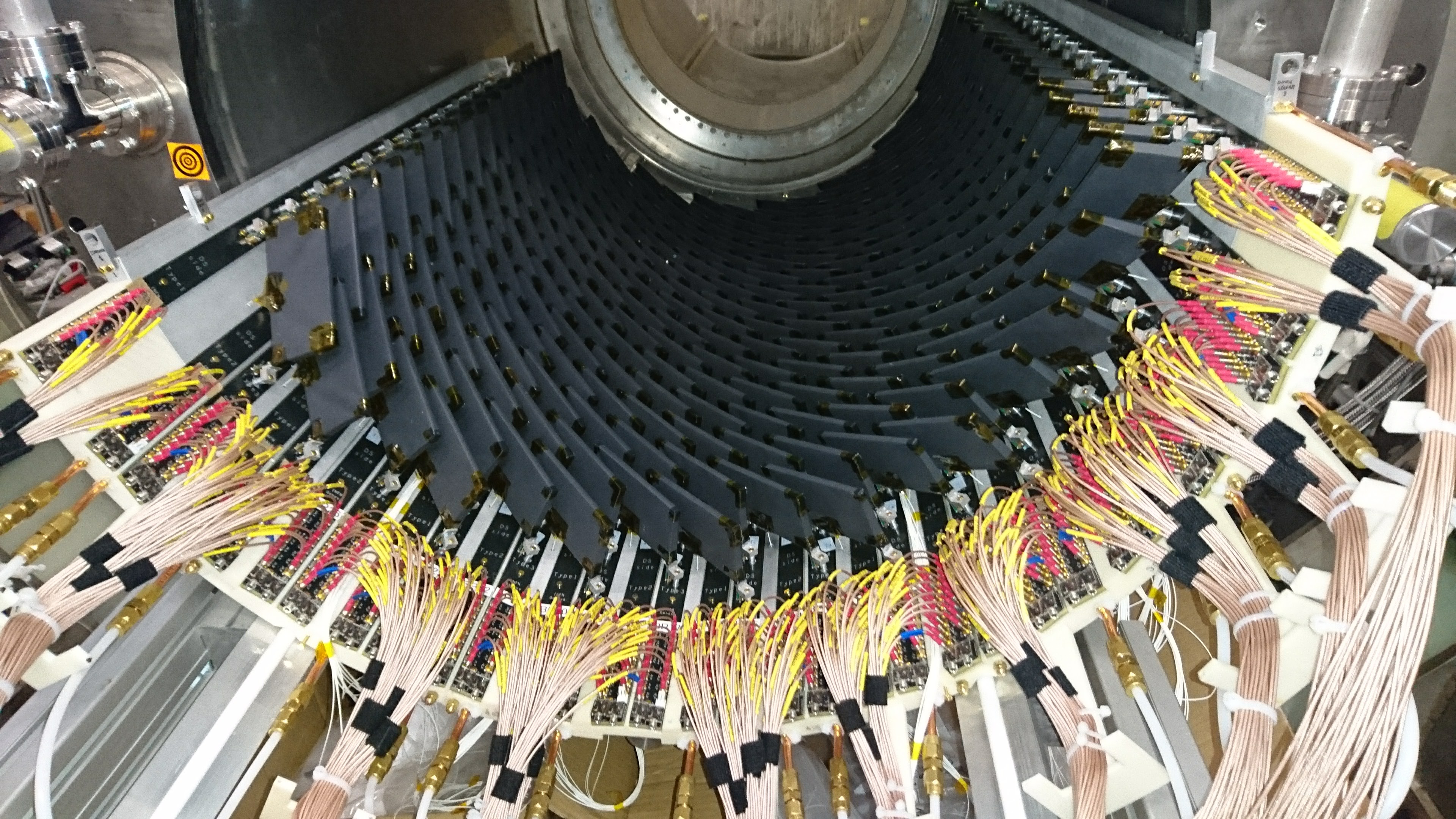}
\caption{The downstream pixelated Timing Counter.}
\label{fig:pTC}
\end{figure*}

The detector consists of two semi-cylindrical sectors arranged mirror-symmetrically upstream and downstream of the target, within the volume between CDCH and the inner wall of the COBRA magnet.
\Fref{fig:pTC} shows an illustration of the downstream sector. 
The support structure consists of a \SI{20}{\mm} thick semi-cylindrical aluminium shell. 
Plastic scintillator counters are mounted on 16 back-planes running in the $z$-direction along the structure. 
To ensure temperature-stable operation, water flows from a cooling system through copper pipes embedded in the structure.

The pTC extends in the $z$--$\phi$ plane $\num{23} < |z| < \SI{117}{\centi\meter}$ and $\ang{-166} < \phi < \ang{5}$ to cover the angular acceptance of \megp\ positron candidates with a back-to-back \photon-ray reaching the fiducial volume of the LXe detector. 

\begin{figure}[tbp]
\centering
\includegraphics[width=1\linewidth]{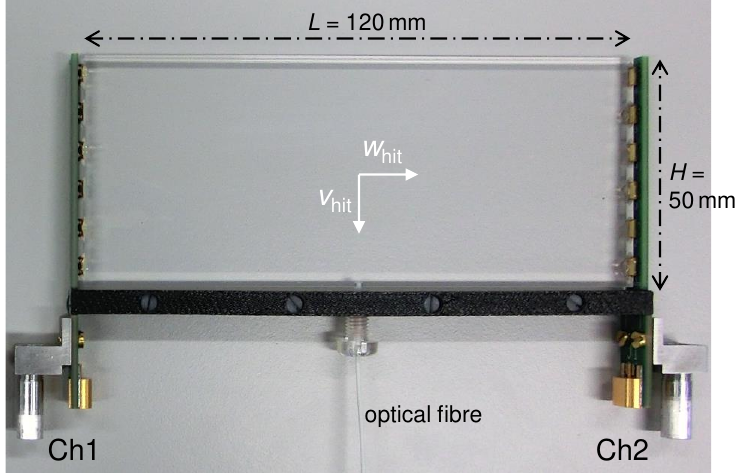}
\caption{A naked counter with $H=\SI{50}{\mm}$. The coordinates $(v_\mathrm{hit},w_\mathrm{hit})$ are the counter local coordinates used in reconstruction in \sref{sec:ptc_rec}.}
\label{fig:counter}
\end{figure}

Each sector is segmented into 256 plastic scintillator tiles (Bicron BC422\textsuperscript{\textregistered} \cite{bc422}), 
coupled to an array of silicon photomultipliers (SiPMs) glued on two opposite sides as shown in \fref{fig:counter}. 
Each array consists of six SiPMs connected in series.  Near-ultraviolet 
sensitive AdvanSiD SiPMs with an active area of \qtyproduct[product-units = bracket-power]{3x3}{\mm} and a pixel pitch of \SI{50}{\micro\meter} (ASD-NUV-SiPM3S-P) are used.
Their spectral response matches well with the \SI{370}{\nano\meter} peak emission of BC422. 
Each tile is wrapped with a highly efficient \SI{35}{\micro\meter} thick polymeric reflector 
(VIKUITI 3M Mirror Film\textsuperscript{\textregistered}) to increase light reflectance at the surface, 
and finally wrapped again with a \SI{30}{\micro\meter} thick black TEDLAR\textsuperscript{\textregistered} film.

The counter sizes \textcolor{black}{(length $L$, height $H$ and thickness $T$)} and locations were optimised using MC simulations. 
The best trade-off between single-counter performance, number of hits for a signal positron (hit multiplicity), detector efficiency and number of channels was found with 
$L\times H\times T = \qtyproduct[product-units = bracket-power]{120x40x5}{\mm}$ and \qtyproduct[product-units = bracket-power]{120x50x5}{\mm} tile sizes (the two different
 $H$ are used in different regions to maximise hit multiplicity while keeping the highest efficiency). 
The counters are arranged at a fixed radius (the top of each counter at $r=\SI{29.7}{\cm}$) in a \numproduct{16 x 16} matrix arranged in the $z$--$\phi$ plane. The longitudinal and angular distances between adjacent counters are \SI{5.5}{\centi\meter} and \ang{10.5}, respectively. 
Each line at fixed $z$ is offset by half a counter from the previous one; 
in addition, the counters are tilted by \ang{45} so that they are nearly perpendicular to the signal positron trajectories. 
This configuration was chosen to maximise the average hit multiplicity for a signal-like positron. 
The idea behind the detector design is to take advantage of both the good single-counter resolution (below 100 ps measured in laboratory tests \cite{Nishimura2020}) and the combination of multiple hit information. 
The mean hit multiplicity per signal event reconstructed from MC, $\langle N_\textnormal{hit} \rangle\sim 9$, 
leads to the expectation of a total time resolution down to $\sigma_{\tpositron,\rm{pTC}} \sim \SI{40}{\pico\second}$.

Most counters (except those at low $|z|$, due to mechanical constraints) are connected to a laser source via optical fibres. 
The signals generated by sending synchronous light pulses with the laser to the counters are used to check detector stability and to calibrate the inter-counter time offsets as described in \sref{sec:ptc_calib}.  

\subsection{Operation}
The R\&D for the pTC single counter began in 2013 \cite{Cattaneo:2014uya}. 
The full detector was developed and finally commissioned in 2017 \cite{Nishimura2020}, when it was tested in the MEG~II spectrometer under experimental conditions. 
Since then, it has always been operational during the MEG~II engineering runs, which were conducted once a year.

During standard operation of the pTC, the circulating water is maintained at the fixed temperature of \SI{9}{\degreeCelsius}, resulting in an effective SiPMs' temperature range of
\qtyrange[range-units = single]{11.0}{14.5}{\degreeCelsius}, depending on their position.
The detector volume is constantly purged with dry air to maintain low humidity and prevent a dew point from being reached. 

The breakdown voltage for each SiPM array was extrapolated from I--V curves recorded at fixed temperature (\SI{30}{\degreeCelsius}). 
Data collection at different temperatures allowed extraction of the coefficient of breakdown voltage as a function of temperature, resulting in \SI{24}{\milli\volt\per\degreeCelsius}/SiPM. 
The optimal operating voltage for each SiPM array was then first determined in laboratory tests by measuring time resolution as a function of overvoltage, and then optimised at the beginning of each run to maximise the signal-to-noise ratio under experimental conditions. 
A typical value for an array of six-SiPM is \SI{\sim 164}{\volt}, i.e., an overvoltage of \SI{3.2}{\volt}/SiPM. 

\begin{figure*}[tbh]
\centering
\includegraphics[width=0.9\textwidth]{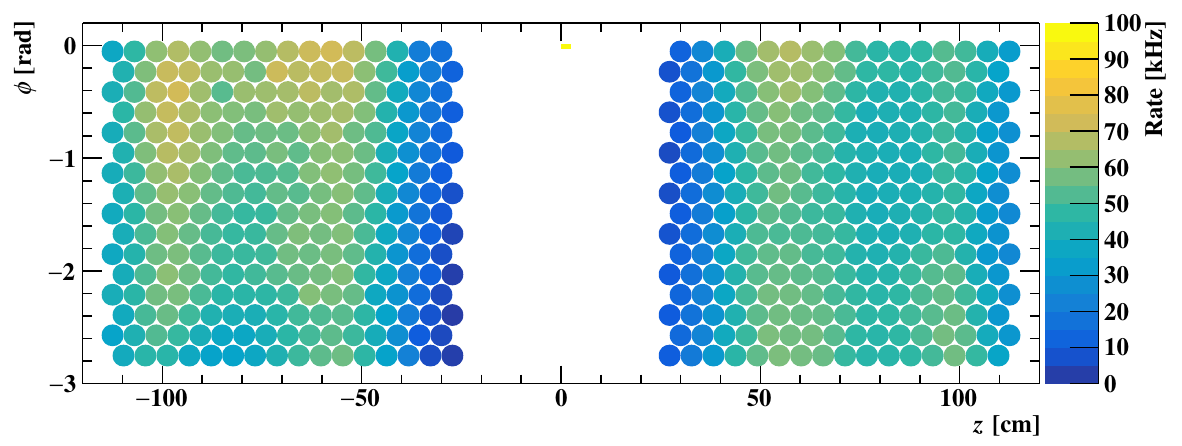}
\caption{Map of counter hit rates measured at $R_{\muon} = \SI{5e7}{\per\second}$ in 2022 run. 
Each circle indicates a counter and the colour shows the hit rate.}
\label{fig:pTC_rate}
\end{figure*}

Since the engineering run in 2017, the detector has been running very stably. 
Only a tiny number of channels (one in 2021, four in 2022) proved dead (i.e., they have no signal). 
The malfunctioning and dead counters have been replaced with spare ones during the maintenance period.
The distribution of pTC hit rate shown in \fref{fig:pTC_rate} can be used as a diagnostic and monitoring tool for beam and background.
For example, the asymmetry between upstream and downstream is due to the polarisation of the muon beam and the muon decays off-target.
The distribution of the hit rate agrees well with that of MC; no unexpected background is observed.
Due to the finely segmented configuration, the hit rate of each counter is \SI{<75}{\kHz} despite of the total pTC hit rate of \SI{\sim3}{\MHz} at $R_{\muon} = \SI{5e7}{\per\second}$.

%

\subsubsection{Issues and problems}

During the first years of operation, some problems related to the detachment of SiPMs from the scintillator surface occurred, probably caused by mechanical stress on the SiPM board and loose adhesion of the optical resin to polished surfaces. 
For this reason, a small subset (a few dozen) of the counters were removed from the detector and glued back on, 
after applying manually some small scratches to the scintillator surfaces glued to the SiPMs, to ensure better adhesion.

\begin{figure}[tb]
\includegraphics[width=1\linewidth]{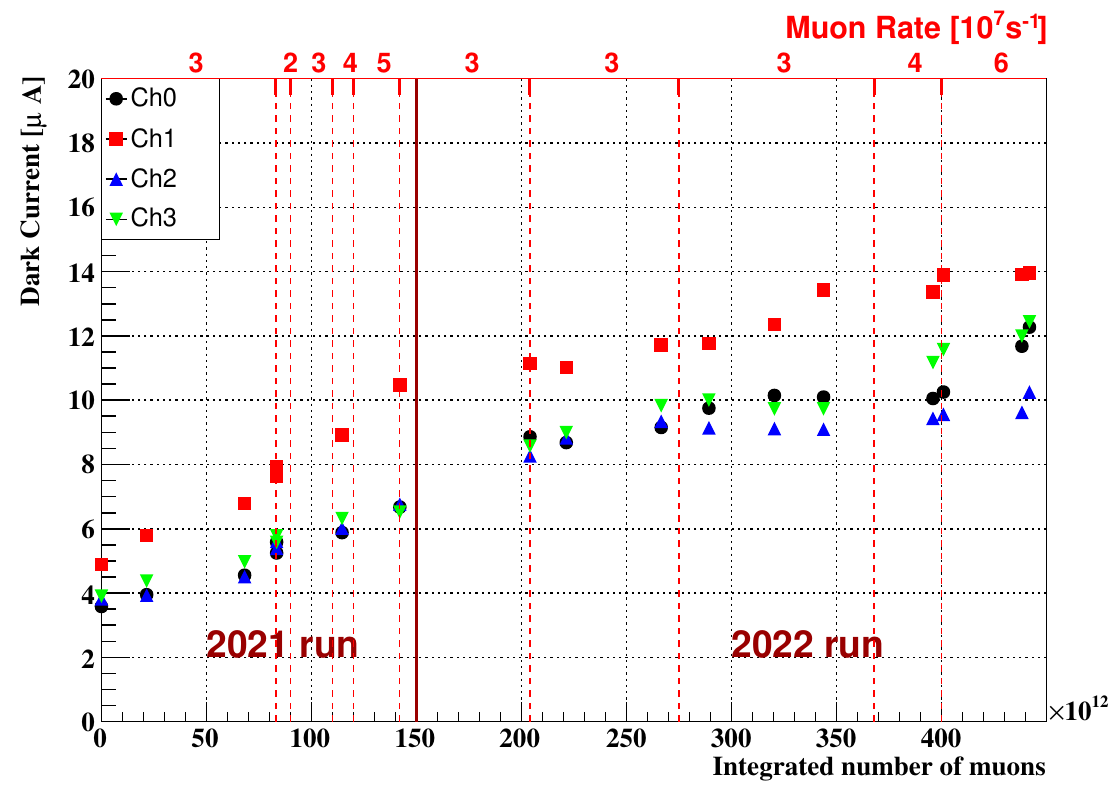}
\caption{The dark currents of some channels compared to the integrated number of stopped muons on target. The vertical dashed lines define the periods with the muon rates shown above.}
\label{fig:darkCurrentpTC}
\end{figure}

An increase in the dark currents of the SiPMs was observed, likely due to radiation damage in the Si bulk.
The \positron\ fluence in 2021--2022 runs is \SI{\sim 1.5e10}{\per\square\cm} at the highest hit rate region.
The behaviour of the dark current as a function of the total muons stopped on target during the last two years is shown for some channels in \fref{fig:darkCurrentpTC}. 
The dark current increment varies between channels due to position-dependent \positron\ hit rates and different operating overvoltages. On average, the increment rate is \SI{9}{\nA/(10^{12}\muup^+)} and  \SI{13}{\nA/(10^{12}\muup^+)} in 2021 and 2022, respectively, which correspond to \qtyrange[range-phrase = --,range-units=single]{\sim 40}{50}{\nano\ampere\per\day}. 

This effect has been studied in detail \cite{Boca:2020elp} and a clear correlation between the increase in dark current and the degradation in time resolution has been demonstrated. 
Based on these studies, we expect a degradation of the overall time resolution of \SI{\sim 13}{\percent} 
after three years of MEG~II running. Although this does not affect the detector performance, 
we decided to refurbish the detector, by replacing \num{\sim 100} counters.

\subsection{Reconstruction algorithms}
\label{sec:ptc_rec}
The \positron\ impact time $t_\mathrm{hit}$ and the position along the long side of the scintillator $w_\mathrm{hit}$
for each counter are reconstructed from the timing of the SiPM signals ($t_\mathrm{Ch1(2)}$ for channel 1 (2) of the counter): 
\begin{align}
&t_\mathrm{hit} = \frac{t_\mathrm{Ch1} + t_\mathrm{Ch2}}{2} 
- \frac{O_\mathrm{Ch1} + O_\mathrm{Ch2}}{2}
- \frac{L}{2v_\mathrm{eff}}, \label{eq:t_hit}\\
&w_\mathrm{hit} = v_\mathrm{eff} \left( \frac{t_\mathrm{Ch1} - t_\mathrm{Ch2}}{2} 
- \frac{O_\mathrm{Ch1} - O_\mathrm{Ch2}}{2}\right), \label{eq:w_hit}
\end{align}
where \textcolor{black}{$L=\SI{120}{\mm}$ is the length of the scintillator,} $O_\mathrm{Ch1(2)}$ is the time offset for the channel, and $v_\mathrm{eff}$ is the effective speed of light in the scintillator; 
$O_\mathrm{Ch1(2)}$ and $v_\mathrm{eff}$ are counter-dependent parameters discussed in \sref{sec:ptc_calib}.
The signal pulse shape after a shaping amplifier\footnote{The shaping is based on a pole-zero cancellation circuit mounted on the WaveDREAM board.} 
has a rise time of \SI{\approx 1.4}{\ns} and a full width at half maximum of \SI{\approx 2.8}{\ns}.
The $t_\mathrm{Ch1(2)}$ is extracted from the pulse using the digital-constant-fraction method, which calculates the crossing time of the signal at a given fraction of the total amplitude. 
In the offline analysis,
different fractions were tested 
and the fraction with the best time resolution was determined separately for each channel (usually 25\%).
The local hit coordinate is transformed into the global coordinates using the counter geometry (position and rotation).

A positron usually leaves hits in multiple counters. 
The series of hits are grouped by a clustering algorithm using the hit timing and position information. 
The same \positron\ can hit counters after exiting the pTC region and travelling another half a turn. These hits are grouped into a different cluster.

The highly granular counter configuration allows estimation of the \positron\ trajectory from the hit pattern of each cluster. 
A look-up table that relates the hit pattern to the radial hit coordinate ($v_\mathrm{hit}$) was created based on the MC simulation and used to infer $v_\mathrm{hit}$ for each hit belonging to the cluster.
The cluster timing and position information is passed to the track-finding algorithm to provide the track time $T_0$ and seed the tracks in CDCH.

The cluster information is next fed into a DAF to fit the trajectory inside the cluster.
Two track fitting procedures are applied to each cluster. 
One is the pTC self-tracking, which uses solely the pTC cluster information and is used for calibration and performance evaluation.
The other uses the track reconstructed with CDCH and combines it with the pTC hits. First, a matching test is made between the CDCH tracks and the pTC clusters, and then,  the track is extended to the end of the last pTC hit in the cluster for each matched combination. 
If multiple clusters matched with a single CDCH track, the first one along the trajectory is adopted.
During the annealing process of DAF, temporally or spatially inconsistent hits are removed. 
This filtering process eliminates not only the contamination of hits by different particles but also hits by the same \positron\ with small turns after the main passage, which are the main cause of the tail in the pTC timing response.

\begin{figure}[tb]
\centering
\includegraphics[width=1\linewidth]{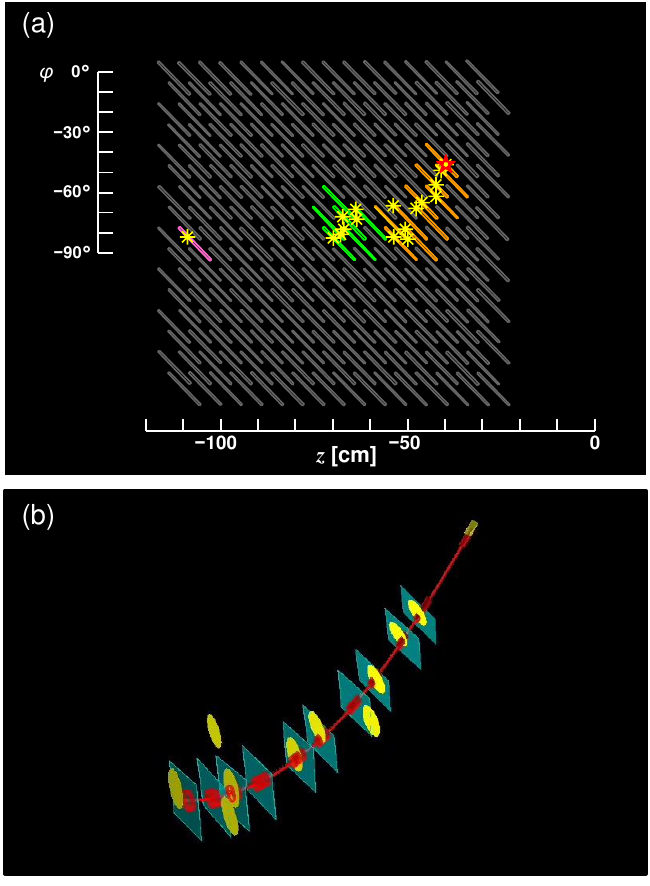}
\caption{An example of pTC hits in an event. (a) The $\phi$--$z$ view of the upstream sector.
Counters with hits are shown in colour; different colours show different clusters. 
The yellow asterisk markers show the reconstructed hit positions and the red star marker shows the point matched with the track from CDCH. The first (orange) and second (green) clusters coincide within \SI{2}{\ns} and thus originate from a single \positron\ in different turns. 
They are separated into different clusters by the spatial information. 
The third (magenta) one originates from another particle well separated in time (+\SI{180}{\ns}).
(b) A 3D view of the first cluster. The cyan squares show the counter planes (arbitrary size), the yellow ellipses show the reconstructed hits, and the red curve shows the fitted track. 
The seventh hit is incompatible with the others, most likely due to a hit of a secondary particle from the main \positron, and the weight becomes zero in the annealing process of DAF.}
\label{fig:pTCrec}
\end{figure}

\Fref{fig:pTCrec} illustrates the clustering and tracking processes with an example event observed in 2021 at $R_{\muon} = \SI{5e7}{\per\second}$.

The timing information in a cluster is combined into the impact time of the first counter:
\begin{align}
t_\mathrm{\positron,pTC} = \sum_{i=1}^{N_\mathrm{hit}}\left(t_{\mathrm{hit},i} - f_{1,i}\right) / N_\mathrm{hit}, 
\label{eq:spx_time}
\end{align}
where $N_\mathrm{hit}$ is the number of hits in the cluster, $t_{\mathrm{hit},i}$ is the reconstructed hit time of the $i$-th hit and 
$f_{1,i}$ is the time of flight from the first hit to the $i$-th hit calculated from the length of the fitted track.
Finally, the time of the \positron\ emission at the target is calculated as,
\begin{align}
t_\mathrm{e^+} = t_\mathrm{\positron,pTC} - f_{0,1},
\end{align}
where $f_{0,1}$ is the time of flight from the target to the first hit calculated from the track length (see \sref{sec:track_rec}).

\subsection{Calibration and alignment} \label{sec:ptc_calib}

The measured signal timing on each channel contains its own time offset \textcolor{black}{($O_\mathrm{Ch1(2)}$)} due to 
different lengths of the readout chain, the variety of scintillator and SiPM responses and differently optimised parameters of the digital-constant-fraction method, 
resulting in the misalignment of the hit times \textcolor{black}{(see \eref{eq:t_hit})} and positions \textcolor{black}{(see \eref{eq:w_hit})} of individual counter measurements.
\textcolor{black}{ From \erefs{eq:t_hit} and (\ref{eq:w_hit}), it is more effective to calibrate the linear combinations of $O_\mathrm{Ch1(2)}$ than to calibrate them separately.}

To align the local $w_\mathrm{hit}$ coordinate, $v_\mathrm{eff}$ and
\begin{align}
&O_\mathrm{intra} = \frac{O_\mathrm{Ch1} - O_\mathrm{Ch2}}{2}\label{eq:intra_offset}
\end{align}
are calibrated using the $w_\mathrm{hit}$ distribution obtained with the Michel positrons.
Since the scintillator length $L$ is precisely controlled ($\mathcal{O}(\SI{10}{\um})$), we use this physical boundary condition. 
The centre of the distribution reflects $O_\mathrm{intra}$ and the width does $v_\mathrm{eff}$. 
The precision of this method was evaluated to be \SI{1.1}{\mm}, which is much better than the  $w_\mathrm{hit}$ resolution of $\sim$\SI{10}{\mm}.

The alignment of each counter relies on two types of optical survey.
One is a three-dimensional scan of the entire pTC structure including individual counters using the FARO EDGE SCANARM. 
This was performed on completion of the assembly prior to installation. 
From the scanned data, the position and rotation of each counter was reconstructed relative to reference points for spherically mounted retro-reflectors (SMRs).
The other is a three-dimensional survey using the Leica Laser Tracker. 
This was performed in situ after the installation and measured the reference points with the SMRs in the MEG~II global coordinate system. 
By connecting the reference points, the position and rotation (six parameters) of each counter were determined to an accuracy of a few \SI{100}{\um}.  

To align inter-counter timing, the following variable is calibrated for each counter:
\begin{align}
 &O_\mathrm{inter} = \frac{O_\mathrm{Ch1} + O_\mathrm{Ch2}}{2}
+ \frac{L}{2v_\mathrm{eff}}. \label{eq:inter_offset}
\end{align}
Two complementary methods were developed: the track-based method and the laser-based one.
The former uses the Michel positron tracks and the latter uses a dedicated laser system. 
Details of the laser system and the method are given in~\cite{laserpTC}.  

The track-based method calculates a set of inter-counter time offsets $O_{\mathrm{inter},k}$ ($k=1,...,512$ is the counter ID) by minimising the following $\chi^2$ for a given data set of Michel positrons,
\begin{align}
\chi^2 = \sum_{j=1}^{N_\mathrm{cluster}} \sum_{i=1}^{N^j_\mathrm{hit}}\left[
 \frac{t^j_{\mathrm{hit},i} - \left(t^j_\mathrm{1st} + f^j_{1,i} + O_{\mathrm{inter},k_i}\right)}{\sigma^j_i} \right]^2,
 \label{eq:millepede}
\end{align}
where $j$ runs for all clusters with successful pTC self-tracking in the data set, $i$ runs for hits in the cluster,
$t^j_\mathrm{1st}$ is the \positron\ impact time at the first hit in the cluster and it is a floating 
parameter in the minimisation (local parameter), the time of flight $f^j_{1,i}$ is computed by 
the self-tracking, and $\sigma^j_i$ is the uncertainty of each measurement represented by the mean 
counter time resolution.
The minimisation is solved with a linear least squares fit using Millepede II \cite{millipede}.

\begin{figure}[tb]
\centering
\includegraphics[width=1\linewidth]{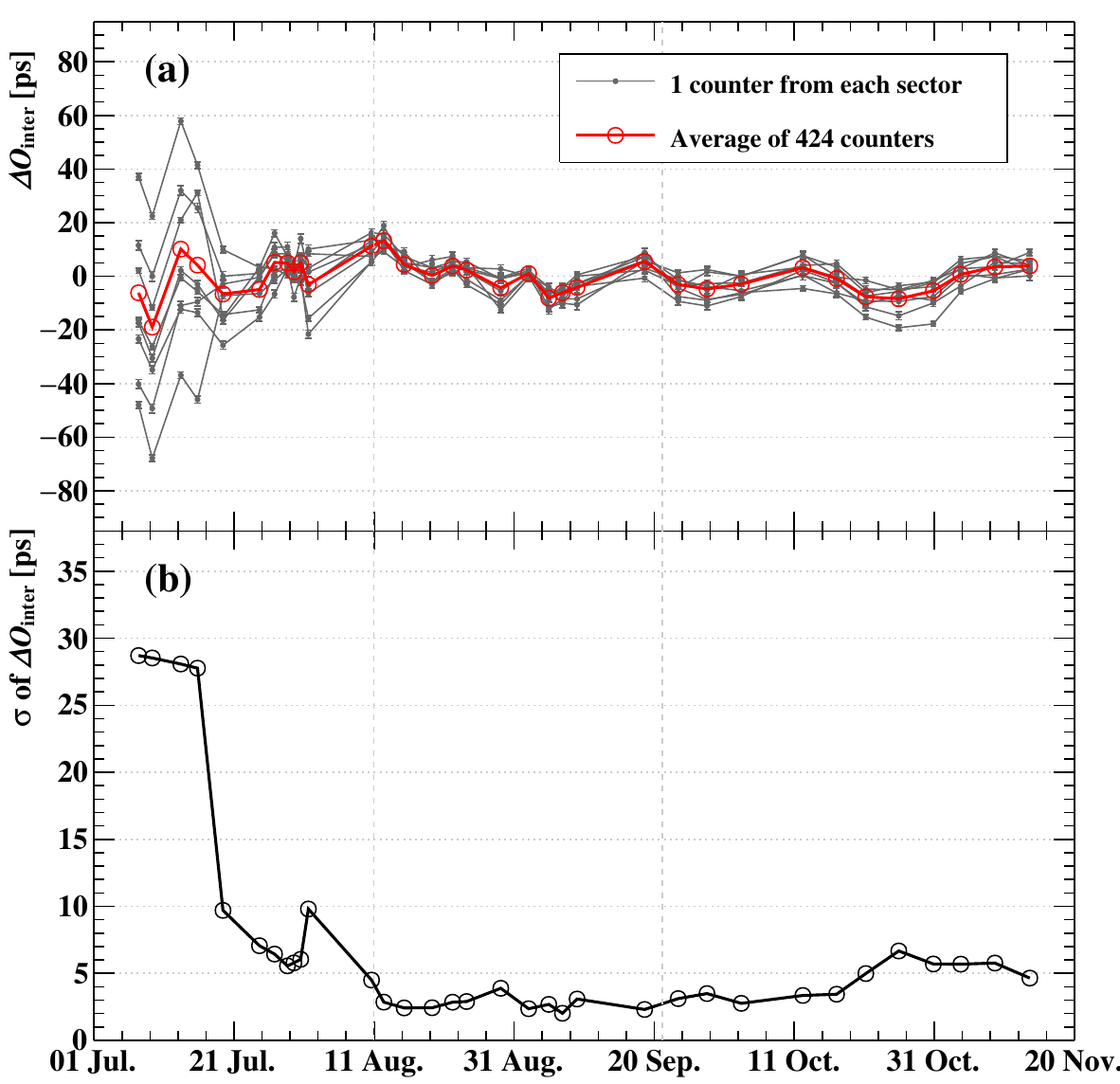}
\caption{(a) Temporal variation of the pTC counter time offsets \textcolor{black}{$O_\mathrm{inter}$} in 2022 traced by the laser calibration. 
For this plot, the offsets were calibrated and aligned with the data between 11th August and 22nd September, \textcolor{black}{and therefore $\Delta O_\mathrm{inter}$ denotes deviation from them}.
The grey graphs are for sampled counters from different groups of optical paths in the laser system and the red one is for the average for all \num{424} laser-equipped counters. 
(b) The dispersion of \textcolor{black}{$\Delta O_\mathrm{inter}$} for the \num{424} counters. 
}
\label{fig:pTCLaser}
\end{figure}

The track-based method can calibrate $O_{\mathrm{inter},k}$ relatively between the counters in a sector for a given data set.
The laser method is used to connect the two sectors and to trace the temporal variation.
\Fref{fig:pTCLaser} shows the temporal variation of $O_{\mathrm{inter},k}$ in 2022 traced by the laser method. 
Until the 10th of August 2022, the commissioning of the run, such as exchange of electronics boards and optimisation of the bias voltages, had been made and hence the whole detector system had been unstable. 
The laser method allows monitoring of the system and calibration of the change in $O_{\mathrm{inter},k}$.
Once the configuration was fixed, the system became stable. The precision of a set of laser calibration with \num{3000} events is evaluated to be \SI{3}{\ps} from the dispersion. 
The dispersion remained stable for about two months and then increased slightly, indicating the need to update the calibration.

The track-based method is subject to a position-dependent systematic bias because a small error in the estimation of $f^j_{1,i}$ can accumulate along the \positron\ path. 
Such a position-dependent bias is detected and corrected by comparing the time offsets from the laser method.

The time offsets obtained by the two methods are in good agreement with a standard deviation of \SI{31}{\ps}, 
which is dominated by the intrinsic uncertainty in the laser method of \SI{27}{\ps} \cite{laserpTC}. We adopt the results from the track-based one with the corrections using the laser-based one. 
The accuracy of the time offset calibration was estimated to be \SI{\sim15}{\ps}, which is negligibly small compared to the single counter time resolution of 
$\sigma_{\tpositron,\mathrm{pTC}}(N_\mathrm{hit}=1)\sim \SI{100}{ps}$ as discussed in \sref{sec:ptc_performance}.

\subsection{Performance}
\label{sec:ptc_performance}
\begin{figure}[tb]
\centering
\includegraphics[width=1\linewidth]{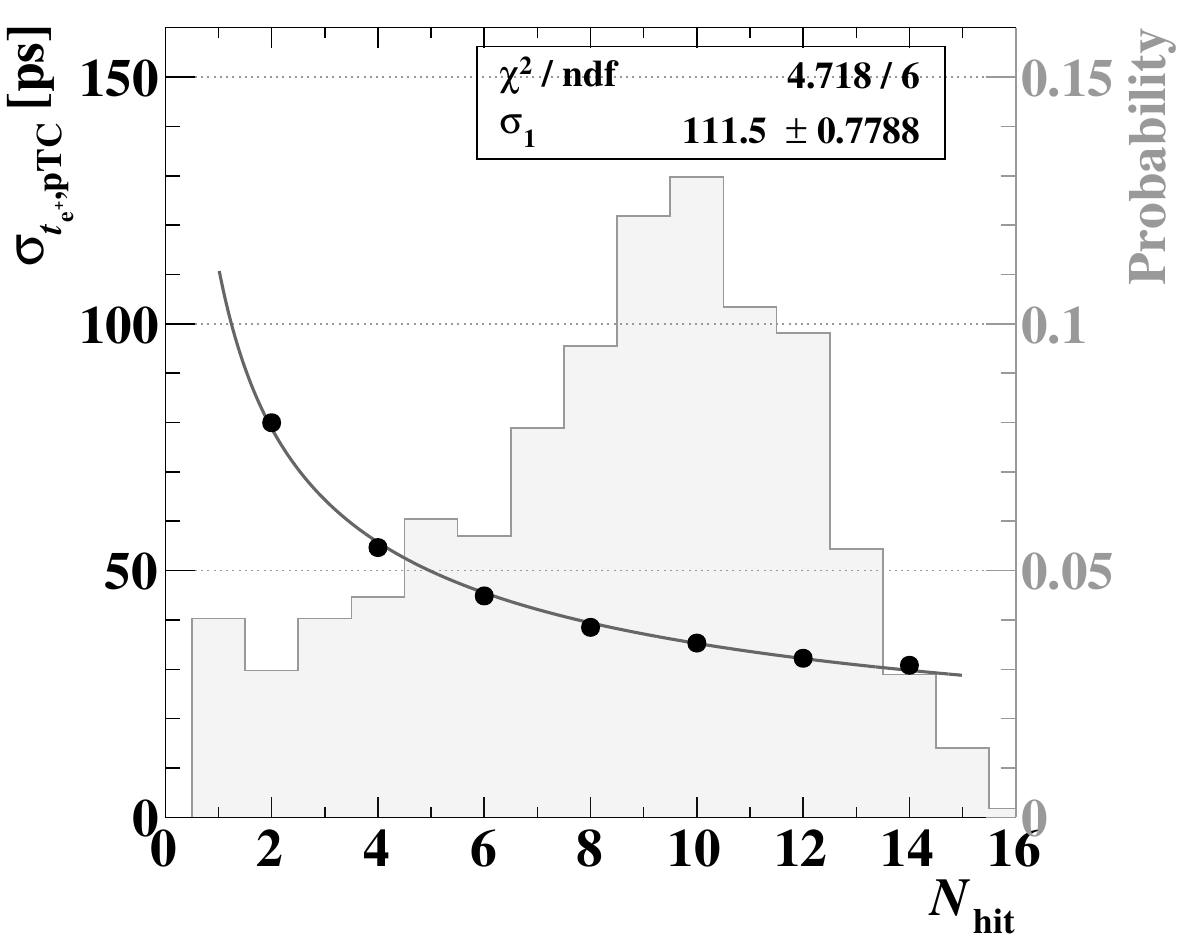}
\caption{Time resolution of pTC as a function of the number of hits evaluated by the even--odd analysis on the 2021 data. 
The curve $\sigma_{\tpositron,\mathrm{pTC}}(N_\mathrm{hit}) = \sigma_1/\sqrt{N_\mathrm{hit}}$ is fitted to the data. \textcolor{black}{The histogram shows the $N_\mathrm{hit}$ distribution for the signal positrons obtained from the MC simulation.}}
\label{fig:pTCresolution}
\end{figure}
The single-counter time resolutions were measured to be $\sigma_{\tpositron,\mathrm{pTC}}(N_\mathrm{hit}=1) \sim$ \SIrange[range-phrase = --,range-units=single]{80}{100}{\ps}
and $\sigma_{\tpositron,\mathrm{pTC}}(N_\mathrm{hit}=1) \sim$ \SIrange[range-phrase = --,range-units=single]{100}{120}{\ps}
for $H=\qtylist[list-units=single]{40;50}{\mm}$ counters, respectively. 
The ranges show the variations among the counters and the resolutions are correlated with the light yield of the counters (the product of the yield of the scintillator and the photon detection efficiency of the SiPMs). 
The resolutions are on average 11\% worse than those obtained in the engineering run in 2017 \cite{Nishimura2020}.  
This is due to several reasons: a lower sampling frequency (from 2.0 GSPS to 1.4 GSPS), increase in the dark count rates due to the radiation damage to the SiPMs, 
and lower light yields due to the scintillator ageing and the detachment of SiPMs from the scintillator on some counters.

The multi-hit time resolution is evaluated with the ``even--odd'' method, in which 
$t_\mathrm{\positron,pTC}$ of \eref{eq:spx_time} is reconstructed independently from two groups  of hits in a cluster 
($i\in\{2k\}$ and $i\in\{2k+1\}$ where $k$ is an integer), and the two results are compared. 
\Fref{fig:pTCresolution} shows the results as a function of $N_\mathrm{hit}$,
reaching an average time resolution of $\sigma_{\tpositron,\mathrm{pTC}}=\SI{43}{\pico\second}$, 
obtained by weighting the multi-hit resolutions with $N_\textnormal{hit}$ distribution from the MC simulation. 
This improves the previous MEG timing counter resolution by almost a factor of two \cite{DeGerone:2011te}.

The efficiency of pTC was studied with a MC simulation. Considering the spread of the muon decay points on the target, the geometrical acceptance is \SI{95}{\percent} for the signal \positron\ with the accompanying \photon-ray in the fiducial volume of the LXe detector. A few percent of \positron s escape from hitting pTC due to multiple Coulomb scattering on the CDCH material, especially the end caps. The total detection efficiency is $\varepsilon_{\positron,\mathrm{pTC}} = \SI{91(2)}{\percent}$.

\section{LXe detector}
\label{sec:LXec}

To measure the energy, position, and timing of the \SI{52.83}{\MeV} \photon-rays from the \megp\ decay, \SI{900}{\litre} LXe is used. 
LXe has many excellent properties such as high light yield, fast response, high stopping power and good uniformity, and is 
therefore used in various fields such as particle physics, nuclear physics, and medicine. 
The LXe detector concept is briefly summarised in \sref{sec:intro}, and the commissioning of the detector and the initial 
operation are presented in \sref{sec:XECoverview}. After the methods used to calibrate the photosensors are described in \sref{sec:XECCalib}, the current performance of the detector is discussed in \sref{subsubsec:LXec-position}--\ref{subsubsec:LXec-BG}.

\subsection{Detector concept} 
\label{sec:intro}

The MEG LXe detector has been upgraded to improve the energy and position 
resolutions for events where the \photon-rays interact close to the inner face of the detector.
The most important improvement is the change in the granularity of the photosensors on the inner face from 216 2-inch round-shaped PMTs to 4092
\qtyproduct[product-units = bracket-power]{15 x 15}{\mm} Multi-Pixel Photon Counters (MPPCs) operating at the LXe temperature ($\SI{\sim 165}{\K}$) to detect the scintillation light emitted 
isotropically from LXe in the VUV range ($\lambda \sim \SI{175}{\nm}$).
The other faces of the detector are equipped with 668 PMTs which were also used in the MEG experiment.
The detailed concept of the MEG~II LXe detector is summarised in \cite{baldini_2018}.

\begin{figure}[tbp]
\centering
\includegraphics[width=1\columnwidth]{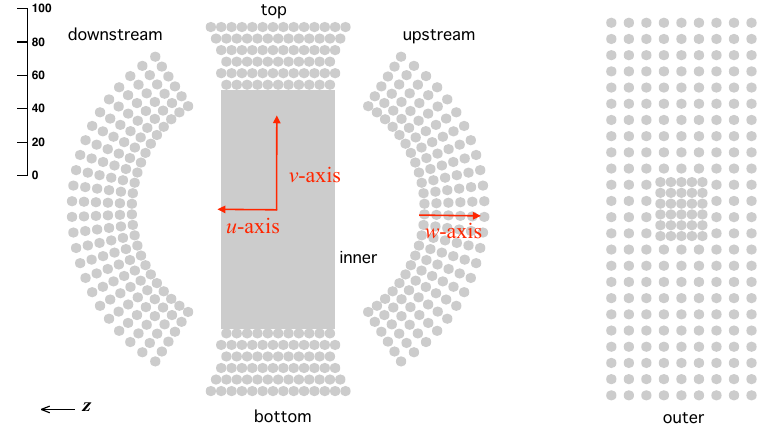}
\caption{Net drawing of the LXe detector and the local coordinate system.}
\label{fig:LXeNet}
\end{figure}

\textcolor{black}{
The local coordinate system $(u, v, w)$ for the LXe detector is defined as follows:
$u$ coincides with $z$ in the cylindrical coordinate
system; 
$v$ is directed along the negative $\phi$-direction at the radius of the inner face of the LXe detector ($r_\mathrm{in}$ = 67.85 cm), which is the direction along the inner face from bottom to top; 
$w = r - r_\mathrm{in}$,
measures the depth from the inner face, as shown in \fref{fig:LXeNet}.
}


\subsection{Commissioning and operation}
\label{sec:XECoverview}
\begin{figure}[tbp]
\centering
\includegraphics[width=1\columnwidth]{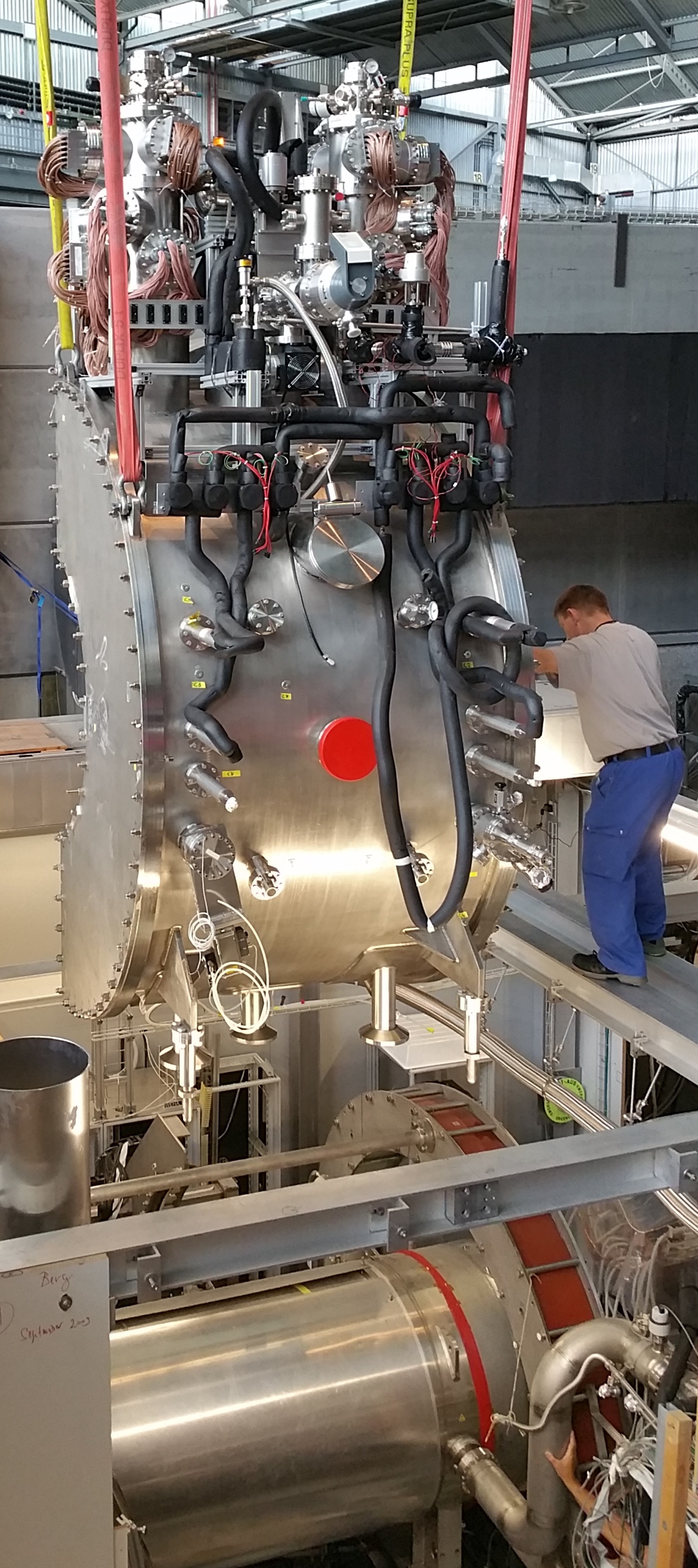}
\caption{The LXe detector installed into the experimental site.}
\label{fig:LXe}
\end{figure}

The upgrade of the LXe detector began in 2016 and was completed in 2017, after which 
it was installed at the experimental site as shown in \fref{fig:LXe}. 
After the photosensors were calibrated, commissioning of the detector with the muon beam was started.
From 2017 to 2020, the number of available channels of readout electronics
was limited, and full readout has been available since 2021.

Some MPPC channels failed due to short circuits or open circuits in the detector, 
and some PMTs due to the failure of HV modules in operation since the MEG experiment;
there are \num{47} dead channels in the MPPCs and \num{34} dead channels in the PMTs during the 2022 beam time.
The number of scintillation photons missing due to these channels is estimated from the solid angle 
subtended by the corresponding photosensors and the number of photons detected in the 
surrounding channels and is taken into account in the reconstruction algorithms to compensate for the effect of the dead channels.

Purification of xenon is necessary to achieve a good light yield from the scintillation. 
Purification in the liquid phase is performed each year before the beam time by circulating the liquid xenon through a molecular sieve.
This method cannot be used during the physics runs because of the noise of the circulation pump.
Purification in the gaseous phase is additionally performed during the whole beam time with a hot metal getter.
The molecular sieve is expected to remove impurities, especially water, and the getter is expected to remove water, oxygen, and nitrogen.

Various types of calibration data, presented in \sref{sec:XECCalib} and \sref{sec:xec_energy}, are collected 
at regular intervals during the physics run to study the stability and uniformity of the detector response.
A detailed description of the calibration equipment can be found in \cite{megdet}.
The entire set of calibration data is collected thrice a week, while a smaller subset for the calibration of the photosensors is collected once a day.
The dead time in the physics run due to calibration was
\SIrange[range-phrase = --,range-units=single]{7}{8}{\percent} in 2021 and was reduced to less than \SI{5}{\percent} in 2022 by optimising the calibration scheme.

Detailed studies of energy and time resolutions have been performed with quasi-monochromatic \qtylist[list-units=single]{55;83}{\MeV} \photon-rays
produced by the $\piup^{-}\mathrm{p}$ charge exchange (CEX) reaction 
($\piup^{-}\mathrm{p}\to \piup^{0}\mathrm{n}$, $\piup^{0}\to 2\photon$) since 2020.
Since this calibration requires a $\piup^-$ beam, a liquid hydrogen (LH$_{2}$) target in 
the COBRA centre, and an additional \photon-ray detector (BGO calorimeter)
to tag back-to-back \photon-rays (see \cite{megdet} for details), 
frequent calibration with this method is unrealistic.
Therefore, a period of two to three weeks is reserved once a year for CEX data acquisition.

The photon detection efficiency (PDE) of the MPPCs for VUV light was measured \num{0.13(1)} on average in the 2017 commissioning run. 
This value is significantly lower than the value of \num{0.20(2)} measured with two MPPC samples in the laboratory \cite{IEKI2019148}.
Surface damage to MPPCs caused by the muon beam was identified as a possible cause for the lower PDE. 
A detailed investigation was carried out in 2019, that demonstrated that the decrease of PDE as a function of muon beam time was due to radiation damage \cite{IEKI2023168365}.

Annealing turned out to restore the reduced PDEs effectively.
Annealing of all MPPCs in the detector was performed for the first time during the 
accelerator shutdown period between runs 2021 and 2022.
The Joule heating of the MPPCs itself served as the heat source.
A custom-built constant-voltage power supply with 30 ports capable of supplying 
\qtyrange[range-units = single]{60}{80}{\volt} and up to \SI{250}{\milli\ampere} per port was used.
Eight MPPCs were connected to each port and 240 MPPCs were biassed simultaneously. 
The MPPCs were illuminated with LED light to induce a current.
A reverse bias voltage of \SI{71}{\volt} was applied to each MPPC with a typical current of \SI{25}{\milli\ampere}, resulting in a power dissipation of \SI{1.775}{\watt}.
Each MPPC was annealed for \SI{\sim28}{\hour}.
PDEs high enough to tolerate radiation damage during the physics run for a full year were achieved as described in \sref{subsec:XECPDE}. 
This procedure (annealing during the shutdown period and continuous physics data taking 
during the beam period) has been and will be repeated every year during the MEG~II physics run.

\subsection{Sensor calibration} 
\label{sec:XECCalib}

The kinematics of the incident \photon-ray is reconstructed from the charges and timings of the signals caused by the LXe scintillation light detected by the photosensors.
The charge \textcolor{black}{$Q_i$ ($i$ is the index of the photosensor)} is measured by integrating the pulse of the digital waveforms in a \SI{150}{\ns} wide window and converting it to the number of photo-electrons $\nphe$ and then to the number of photons incident on the photosensor $\npho$, as follows:
\begin{align}
\nphe &= Q_i/\left(e \cdot G_i \cdot F_{\mathrm{EC},i}\right), \nonumber \\
\npho &= \nphe / \mathcal{E}_i,
 \label{eq:NPho}
\end{align}
where $e$ is the elementary charge, $G_i$ is the gain of the photosensor, $F_{\mathrm{EC},i}$ is the excess charge factor (ECF), and $\mathcal{E}_i$ is the quantum efficiency (QE) of PMT or the PDE of MPPC.
The ECF for MPPC is the ratio between the measured charge and the charge due to the primary photoelectrons,
and it is larger than 1 because of cross-talk and after-pulse; for PMTs $F_{\mathrm{EC},i}= 1$ is assumed.
The results of the calibration for these parameters are described here mainly for 2021 data.
The calibration of the timing parameters is described in \sref{sec:xec_timing}.

\subsubsection{Noise reduction}
The waveform for each channel is read out by the WaveDREAM board and digitised at \SI{1.4}
{GSPS} by \textcolor{black}{the Domino Ring Sampler (DRS4) chip mounted on the board}, as described in \sref{sec:rtdaq}.
Calibration and noise reduction are applied to the digitised waveform data by subtracting the 
following four types of noise templates, created from pedestal events acquired with periodically output triggers.  

The first noise template is dedicated to correct the voltage offset of each sampling capacitor cell in the DRS4 chip.
Voltage calibration is first performed in the online analysis (see \sref{DAQ_design}), but 
some offsets that cause a low-frequency noise remain and must be removed.
The remaining voltage offset, corresponding to each physical capacitor cell, is extracted by 
averaging the voltage for each cell for pedestal events.
The second is used to compensate for high-frequency noise caused by the cross-talk 
of a clock signal distributed to the WaveDREAM boards for time synchronisation.
Since it is synchronous with the clock signal, it can be extracted by averaging the pedestal 
waveforms after adjusting the timing to the clock phase.
The third is to correct the temperature dependence of the slope of the baseline.
Each capacitor cell has a small leakage current that depends on the temperature and results in a slope of the baseline. 
The relation between the temperature and the slope for each channel is extracted from the 
pedestal events acquired at different temperatures.
The fourth is to subtract the noise correlated with 
\textcolor{black}{the sampling cell corresponding to the first point of the waveform data in the time order.
The sampling process of DRS4 is running continuously and cyclically, and when a trigger signal comes it stops sampling and starts readout from the stopped cell.
Therefore, the physical cell corresponding to the first point in the waveform changes for each event and the readout time of each cell also changes.
It causes a variation of the voltage offset. 
This effect is reduced by creating noise templates depending on the first sampling cell.}

\begin{figure}
\centering
\includegraphics[width=1\columnwidth]{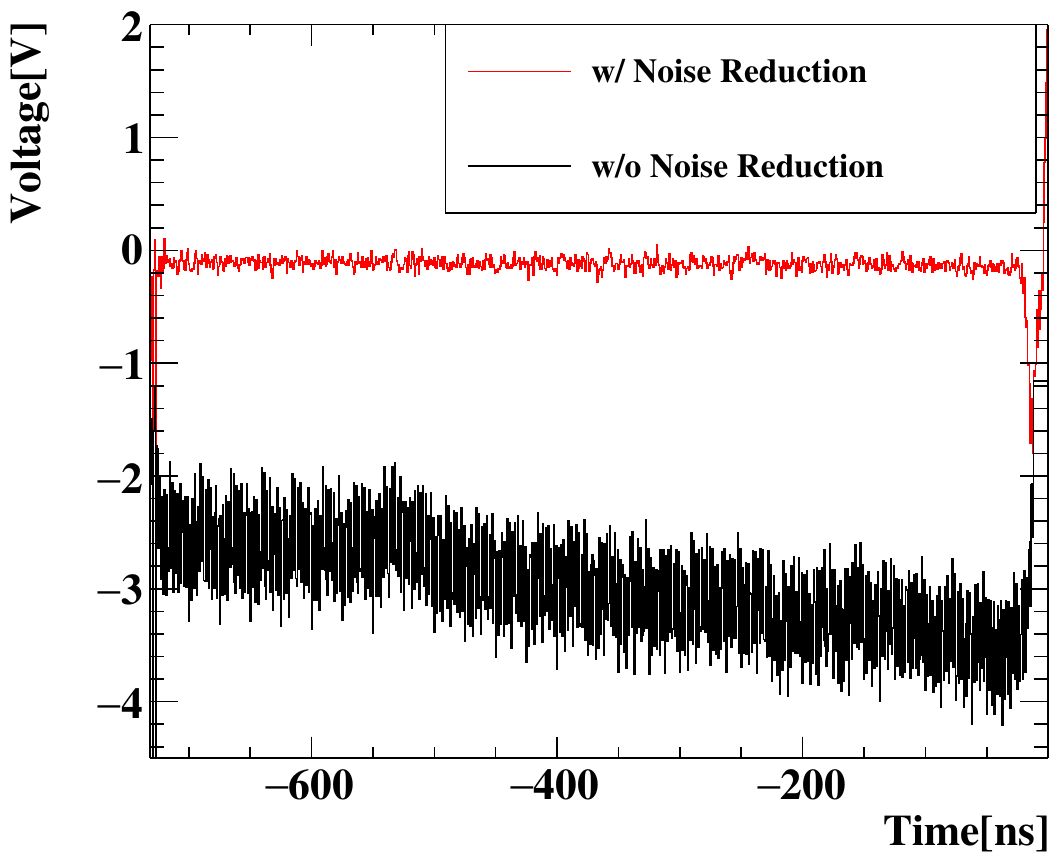}
\caption{The sum waveforms of all MPPCs. The black and red ones are before and after noise reduction, respectively \cite{kobayashi_2022}.}
\label{fig:LXe_waveform_noise}
\end{figure}

\Fref{fig:LXe_waveform_noise} shows the waveforms summed for all MPPC channels before and after noise reduction. 
The offset is corrected, and the slope and high-frequency noise are reduced.
The dispersion of the reconstructed energy of pedestal events is reduced to $\sigma_{\mathrm{noise},\mathrm{LXe}}=\SI{0.15}{\MeV}$, 
which is sufficiently small compared to the energy resolution of the LXe detector (\sref{subsubsec:LXec-energy}).

\subsubsection{PMT gain}
\begin{figure}
\centering
\includegraphics[width=1\columnwidth]{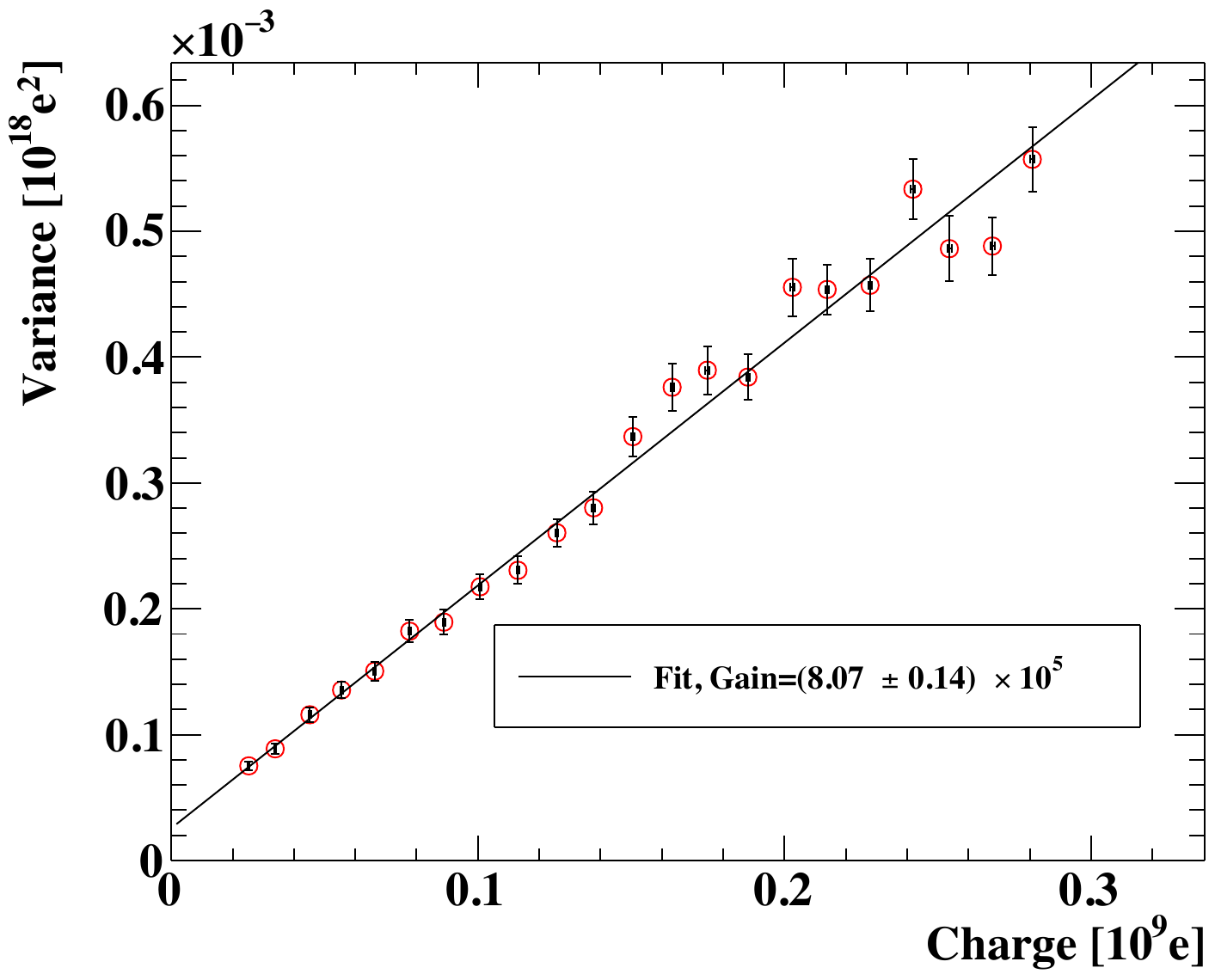}
\caption{The correlation between the mean and the variance of the PMT charge at different LED intensities. 
The gain is calculated from the slope of the linear fit.}
\label{fig:pmtgain_intensityscan}
\end{figure}

The absolute gain in each PMT is determined from the correlation between the mean 
$\overline{Q}_i$ and the variance $\sigma_{Q_i}^2$ of the charge for a constant light source:
\begin{align}
\sigma_{Q_i}^2 = G_i \cdot e \cdot \overline{Q}_i + \sigma_{\mathrm{noise},i}^2,
 \label{eq:PMTgain}
\end{align}
where $\sigma_{\mathrm{noise},i}$ is a noise term. 
This correlation is measured by flashing the blue LEDs installed in the LXe detector at 22 
different intensities. The gain is determined by fitting \eref{eq:PMTgain}
as shown in \fref{fig:pmtgain_intensityscan}.

\begin{figure}
\centering
\includegraphics[width=1\columnwidth]{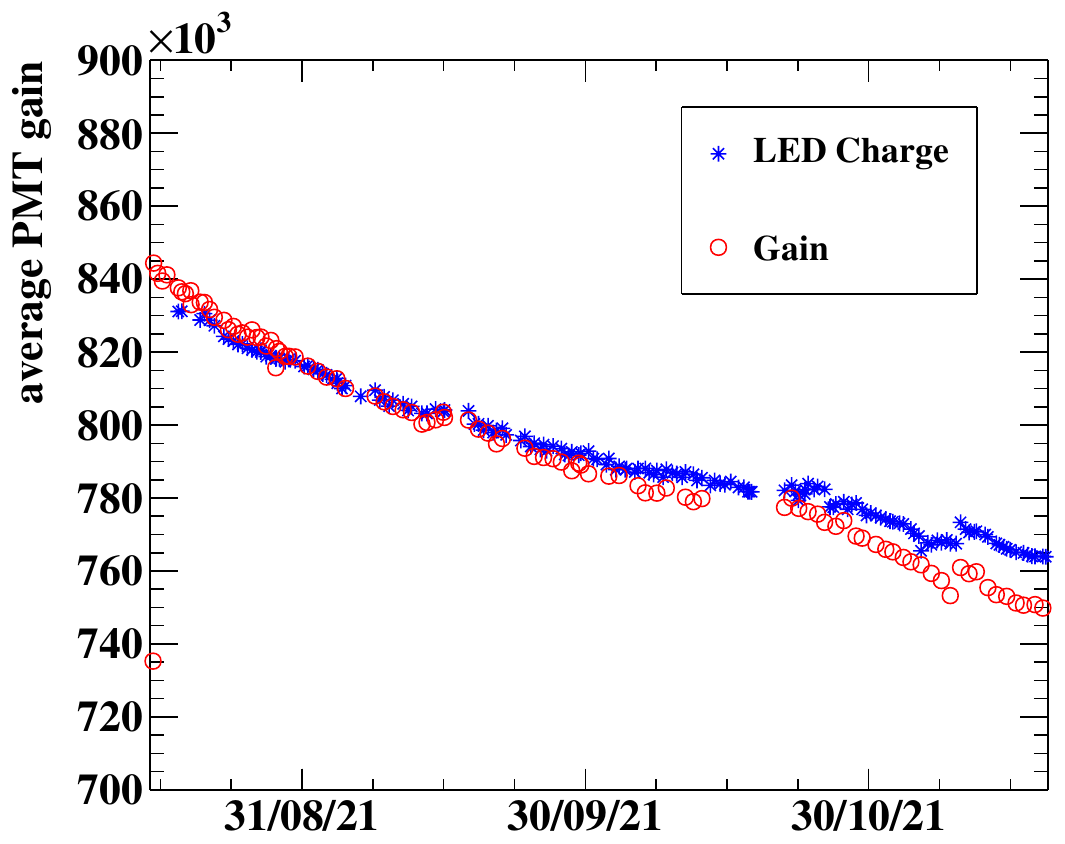}
\caption{Temporal evolution of the average gain of all PMTs during the 2021 run. 
The red plot shows the average of the absolute gains and the blue one shows the average of the 
mean charges for constant LED intensity scaled to the gain values with the data on 15th September.}
\label{fig:pmtgain_history}
\end{figure}

The gains of all PMTs were set to $\sim$\num{0.8e6} at the beginning of each run 
with \SI{3}{\percent} accuracy by adjusting HV for each PMT.
The average gain decreased continuously during the run as shown in \fref{fig:pmtgain_history}.
This phenomenon was already observed in the MEG experiment.
While a single set of HVs was used throughout the 2021 run, the HVs were 
readjusted twice during the 2022 run to restore the reduced gains when the average gain decreased by \qtyrange[range-phrase = --,range-units=single]{10}{20}{\percent}.

The temporal evolution is traced by the absolute gain measurements.
Their fluctuation is smoothed periodically 
in the offline reconstruction using 
the mean charge for fixed intensity LED events, as the accuracy of the charge 
measurement (statistical uncertainty \SI{<0.1}{\percent}) is higher than that of the determination of the absolute gain.
The discrepancy between the temporal evolution of the two measurements (red and blue points in \fref{fig:pmtgain_history}), \textcolor{black}{that does not introduce a significant uncertainty,} 
could be due to the change in the transmittance of the LED light ($\lambda \sim \SI{460}{\nm}$) in LXe or to the LED light instability, which only affect the measurement of the mean charge.
The former contribution is likely dominant because
a correlation is observed between this discrepancy and the change in the light yield of the VUV scintillation light, sensitive to the purity of the LXe,
which improves over time due to continued
purification in the gaseous phase.

\subsubsection{MPPC gains and ECFs}
 
MPPC gains and ECFs are calibrated using low-intensity LED data. 
The intensity of the LEDs is adjusted for the MPPCs nearest to each LED to detect one photo-electron on average.
The gain is obtained from the distance between two peaks corresponding to zero and one photo-electron. 
However, it is difficult to separate the two peaks with the standard integration range of \SI{150}{\ns} in presence of noise. 
Therefore the gains and the ECFs are extrapolated from the results with several shorter integration ranges.

\begin{figure}
\centering
\includegraphics[width=1\columnwidth]{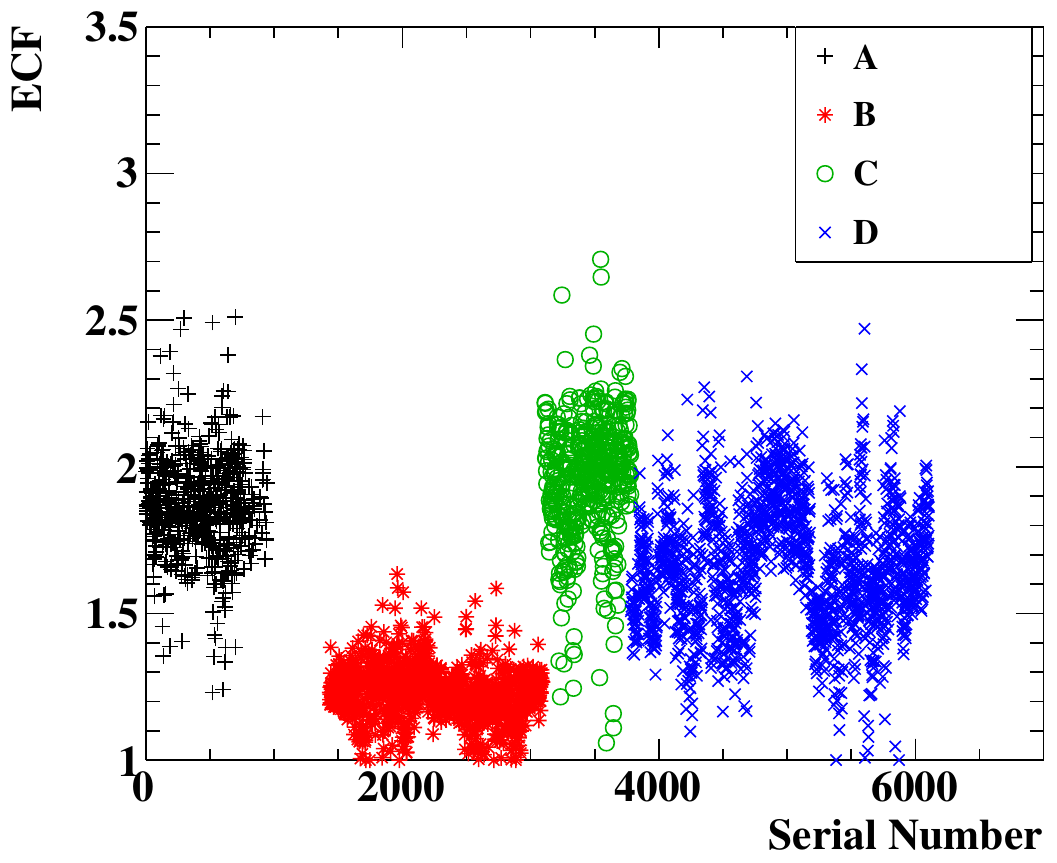}
\caption{The ECFs of the MPPCs versus the serial number \cite{kobayashi_2022}. Different markers show MPPCs produced in different production lots.}
\label{fig:ecf_channel}
\end{figure}

The ECF is measured as follows:
\begin{align}
F_{\mathrm{EC},i} = \overline{Q}_i/\left(e \cdot G_i \cdot \langle{\nphe}\rangle \right),
 \label{eq:ECF}
\end{align}
where 
$\langle{\nphe}\rangle$ is the expected number of photo-electrons for the LED that
can be calculated from the number of zero photo-electron events $n_{\mathrm{zero},i}$ and the total number of events $n_{\mathrm{total},i}$ as
\begin{align}
\langle{\nphe}\rangle = -\log \left(n_{\mathrm{zero},i}/n_{\mathrm{total},i}\right).
 \label{eq:lambda}
\end{align}
\Fref{fig:ecf_channel} shows the ECFs as a function of the serial number of the MPPCs. 
The MPPCs were manufactured in four different periods, each labelled with different markers. 
A dependence of the ECFs on the production lot was found.
This dependence is still acceptable as it can be measured and corrected.

The uncertainty of the absolute gain measurement is \SI{2.5}{\percent} from statistics and that of the ECF is 
\SIrange[range-phrase=--,range-units = single]{1}{5}{\percent}, which mainly comes from the statistical uncertainty of $\langle{\nphe}\rangle$.

\subsubsection{PDEs of MPPCs and QEs of PMTs}
\label{subsec:XECPDE}
The MPPC PDEs and the PMT QEs for VUV light are calibrated with $\alphaup$-particles from $^{241}$Am 
sources deposited on thin wires mounted inside the active volume of the detector.
Since the positions of the $\alphaup$-particle sources are known, the expected number of photons arriving 
at each photosensor can be estimated with a MC simulation.
Therefore, the PDEs and QEs are calculated by comparing the mean number of measured photo-electrons 
$\overline{N}_{\mathrm{phe},i}$ with the number $\overline{N}_{\mathrm{phe},i}^\mathrm{MC}$ in the MC simulation as follows:
\begin{align}
\mathcal{E}_i=\mathcal{E}_{i}^\mathrm{MC}\times \frac{\overline{N}_{\mathrm{phe},i}}{\overline{N}_{\mathrm{phe},i}^\mathrm{MC}}\times F_{\mathrm{LY}},
\label{eq:PDE}
\end{align}
where $\mathcal{E}_{i}^\mathrm{MC}$ is the PDE (QE) assumed in the MC simulation, and $F_{\mathrm{LY}}$ is \textcolor{black}{a light yield correction factor common to all photosensors  so that $F_{\mathrm{LY}}=1$ corresponds to the fact that the mean value of the PMT QEs is 0.16.} This is the value on the sheet supplied by Hamamatsu Photonics K.K. and is the same condition as in the MC simulation.

Since the main background source in the calculation of  PDE (QE) is cosmic rays, 
$\alphaup$-particle events are separated from cosmic ray events by using a pulse shape discrimination 
technique that distinguishes highly ionising particles (whose waveform has a shorter time component) 
from minimum ionising particles (whose waveform has a longer time component) based on the relation
between the charge and the amplitude of all PMTs.

The systematic uncertainty in the absolute value of the PDEs (QEs) estimated with this method is \SI{10}{\percent}; 
the main contributions are the uncertainty in the scintillation light yield for 
$\alphaup$-particles (\SI{5}{\percent}) and the uncertainty in the MC simulation for the effect 
of reflection at the inner surface of the detector (\SI{5}{\percent}) \cite{IEKI2023168365}.

\begin{figure}
\centering
\includegraphics[width=1\columnwidth]{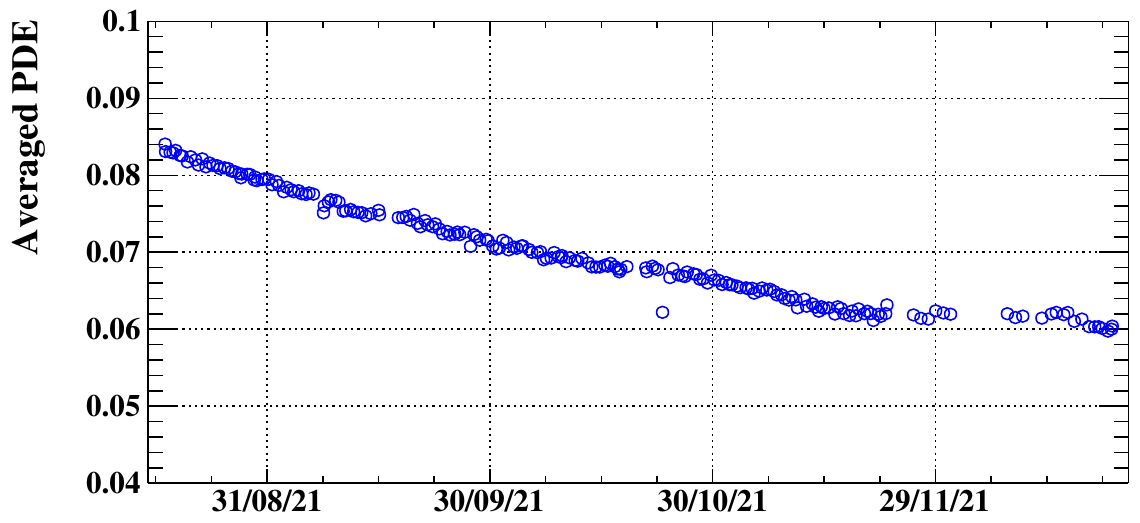}
\caption{Temporal evolution of the average PDE of all MPPCs.}
\label{fig:mppcpde_history}
\end{figure}
\begin{figure}[tbp]
\centering
\includegraphics[width=1\columnwidth]{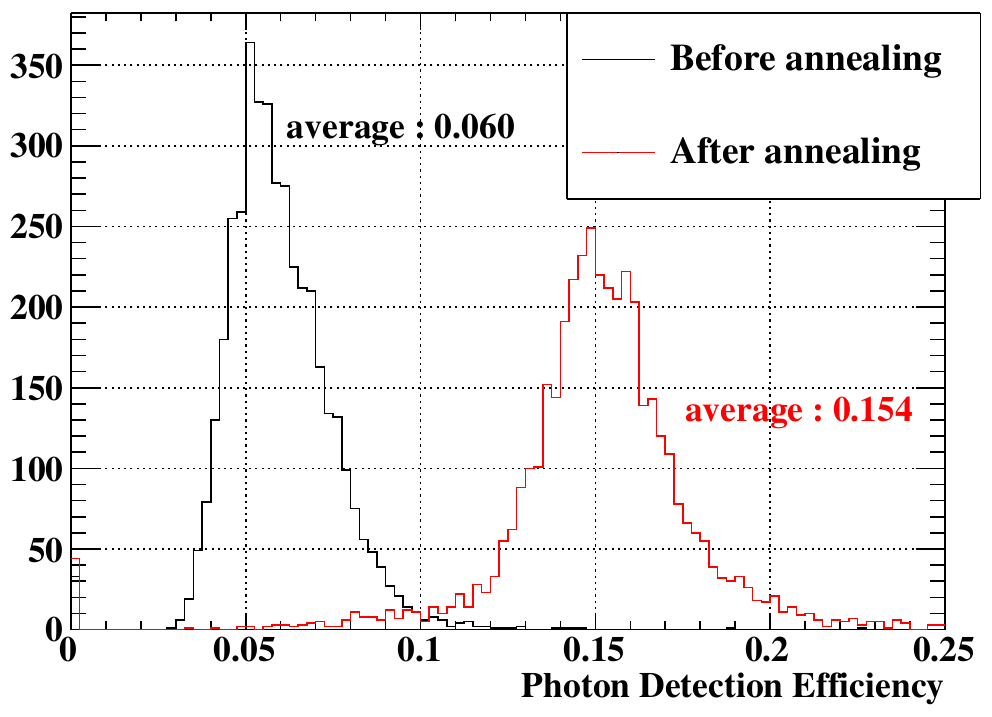}
\caption{Distribution of MPPC PDEs before and after the annealing in 2022.} 
\label{fig:PDEafterannealing}
\end{figure}

\Fref{fig:mppcpde_history} shows the temporal evolution of the average PDE of all MPPCs. 
The average PDE decreased from 0.082 to 0.060 in the 2021 run due to radiation damage.
As explained in \sref{sec:XECoverview}, full channel annealing was carried out after the beam time in 2021.
The average PDE was increased to \num{0.154} by the annealing as shown in \fref{fig:PDEafterannealing}.

\subsection{Position reconstruction and resolution}
\label{subsubsec:LXec-position}
The first conversion position of the incident \photon-ray in the LXe detector 
$\Vec{x}_{\photon,\mathrm{LXe}}$ is reconstructed from the light distribution viewed by the MPPCs.
%
%
The following $\chi_{\mathrm{pos}}^{2}$ is minimised:
\begin{align}
  &\chi_{\mathrm{pos}}^{2}\left(\Vec{x}_{\photon,\mathrm{LXe}}\right) =
  \sum_{i \in \mathrm{region}}\left[\frac{N_{\mathrm{pho},i}-C\times \Omega_{i}\left(\Vec{x}_{\photon,\mathrm{LXe}}\right)}{\sigma_{\mathrm{pho}, i}}\right]^{2} \label{chi2pos}, \\
 &\sigma_{\mathrm{pho}, i}  =
 N_{\mathrm{pho}, i} / \sqrt { N_{\mathrm{phe}, i} } \label{nphosigma},
\end{align}
where 
$\Omega_{\mathrm{i}}(\Vec{x}_{\photon,\mathrm{LXe}})$ is the solid angle 
at position $\Vec{x}_{\photon,\mathrm{LXe}}$ subtended by the MPPC, 
\textcolor{black}{$C$ is a floating parameter of the fit to convert the solid angle to the number of photons},  and
$\sigma_{\mathrm{pho},i}$ is the uncertainty of 
$ N_{\mathrm{pho}, i}$
for each MPPC defined in \eref{nphosigma}.
The MPPCs in a circular region around the peak position in 
the light distribution are used for the fit.


This fit is based on the assumption that the number of photons detected by each MPPC is 
proportional to the solid angle at the interaction point subtended by the MPPC.
In reality, the source of the scintillation photons has a finite size because an electromagnetic 
shower forms in a direction that correlates with the direction of the incident \photon-ray. 
Therefore, the result of the fitting, $\Vec{x}_{\gamma,\mathrm{fit}}=(u_{\gamma,\mathrm{fit}}, v_{\gamma,\mathrm{fit}}, w_{\gamma,\mathrm{fit}})$, 
is biassed in the direction of the shower evolution. 
To account for this effect, two corrections are made based on the MC simulation studies.
The first is a correction depending on $u_\mathrm{fit}$. 
The position is reconstructed further outward when the incident angle is large 
(large $|u|$), and, therefore, a larger correction is required.
The second is a correction for the event-by-event fluctuation toward the 
shower evolution which can be estimated from the difference of $\Vec{x}_\mathrm{fit}$ 
reconstructed with different radii of the circular region.

A systematic error in position reconstruction results from errors in MPPC positions.
To take full advantage of the improved granularity of the readout, we measured the MPPCs' positions using two complementary methods.
The first is a direct optical survey using a 3D laser scanner, which was carried out at room temperature during 
the construction phase. The second is a measurement with a well-aligned collimated X-ray beam, 
which was carried out after the detector had been installed on site and filled with LXe.
The latter is necessary because thermal 
contraction and deformation during LXe filling affect the sensor positions. 
The methods and results are described in detail in \cite{Libeiro:2023qhd}. 
The uncertainty of the first method is \SI{\sim 0.1}{\mm}.
The largest uncertainty results from the
reproducibility in the second method. It was checked by moving and repositioning the LXe detector and the \photon-ray beam and by repeating the optical survey. It resulted to be \SI{0.57}{\mm} in $z$.
The total uncertainties were reduced to below \SI{0.6}{\mm} in $z$ and \SI{0.7}{\milli\radian} 
in $\phi$, which are sufficiently small compared to the position resolution described below. 
The global position of the detector is also aligned with the cosmic ray events passing through 
both the LXe detector and the CDCH, as described in \sref{sec:dchxcali}.

\begin{figure}[tb]
\centering
\includegraphics[width=1\linewidth]{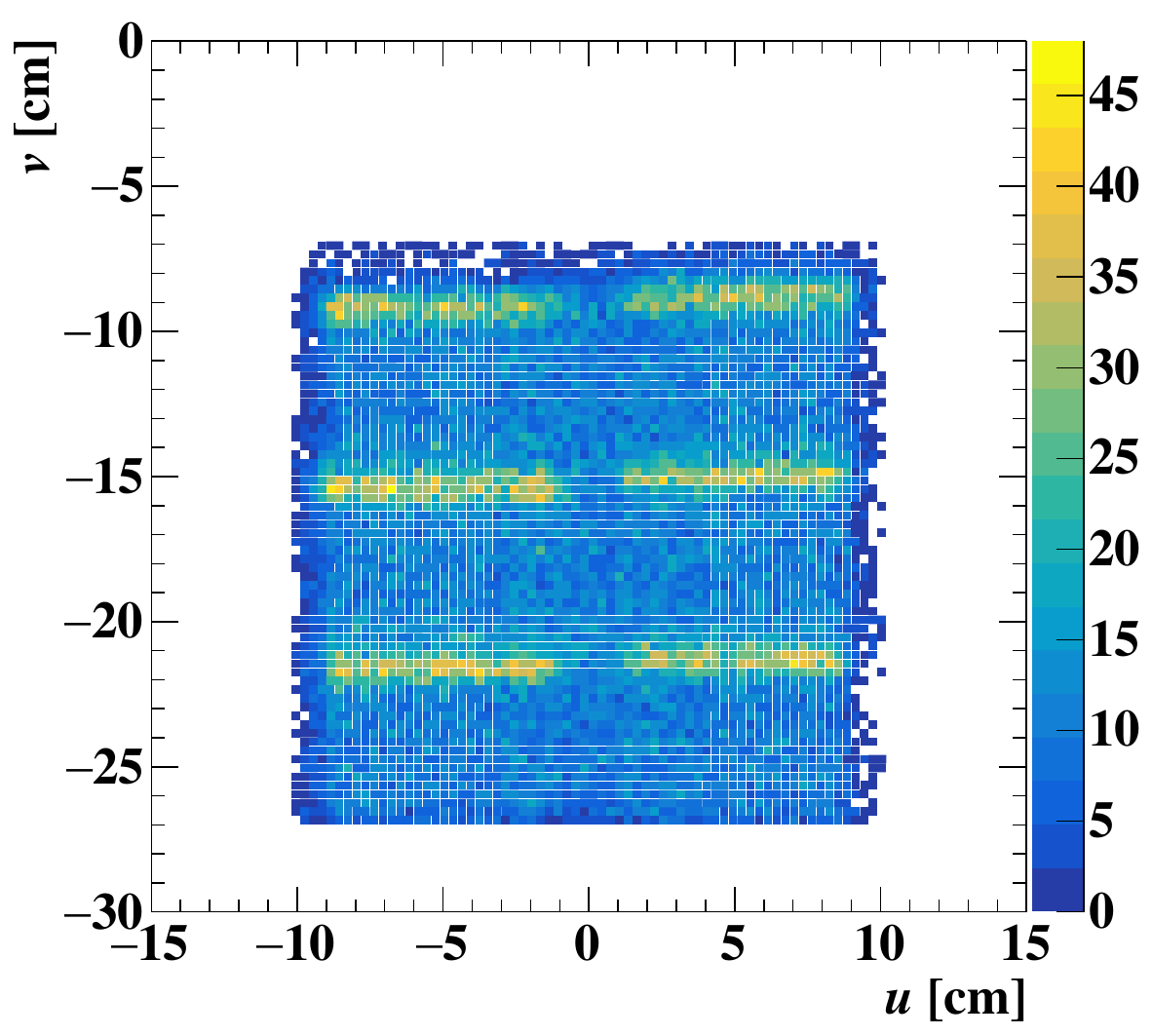}
\caption{The position distribution of $^{7}\mathrm{Li}(\mathrm{p},\photon)^{8}\mathrm{Be}$ 
line events in $u$--$v$ plane with the collimator for the $v$-resolution measurement. 
The readout region covers six out of eight slits, which are imaged as horizontal lines.}
\label{fig:LXecollimator}
\end{figure}

The position resolutions are evaluated by imaging a lead collimator with 
\photon-rays from the $^{7}\mathrm{Li}(\mathrm{p},\photon)^{8}\mathrm{Be}$ reaction produced by a proton beam accelerated by a Cockcroft--Walton (CW) accelerator \cite{MEG:2011rgj}. 
A \qtyproduct[product-units = bracket-power]{240x240x25}{\mm} collimator 
with eight slits, each \SI{5}{\mm} wide and \SI{80}{\mm} long
with \SI{50}{\mm} spacing between the slits,
was installed between the detector and the COBRA magnet in a dedicated run.
\Fref{fig:LXecollimator} shows the two-dimensional position distribution of the 
\photon-rays that passed through the collimator. 
The sharpness of the reconstructed slit images represents the $v$ resolution.
The $u$-resolution is measured by rotating the collimator by \ang{90}.
The resolutions are estimated by fitting the MC simulation
model, which was smeared by the resolution, to the data, resulting in $\sigma_{u_\gamma,v_\gamma}=\SI{2.5}{\mm}$ 
(\SI{4.0}{\mm}) for $w<\SI{2}{\cm}$ ($>\SI{2}{\cm}$)
in $u$ and $v$. The position resolution in $w$ is estimated to be 
$\sigma_{w_\gamma}=\SI{5.0}{\mm}$ based on the MC simulation.

\subsection{Time reconstruction and resolution}
\label{sec:xec_timing}
The time of the \photon-ray first interaction with LXe is reconstructed by minimising the following chi-square function: 
\begin{align}
\chi^2(t_{\photon,\mathrm{LXe}}) = \sum_{i} \left( \frac{t_{\mathrm{pm},i}-t_{\mathrm{prop},i}-t_{\mathrm{walk},i}-t_{\mathrm{offset},i}-t_{\photon,\mathrm{LXe}}}{\sigma_{\mathrm{pm},i}} \right)^2,
 \label{eq:tgamma}
\end{align}
where $t_{\mathrm{pm},i}$ is a time detected at each sensor calculated from the waveform data 
using a constant fraction method to mitigate the time walk effect;
$t_{\mathrm{prop},i}$ is the travel time of the scintillation light from the reconstructed first interaction point to each sensor; 
$t_{\mathrm{walk},i}$ is the effect of the remaining time walk, and $t_{\mathrm{offset},i}$ is the time offset of each sensor.
Photosensors that have detected $\nphe > 50$ are used in the fit.

The two \photon-rays from $\piup^0$ decay are used to calibrate $t_{\mathrm{walk},i}$ and $t_{\mathrm{offset},i}$ and evaluate the time resolution.
A pre-shower counter, consisting of a \SI{4}{\mm} thick lead converter and two plastic plates read 
out by MPPCs from both ends, was installed in the CEX run on the opposite side of the LXe detector across the LH$_2$ target to measure time with high precision.
The pre-shower counter detects one of the two \photon-rays and the 
detection time \textcolor{black}{$t_{\photon,\rm{ps}}$} is used as a reference for the hit time of the other \photon-ray in the LXe detector. 
The parameter $t_{\mathrm{walk},i}$ is obtained as a function of $\nphe$ for the remaining time from the reference time, and a common function is used for photosensors on the same face of the detector,
while $t_{\mathrm{offset},i}$ is obtained for each channel as a constant offset remaining after the $t_{\mathrm{walk},i}$ correction.

Since the position of the $\piup^0$ decay vertex is unknown, the spread $\sigma_{\rm{vertex}}$ contributes to the dispersion of the time difference of the two \photon-rays.
The time dispersion due to $\sigma_{\rm{vertex}}$ was measured with two plastic counters, a pre-shower counter and another counter of identical structure placed in front of the LXe detector in a dedicated CEX run.

\begin{figure}
\centering
\includegraphics[width=1\columnwidth]
{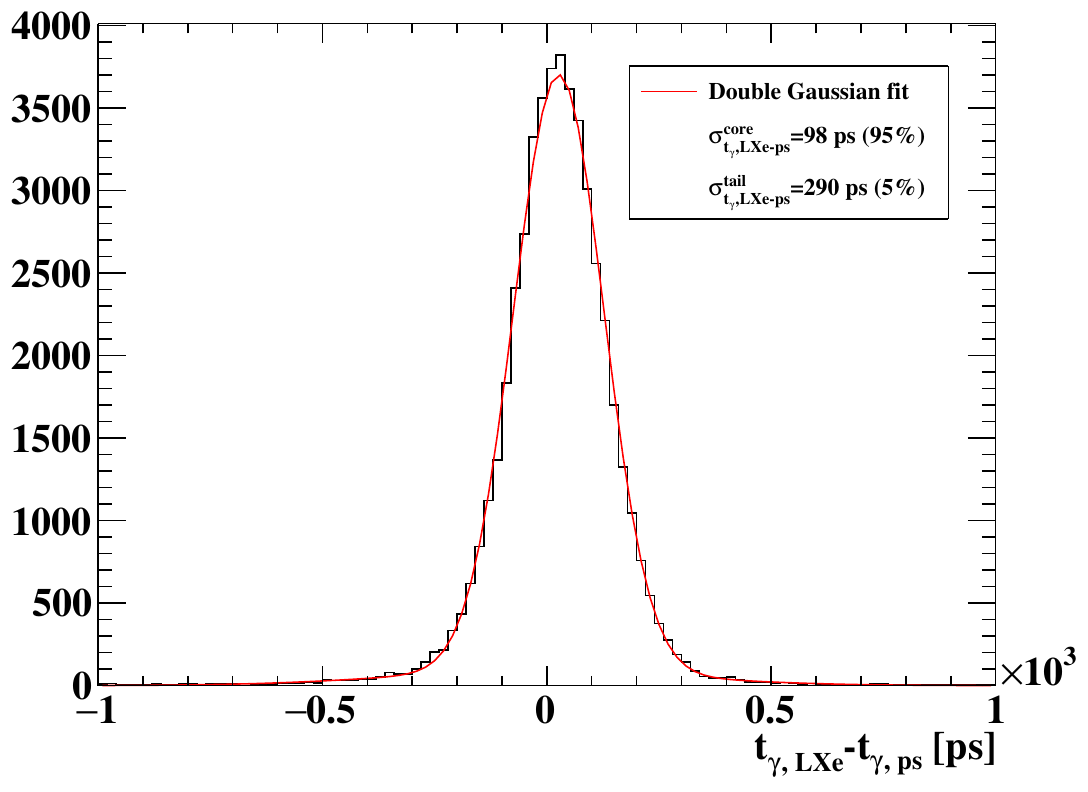}
\caption{The time difference between the reconstructed \photon-ray timing in LXe detector and that on the pre-shower counter for $\egamma=\SI{55}{\MeV}$ fitted with a double Gaussian function.}
\label{fig:LXe_timeresolution}
\end{figure}

\Fref{fig:LXe_timeresolution} shows the distribution of $t_{\photon,\rm{LXe-ps}} = t_{\photon,\rm{LXe}} - t_{\photon,\rm{ps}}$ with
the times at the centre of the vertex distribution converted and corrected for their times of flight.
A double Gaussian function was fitted to the distribution, 
resulting in $\sigma^{\rm{core}}_{t_{\photon,\rm{LXe-ps}}} = \SI{98}{\pico\second}$ in the core part 
(\SI{95}{\percent}) and $\sigma^{\rm{tail}}_{t_{\photon,\rm{LXe-ps}}}=
\SI{290}{\pico\second}$ in the tail part (\SI{5}{\percent}).
The time resolution of the pre-shower counter was assessed by the difference between the time 
of the incident \photon-ray reconstructed by each plate:
$\sigma_{t_{\photon,\rm{ps}}}= \qty[separate-uncertainty-units= single]{28.2(0.2)}{\pico\second}$. 
The time dispersion due to $\sigma_{\rm{vertex}}$ is $\sigma_{t_{\photon},\mathrm{vertex}} = \qty[separate-uncertainty-units = single]{68(6)}{\pico\second}$.
Finally, the time resolution of the LXe detector is $\sigma_{t_{\photon,\rm{LXe}}} = \qty[separate-uncertainty-units = single]{65(6)}{\pico\second}$ at $\egamma=\SI{55}{\MeV}$. 
The large uncertainty in $\sigma_{t_{\photon,\rm{LXe}}}$ results from the 
large statistical uncertainty in $\sigma_{t_{\photon},\mathrm{vertex}}$, due to the instability of the LH$_2$ target in the 2021 CEX run. 

\subsection{Energy reconstruction and resolution}
\label{subsubsec:LXec-energy}
The energy of the incident \photon-ray, $E_{\photon}$, is reconstructed by adding
the number of scintillation photons collected by each photosensor $N_{\mathrm{pho}, i}$ and scaling it as 
\begin{align}
  &E_{\photon} = S \times T(t) \times F(u,v,w) \times N_{\mathrm{sum}},
    \label{eq:LXe-energy} \\
  &N_{\mathrm{sum}} = N_{\mathrm{MPPC}} \times k(t) + N_{\mathrm{PMT}},
    \label{eq:LXe-nsum} \\
  &N_{\mathrm{MPPC(PMT)}} = \sum_{i\in \mathrm{MPPC(PMT)}} w_i(u, v) \times N_{\mathrm{pho}, i},
   \label{eq:LXe-nmppc}
\end{align}
where $S$ is the scaling factor of $N_{\mathrm{sum}}$ to the energy,
$T(t)$ and $k(t)$ are correction functions for the temporal evolution described in \sref{sec:XECLightYield},
$F(u,v,w)$ is a function to correct for position dependence described in \sref{sec:XEC-Uniformity},
and $w_i(u,v)$ is a weight for each photosensor.
The weight $w_i(u,v)$ is a product of the photosensor coverage, the dead channel compensation factor (see \sref{sec:XECoverview}), and the light collection efficiency factor as a function of the reconstructed \photon-ray position $(u, v)$ and the face of the detector.

Pile-up of low-energy \photon-rays from $\muonp$ decays affect the energy reconstruction.
To mitigate this effect, 
pile-up identification and unfolding algorithms are applied as described in \sref{subsubsec:LXec-BG}.

\subsubsection{Time dependence} 
\label{sec:XECLightYield}

\begin{figure}[tbp]
\centering
\includegraphics[width=1\columnwidth]{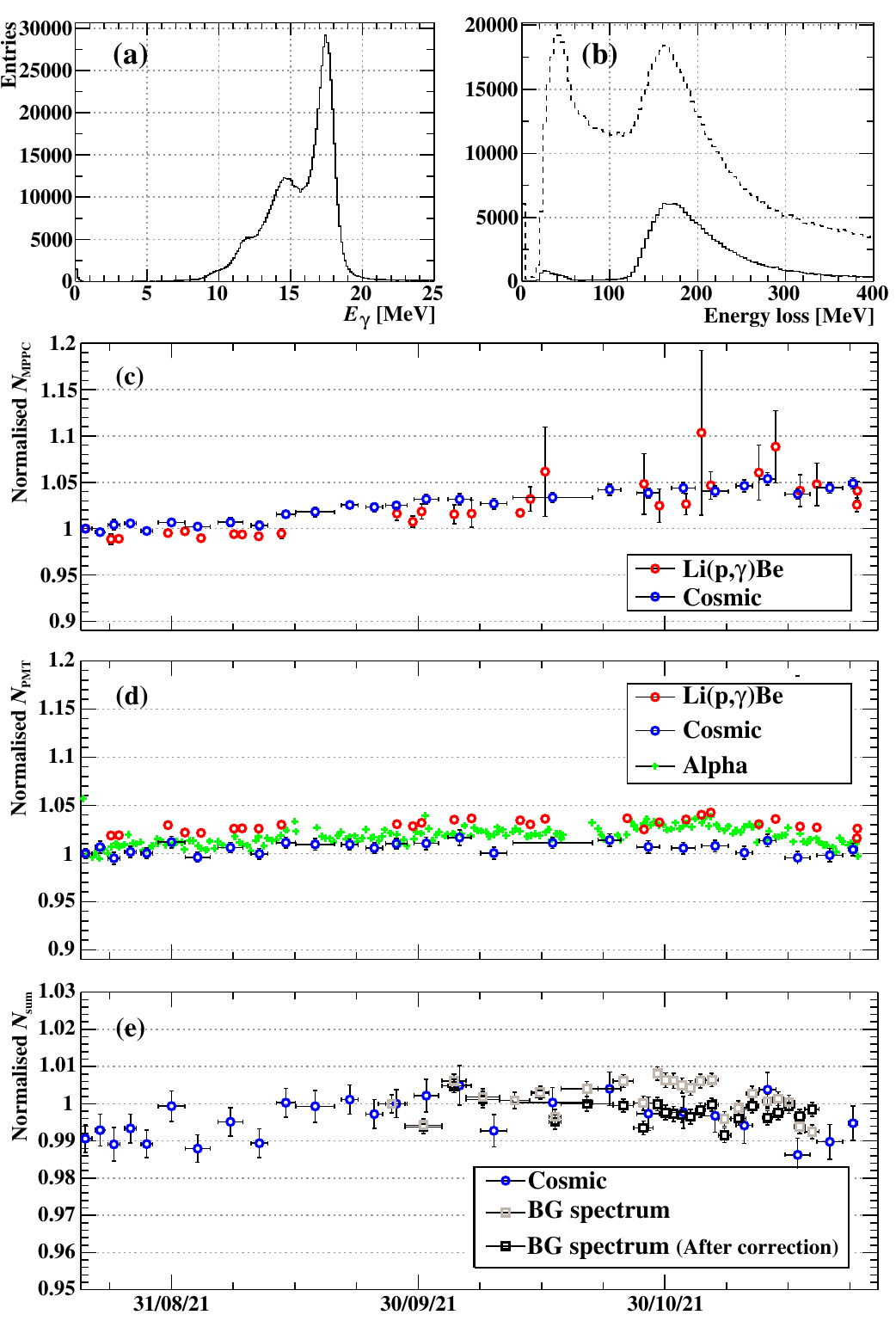}
\caption{(a): Energy spectrum of the $^{7}\mathrm{Li}(\mathrm{p},\photon)^{8}\mathrm{Be}$ line.
(b): Energy spectrum of the cosmic ray data. The dotted line shows the raw 
spectrum while the solid line shows the spectrum after the cosmic ray event selection. (c) and (d): 
Temporal variation of the light yield for each calibration source during 2021 run measured by MPPCs and PMTs, respectively. 
The data are normalised to 1 at the beginning of the beam time. (e): Temporal variation of $N_{\textrm{sum}}$. The data are normalised at 1 at the beginning of physics run.}
\label{fig:XECcalibration_LY}
\end{figure}

The stability of the detector is monitored with the $^{7}\mathrm{Li}(\mathrm{p}, \photon)^{8}\mathrm{Be}$ reaction (\SI{17.6}{\MeV} line),
$\alphaup$-particles and cosmic rays.
\Fref{fig:XECcalibration_LY} (a) and (b) show typical energy spectra for the $^{7}\mathrm{Li}(\mathrm{p},\photon)^{8}\mathrm{Be}$ reaction and cosmic rays, respectively.
The $^{7}\mathrm{Li}(\mathrm{p},\photon)^{8}\mathrm{Be}$ spectrum shows a clear peak at \SI{17.6}{\MeV} and a
secondary peak at \qtylist{\sim 14}{\MeV} as expected \cite{Zahnow_1995}.
For the cosmic ray spectrum, event selection based on the reconstructed position and the ratio of $N_{\mathrm{MPPC}}$ to the number of photons detected on the outer faces is applied.
After the selection, the distribution of energy release follows a Landau distribution with a broad peak at \SI{\sim 170}{\MeV}.

The temporal evolution of the normalised numbers of detected photons on the MPPC and PMT surfaces for these data are shown in \fref{fig:XECcalibration_LY}(c)--(d). 
They show common trends for all calibration sources,
but different trends were observed between MPPCs and PMTs. 
This difference results from the LXe purity and the different distances to the $\alphaup$-particle sources used in the
calculation of PDE (QE).
The cosmic ray data are used to correct for the difference as the time-dependent 
weights of MPPCs and PMTs $k(t)$ in \eref{eq:LXe-nsum}.

\textcolor{black}{The temporal variation can also be evaluated using the energy spectrum of the background \photon-rays, with an endpoint energy of \SI{52.83}{\MeV}, measured during the physics run  (see also \sref{subsubsec:LXec-BG}).}
\Fref{fig:XECcalibration_LY}(e) shows the temporal evolution of the normalised $N_{\textrm{sum}}$ for the cosmic rays and the background \photon-rays.
A function $T(t)$  to correct for this time dependence of $N_{\textrm{sum}}$ is constructed using 
cosmic rays for a finer structure of temporal evolution and background \photon-rays for a coarser structure.
The estimated accuracy of the temporal evolution correction is \SI{0.3}{\percent} from the residual variations of the scales for the background \photon-rays. 

\subsubsection{Position dependence}
\label{sec:XEC-Uniformity}

\begin{figure}[tbp]
\centering
\includegraphics[width=1\columnwidth]{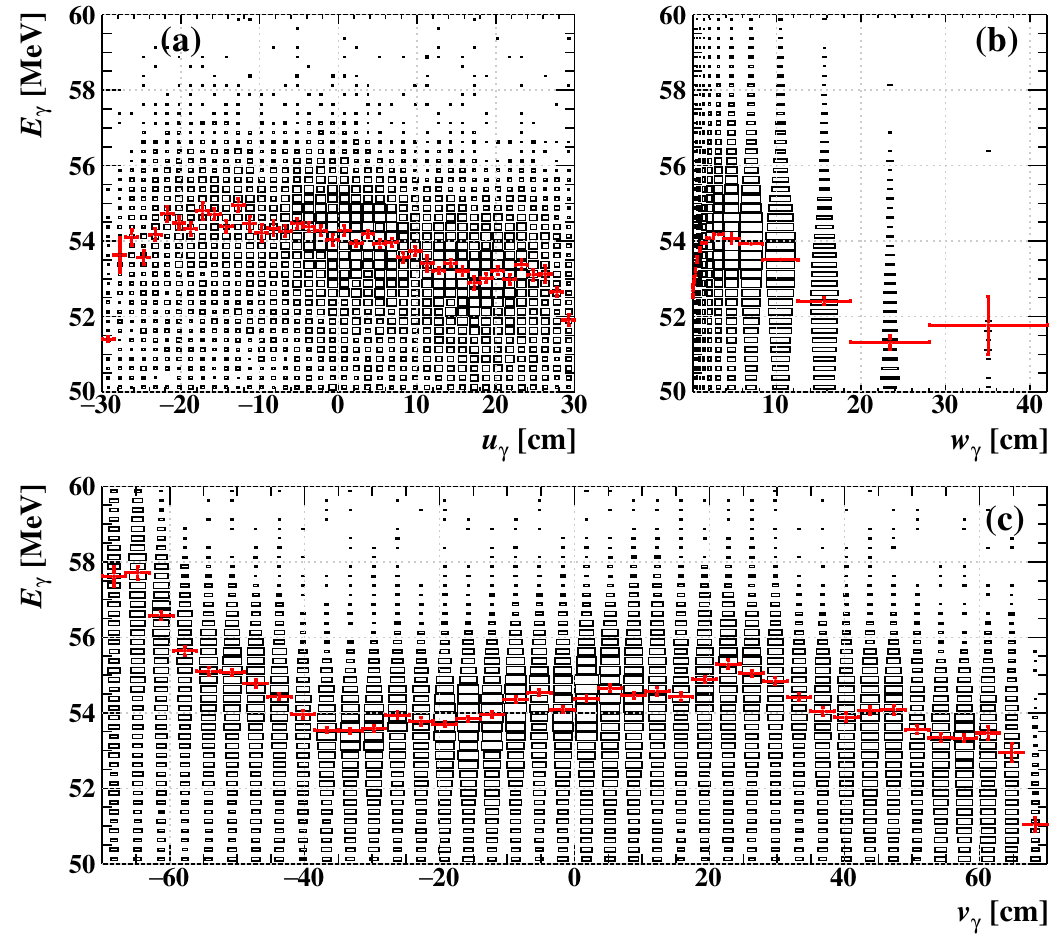}
\caption{Position dependence of the reconstructed \SI{55}{\MeV} \photon-rays from $\piup^{0}\to \photon\photon$ decays. The red points are the peak energy in slices projected along $u_{\photon}$, $v_{\photon}$, or $w_{\photon}$.}
\label{fig:XECcalibration_NonUniformity}
\end{figure}

A non-uniform response of the energy scale is visible in \fref{fig:XECcalibration_NonUniformity}.
A three-dimensional correction function $F(u,v,w)$ was created using the \SI{55}{\MeV} \photon-rays of $\piup^{0}\to \photon\photon$ decays, the $^{7}\mathrm{Li}(\mathrm{p},\photon)^{8}\mathrm{Be}$ \SI{17.6}{\MeV} line and the background \photon{}-rays to smooth the response.

The physics run in 2021 was divided into five periods based on the intensity of the muon beam, and $F(u,v,w)$ was calculated for each period.
The period-dependent $F(u,v,w)$ can not only complement the correction for the temporal evolution due to the change in LXe purity but also contribute to the stability of the energy resolution.

The accuracy of the non-uniformity correction is estimated to be \SI{0.2}{\percent} by mean of the uncertainty of the three-dimensional correction factors.

\subsubsection{Energy scale} 

\begin{table}[tb]
  \centering
   \caption{Breakdown of the energy scale uncertainty in 2021.}
   \begin{tabular}{l r} 
   \hline
   Element & Uncertainty (\%) \\
   \hline
   Temporal evolution & 0.3 \\
   Position dependence & 0.2 \\
   Non-linearity between \SI{55}{\MeV} and \SI{52.83}{\MeV} & 0.1 \\
   \hline
   Total & 0.4 \\
   \hline
   \end{tabular}
   \label{tab:EGammaScaleUncertainty}
  \end{table}

The scale factor $S$ is determined from the peak of the $N_{\mathrm{sum}}$ distribution for the \SI{55}{\MeV} \photon-rays and the energy scale of the background \photon{}-rays.
The uncertainty for the $\egamma$ energy scale in 2021 is \SI{0.4}{\percent},
\tref{tab:EGammaScaleUncertainty} reports the breakdown in separate contributions.

\subsubsection{Energy resolution} 
\label{sec:xec_energy}
\begin{figure}[t]
  \centering
  \includegraphics[width=1\linewidth]{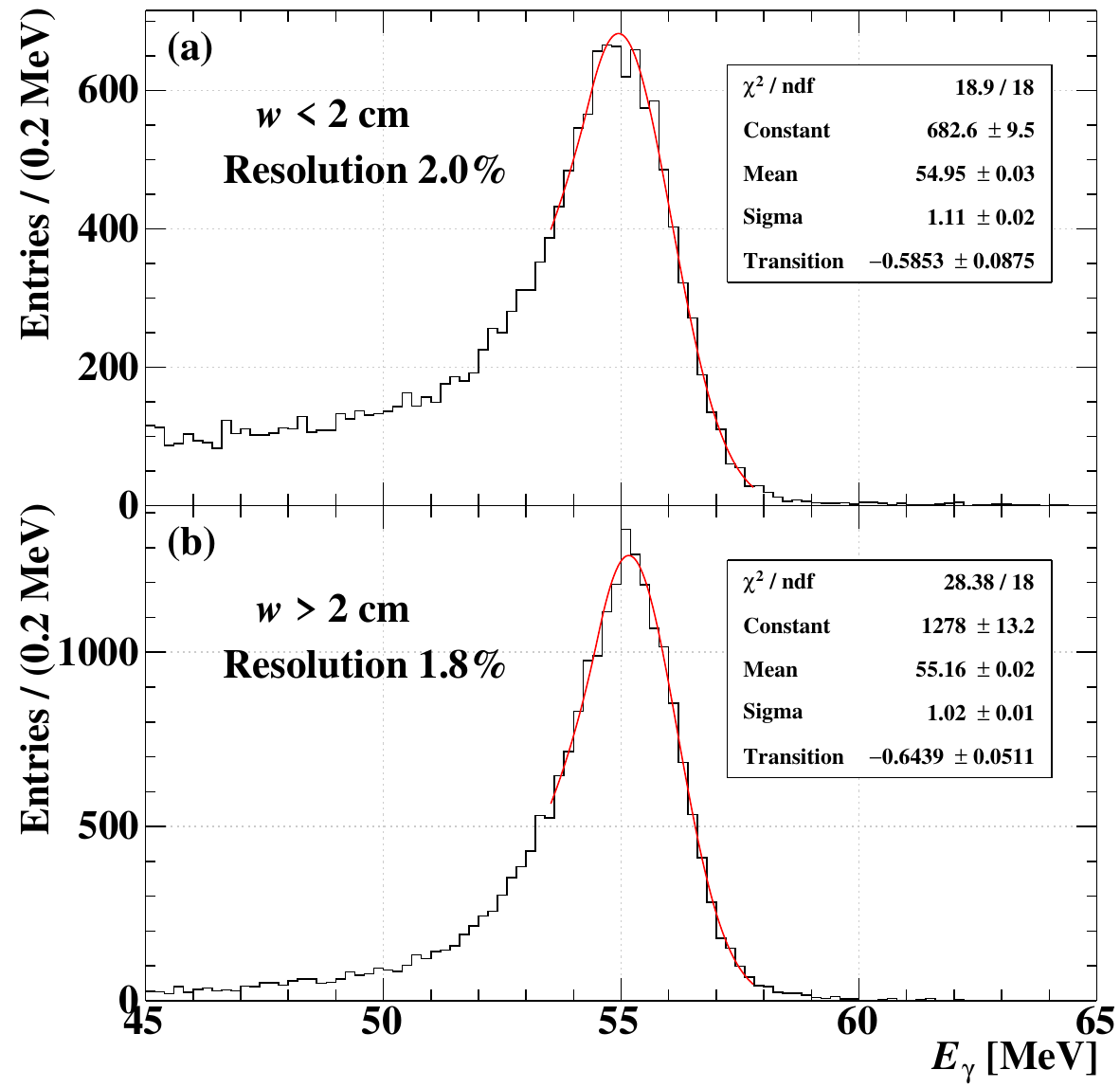}
    \caption{Energy response to \SI{55}{\MeV} \photon-rays hittinh a central area of the detector 
($u \in [\SI{-10}{\cm}, \SI{10}{\cm}] \land v \in [\SI{-30}{\cm}, \SI{-10}{\cm}]$) in different $w$ ranges. The fitting function of \eref{eq:LXe-ExpGaus} is shown in red. 
    }
  \label{fig:LXe-55MeVEGamma}
\end{figure}

The energy response was investigated with the quasi-monochromatic \SI{55}{\MeV} \photon-rays.
\Fref{fig:LXe-55MeVEGamma} shows the energy response in a central part of the detector in different $w$ regions.
Asymmetric spectra were observed mainly at $w < \SI{2}{cm}$; the low-energy tail originates 
from \photon-rays interacting in front of the detector fiducial volume and shower leaks from the incident face.
The energy resolution is evaluated by fitting the following function to the data:
\begin{equation}
  f(x) =
  \begin{cases}
    A \exp{\left[ -\frac{\left(x - \mu_{\egamma}\right)^2}{2 \sigma_{\egamma}^2} \right]} & \quad (\text{if }x > \mu_{\egamma} + \tau), \\
    A \exp{\left[ \frac{\tau\left(\tau/2 - x + \mu_{\egamma}\right)}{\sigma_{\egamma}^2} \right]} & \quad (\text{if }x \le \mu_{\egamma} + \tau),
  \end{cases}
  \label{eq:LXe-ExpGaus}
\end{equation}
where $A$, $\mu_{\egamma}$, $\sigma_{\egamma}$, and $\tau$ are the constant, mean, sigma, and 
transition parameters in \fref{fig:LXe-55MeVEGamma}, respectively.
The relative energy resolution is equal to $\sigma_{\egamma}/ \mu_{\egamma} = \SI{2.0}{\percent}$ $(\SI{1.8}{\percent})$ 
for $w<\SI{2}{\cm}$ ($w>\SI{2}{\cm}$).

The expected relative energy resolution for MEG~II was estimated to be \SIrange[range-phrase=--,range-units = single]{1.0}{1.7}{\percent} based on the 
difference between the measured and simulated values in MEG \cite{baldini_2018}
as the reason for the difference was not fully understood.
The obtained resolution of \SI{1.8}{\percent} for $w > \SI{2}{\cm}$ corresponds to the worst case where the difference is completely preserved.
The reason for this is not entirely clear, but one of the possible causes is the optical properties of LXe, as discussed in \cite{baldini_2018}, as the difference was observed in both MEG and MEG~II experiments. The hypothesis that the cause of this difference is due to the PMT behaviour is ruled out since it remains despite the change from PMTs to MPPCs on the front face in MEG~II.

\subsection{Background rejection and efficiency}
\label{subsubsec:LXec-BG}

Pile-up \photon-rays at high beam intensities increase the background events near the signal energy.
Therefore, a dedicated analysis is applied to reduce this contamination.
%
Pile-up events can be identified from the light distribution in the MPPCs.
If the pile-up \photon-rays are temporally separated from the main \photon-ray, 
multiple \photon-rays are unfolded in the weighted sum waveforms of the MPPCs and PMTs by fitting a superposition of $n$ template pulses, where $n$ is the number of detected \photon-rays.
If, on the other hand, the pile-up \photon-rays are too close in time with the main \photon-ray to be unfolded,  
such events are discarded from the analysis sample.
If the reduced chi-square values of the waveform fitting are large ($\chi_\mathrm{PMT}^2/N_\mathrm{dof} >8$  or $\chi_\mathrm{MPPC}^2/N_\mathrm{dof}>20$), 
such events are also discarded. 

\begin{figure}[t]
  \centering
  \includegraphics[width=1\columnwidth]{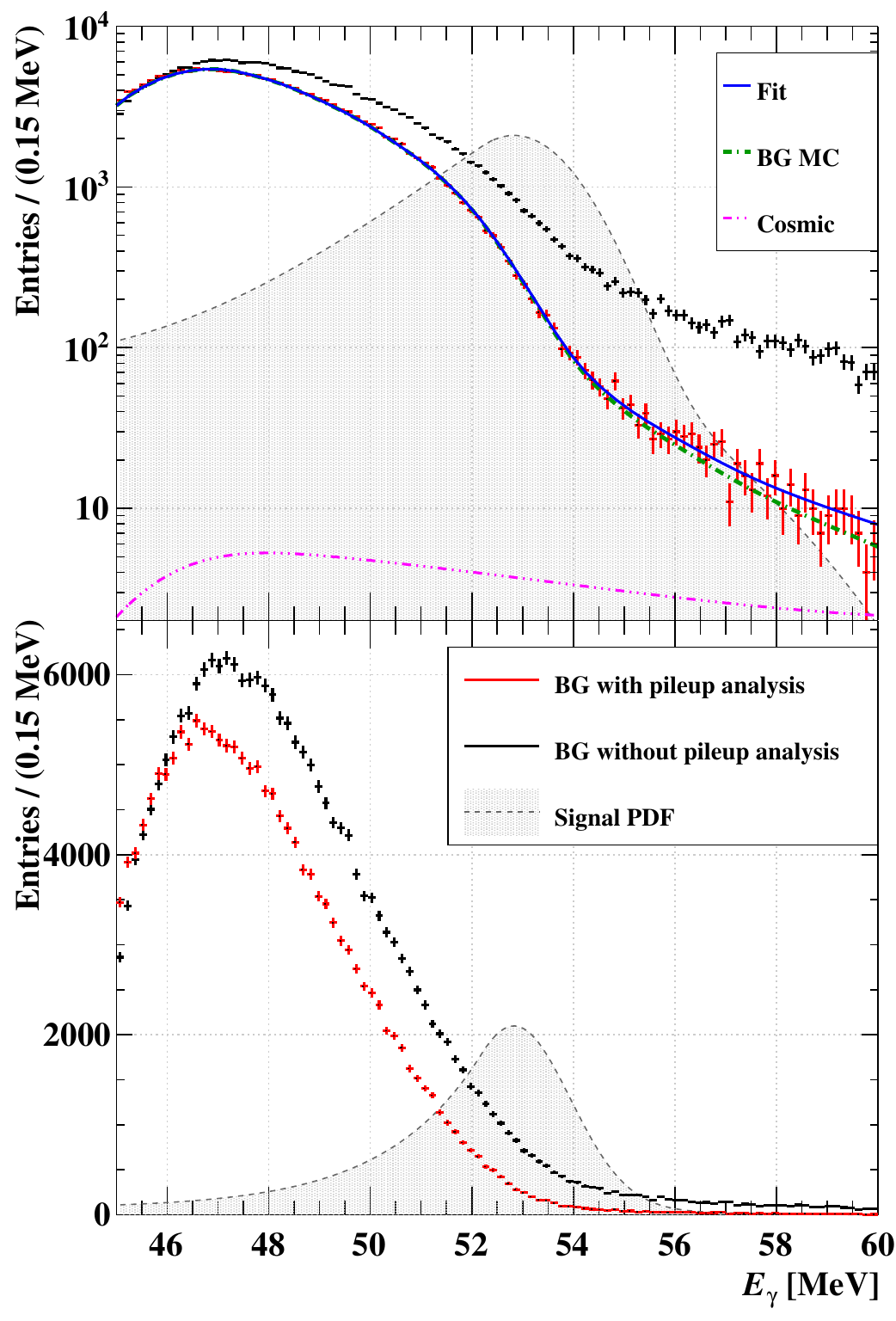}
  \caption{Background $\egamma$ spectra with (red) and without (black) pile-up 
  analysis 
  at $R_{\muon} = \SI{5e7}{\per\second}$ in 
  log scale (top) and in linear scale (bottom). The drop below \SI{\sim47}{\MeV} is a 
  trigger threshold effect (see \sref{sec:meg_trigger}). 
  The blue solid curve in the top panel is the best fit to the spectrum with pile-up 
  analysis, resulting from the sum of the spectrum obtained with a dedicated MC 
  simulation \textcolor{black}{smeared for the worse resolution in the data} (green dashed) and the measured cosmic ray spectrum (magenta dashed).
  The expected spectrum for the signal \photon-rays, arbitrarily normalised to 
  \num{50000} events in $\egamma \in [\SI{48}{\MeV},\SI{58}{\MeV}]$, is also superimposed (grey shaded).}
  \label{fig:LXe-BGEGamma}
\end{figure}

The background $E_{\photon}$ spectra with and without this pile-up analysis at $R_{\muon} = \SI{5e7}{\per\second}$ are shown in \fref{fig:LXe-BGEGamma}. 
The background events with $\egamma \in [\SI{48}{\MeV},\SI{58}{\MeV}]$ 
are reduced by \SI{35}{\percent} by the pile-up analysis. 
The reduction around the signal energy is apparent.
The analysis efficiency for signal \photon{}-rays after applying this pile-up analysis is estimated to be \SI{92(2)}{\percent} at 
$R_{\muon} = \SI{3e7}{\per\second}$ using a dedicated MC simulation of signal events with pile-up \photon-rays and RMD-enhanced data tagged by the RDC.

The cosmic ray background events in the signal region have different depth distributions and MPPC/PMT charge ratios than the signal events originating from the target. Using this information, events identified as cosmic rays are removed. The inefficiency of the signal events due to this cut is negligible.


The detection efficiency for signal \photon-ray is estimated from the MC simulation 
$\varepsilon_\photon = \SI{69}{\percent}$ \cite{baldini_2018}.
To check and correct this MC estimate, we used the $\piup^{0}\to 2\photon$ data. 
Once the BGO calorimeter detects an \SI{83}{\MeV} \photon-ray, we know that the other 
\SI{55}{\MeV} \photon-ray is emitted in the opposite direction (namely, towards the LXe detection).
Therefore, by counting the events where the LXe 
detector detects \photon-rays in the self-triggered BGO events, we 
can measure the detection efficiency of the LXe detector at \SI{55}{\MeV}.
To account for the different measurement conditions compared to the standard 
physics run, such as the additional material for the LH$_2$ target 
system, a dedicated MC simulation was performed for this measurement. 
The estimated MC efficiency for this setup is 
$\varepsilon^{\mathrm{MC}}_{\photon,\mathrm{CEX}} = \SI{64}{\percent}$,
while the efficiency from the data is $\varepsilon^\mathrm{data}_{\photon,\mathrm{CEX}} = \SI{61 (1)}{\percent}$.
This discrepancy has not yet been resolved and is accounted for
in the systematic uncertainty. The best estimate of the detection efficiency for the 
signal \photon-rays is $\varepsilon_\photon = \SI{67(2)}{\percent}$.

The overall efficiency for signal \photon-rays is the product of the detection efficiency 
and the analysis efficiency $\varepsilon_\photon = \SI{62 (3)}{\percent}$.

\section{Radiative decay counter}
\label{sec:rdc}

One of the main sources of high-energy \photon-rays are the RMD (\radiative) events.
If one of these events coincides in time with a 
high-energy positron from a Michel decay 
of another muon, it can become an accidental background event.
The RDC is a new subdetector introduced in the MEG~II experiment to identify such 
high-energy \photon-rays by detecting the low-energy positrons emitted in the same RMD.
When an RMD emits a \photon-ray with an energy above \SI{48}{\mega\eV}, 
the energy of a large fraction (\SI{\sim 66}{\percent}) of the accompanying positrons falls in the
\qtyrange[range-units = single]{1}{5}{\mega\eV} range.
The trajectories of these low-energy positrons in the COBRA magnetic field have too small radii 
to be detected by CDCH or pTC. 
Therefore, dedicated positron detectors located close to the beam axis 
are required.

Time-coincidence measurement can be used to identify the accidental background events with RMD \photon-rays, but
standard Michel positrons happen to hit accidentally the RDC; they usually have higher energies than RMD positrons.
Hence, measurement of positron energy is an additional effective tool to select efficiently the RMD positrons.


In 2017, an RDC was installed at the downstream beamline (Downstream-RDC, DS-RDC). It measures both energy and timing.
An additional RDC that measures timing only is currently being 
developed for installation upstream (Upstream-RDC, US-RDC).

\subsection{Downstream RDC}
\label{sec:rdcdownstream}

\begin{figure}[tbp]
\centering
\includegraphics[width=1\columnwidth]{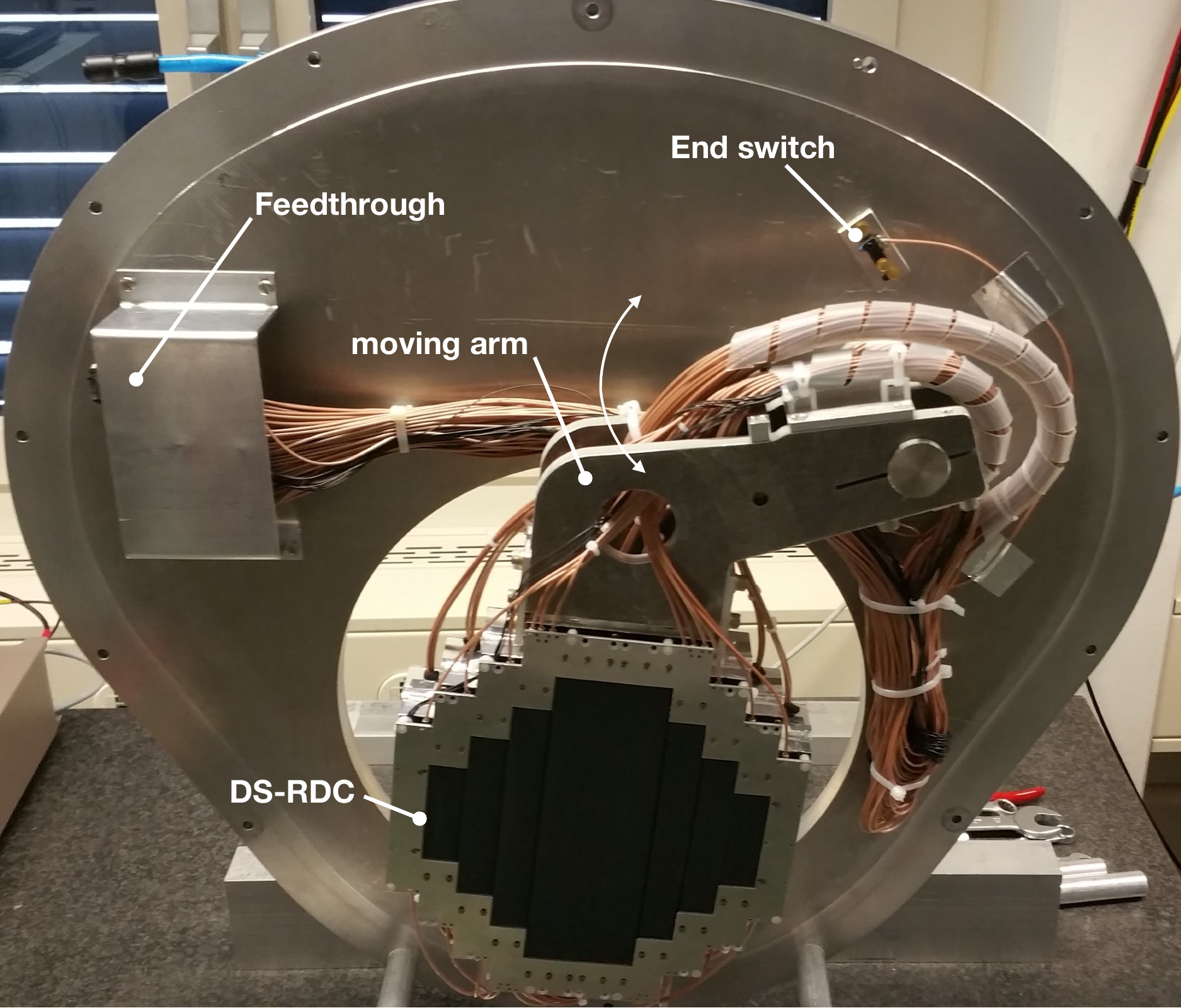}
\includegraphics[width=0.7\columnwidth]{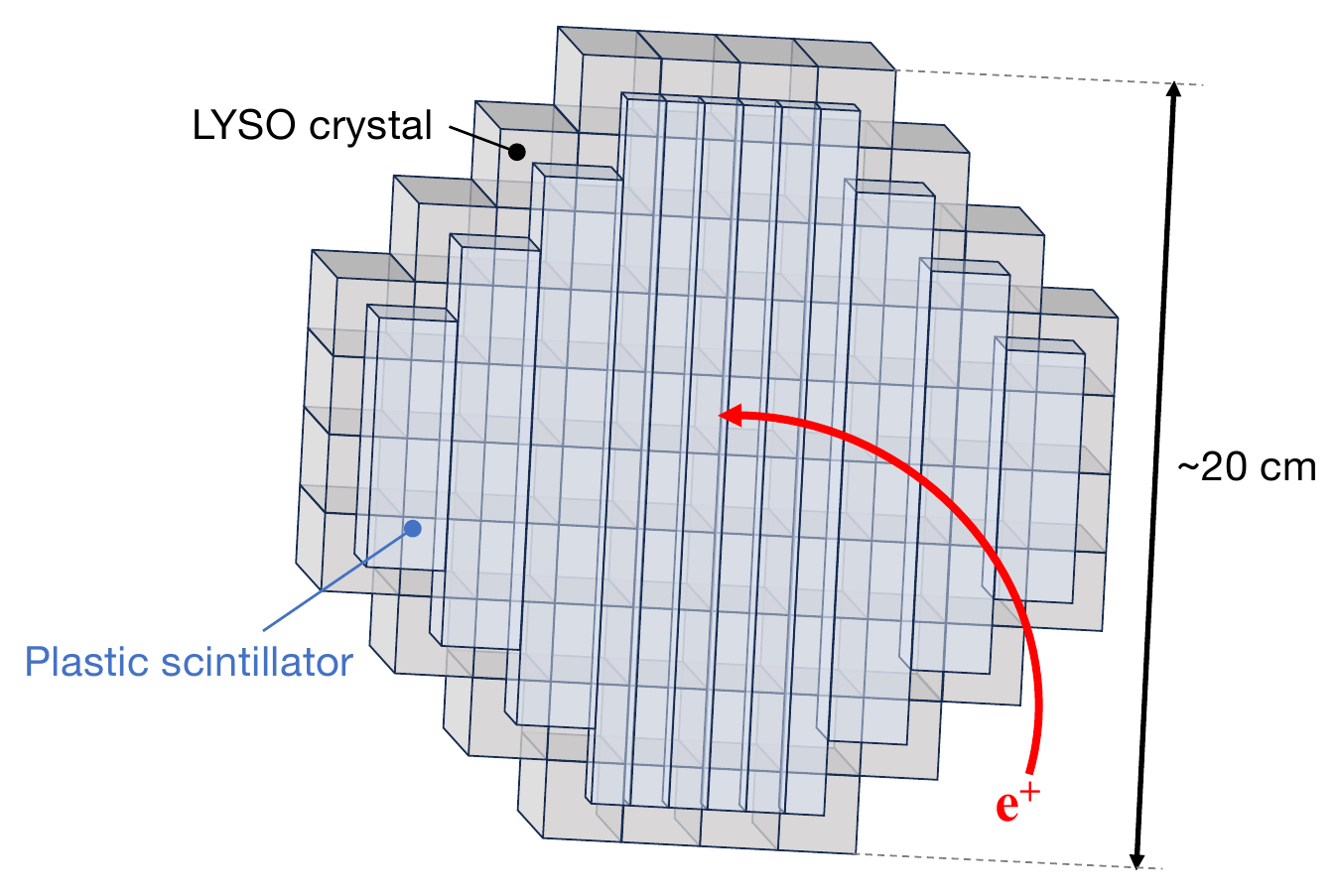}
\caption{The downstream RDC.}
\label{fig:DS-RDC}
\end{figure}
The DS-RDC is located \SI{142}{\cm} downstream of the target on the beam axis.
It consists of 76 LYSO crystals for energy measurement, arranged in an octagonal plane, and 12 plastic scintillator bars (PSs) for timing measurement, arranged in front of the crystals in a plane covering the same area.
The size of each LYSO crystal is \qtyproduct[product-units = bracket-power]{2x2x2}{\cm}.
Each crystal is covered with a \SI{65}{\micro\meter} thick reflective 
sheet (ESR, 3M) to increase efficiency of the light collection.
One MPPC with pixel size of \SI{25}{\micro\meter} (S125720-25, Hamamatsu Photonics) is attached to each LYSO crystal with a spring.
Six \SI{1}{\cm} wide PS bars are used in the central region; two MPPCs are glued on each side of each bar with optical cement.
The other six PS bars located in the outer regions are \SI{2}{\cm} wide with varying lengths; three MPPCs are glued on each side.
The MPPCs on the same side of each bar are connected in series and their signals are read out by one channel. 

The detector is supported by a movable arm with a pneumatic cylinder so that it can be
moved out of the beam axis to insert the beamline for the proton beam from the CW 
accelerator and the Li target to calibrate the LXe detector.
\Fref{fig:DS-RDC} shows the DS-RDC and the moving arm system.
The position of the DS-RDC is monitored by physical switches. 
To avoid collisions with the CW beamline, 
the two systems are software-locked so that one cannot be deployed while the other is installed.

\subsubsection{Operation of the downstream RDC}

The DS-RDC is expected to be on the beam axis (measurement position) during the physics run,
but this was not always the case as the interlock system was not yet fully functional. 
In 2021(2022) the DS-RDC was in the measurement position for \SI{74}{\percent} (\SI{89}{\percent}) of the physics run.

\begin{figure}[tbp]
  \centering
  \includegraphics[width=1\columnwidth]{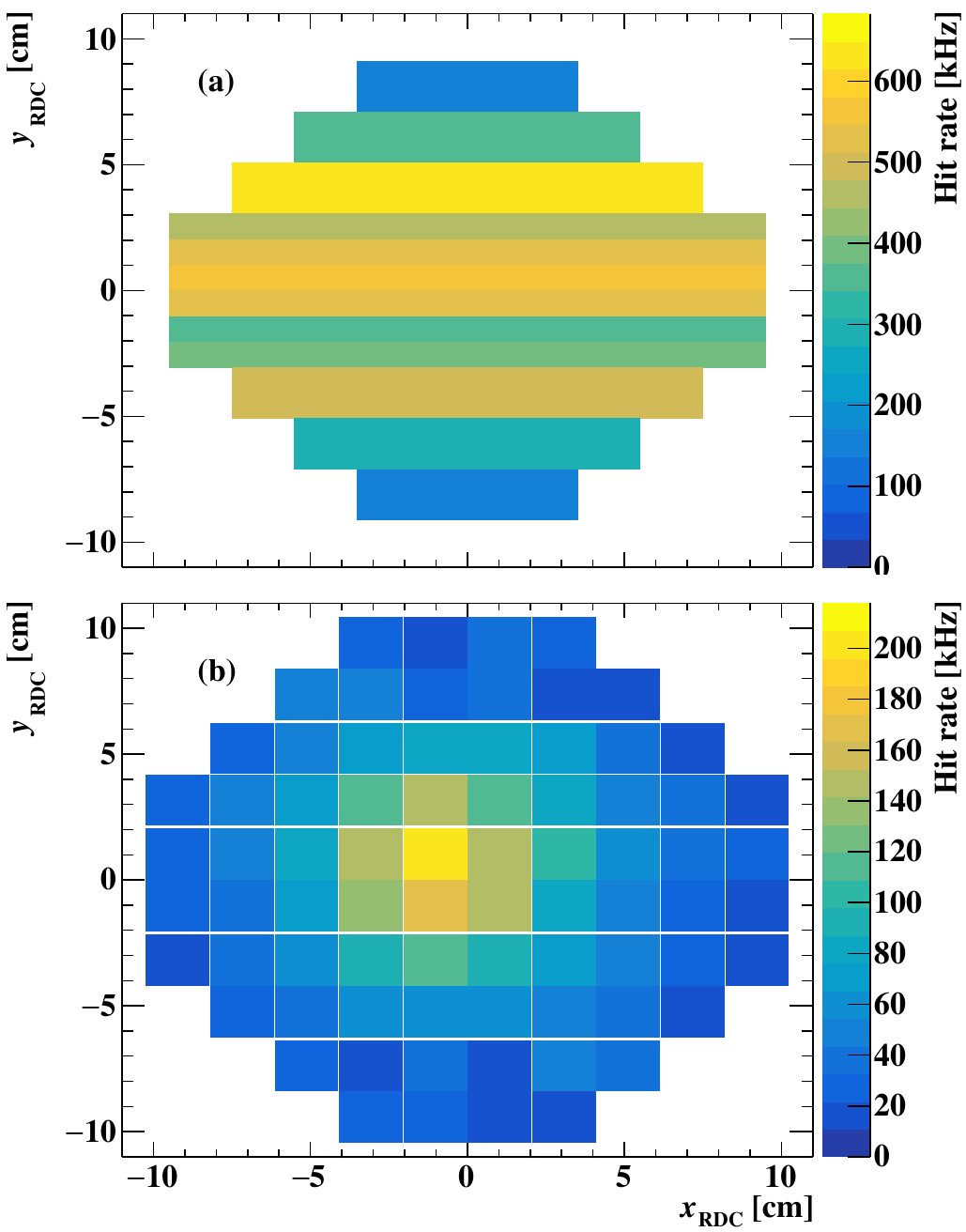}
  \caption{Hit rate distributions of PSs (a) and crystals (b) at $R_{\muon} = \SI{3e7}{\per\second}$ in 2021. The axes are the DS-RDC local coordinates defined by a rotation of \SI{83}{\degree} around the beam axis.}
  \label{fig:rdc-hitratemap}
\end{figure}

\Fref{fig:rdc-hitratemap} shows the distributions of hit rates of PSs and crystals at $R_{\muon} = \SI{3e7}{\per\second}$ in 2021.
As expected, the distributions peak near the centre.
The shift of the peak from the centre is due to the off-centre beam position on the target shown in \fref{fig:PositronTGTDist2D}.
The total hit rates in the DS-RDC are \SI{4.3}{\MHz}, \SI{5.5}{\MHz}, and \SI{6.5}
{\MHz} at $R_{\muon} = \text{3, 4, and } \SI{5e7}{\per\second}$, respectively.
The ratios of the overall hit rates to $R_{\muon}$ are independent of $R_{\muon}$ and are 
always \numrange[range-phrase=--,range-units=single]{0.13}{0.14}.
Therefore, the DS-RDC serves as a real-time diagnostic tool for the beam using the hit rate and distribution.

The energy deposition in each LYSO crystal is reconstructed as
\begin{align}\label{EneRec}
  E_{i,\mathrm{LYSO}} & = A_i \times Q_{i,\mathrm{LYSO}} \ \ (i=0,1,\cdots, 75),
\end{align}
where $Q_{i,\mathrm{LYSO}}$ is the integrated charge of the LYSO signal for a \SI{250}{\ns} window 
and $A_i$ is an energy scale factor for each crystal.
The factor $A_i$ was calibrated before the start of the physics 
run using the \SI{597}{\keV} (\SI{88}+\SI{202}+\SI{307}{\keV}) \photon-ray emission 
following the $\beta$-decay of $^{176}$Lu, which is naturally present in LYSO crystals (self-luminescence).

\begin{figure}[tbp]
\centering
\includegraphics[width=1\columnwidth]{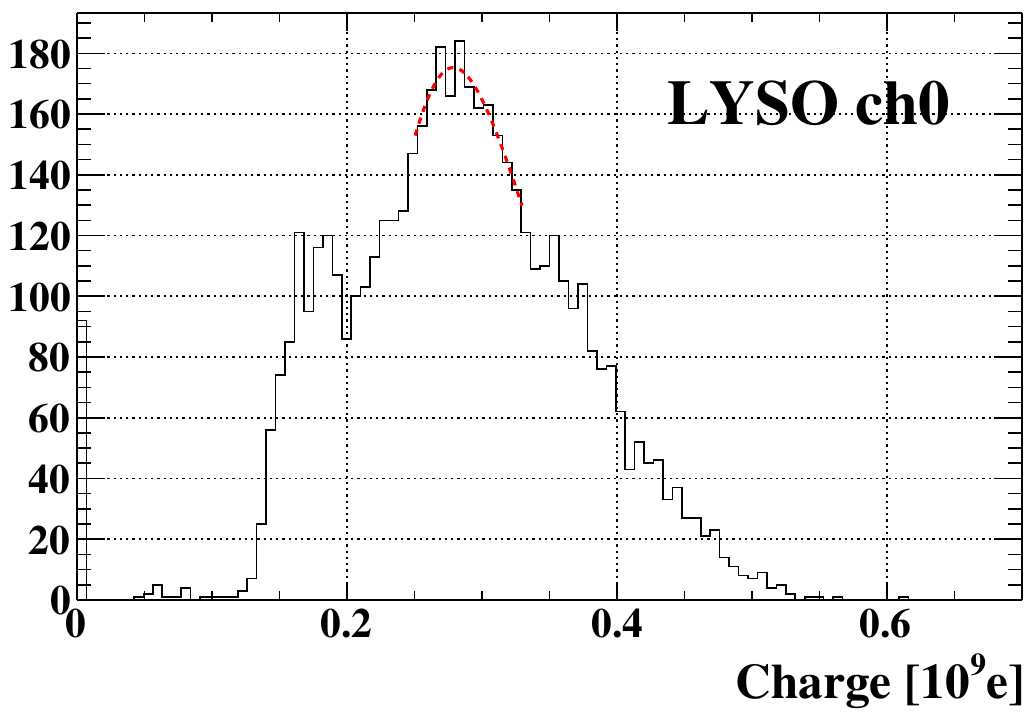}
\caption{Charge spectrum for the LYSO self-luminescence and fit result with \eref{eq:fitfunctionRDC} (red curve)}
\label{fig:CalibrationRDC2021}
\end{figure}

\Fref{fig:CalibrationRDC2021} shows an example of the charge distribution obtained by self-luminescence.
The following function is fitted to the main peak to obtain $A_i$:
\begin{equation}
    \int \mathcal{G}\left(x/A_i-E, E\times\sigma_E+\sigma_{\textrm{noise}}^2\right)\times \mathcal{D}\left(E\right)dE, \label{eq:fitfunctionRDC}
\end{equation}
\begin{eqnarray}
    \mathcal{G}\left(x,\sigma\right) &=& \textrm{exp}\left(-x^2/\sigma^2\right), \\
    \mathcal{D}\left(x\right) &=& \sum^{6}_{j=0} p_j\times \beta\left(x/A_i-E_j\right), \\
    \beta\left(x\right) &=& C\left(Q-x\right)^2\left(x+m_\mathrm{e}c^2\right)\sqrt{x^2+2x m_\mathrm{e}c^2},
\end{eqnarray}
where $E_{0-6}$
are the energies of the LYSO self-luminescence \photon-ray peaks, $p_{0-6}$ are the fitting 
parameters representing the amplitudes of the \photon-ray peaks, $C$ is a constant, $Q=\SI{596}{\keV}$ 
is the Q-value of the $^{176}$Lu $\beta$-decay, and $m_\mathrm{e} c^2$ is the rest energy of the electron.

\subsubsection{Performance}
\label{sec:rdcperformance}
\begin{figure}[tbp]
  \centering
  \includegraphics[width=1\columnwidth]{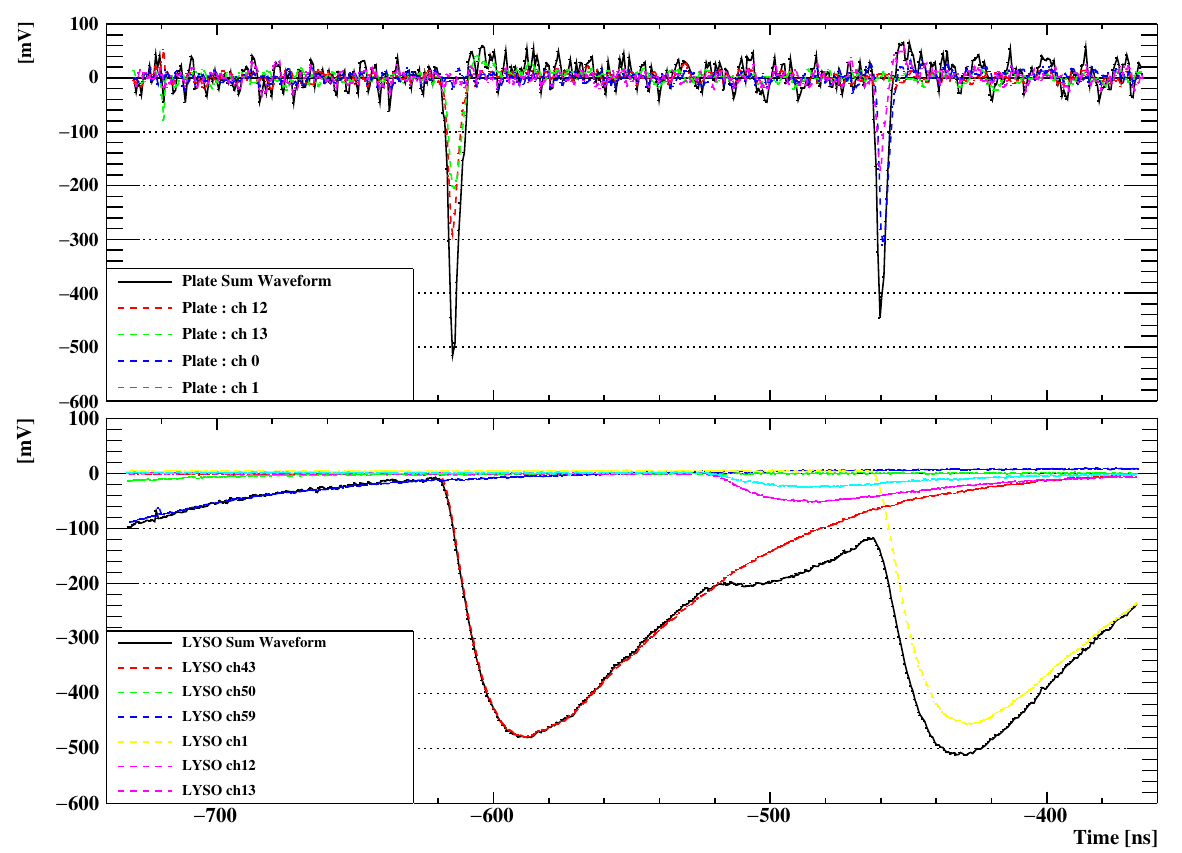}
  \caption{Typical waveforms of the PSs (top) and the LYSO crystals (bottom).}
  \label{fig:typicalWF_RDC} 
\end{figure}
Typical waveforms of the DS-RDC are shown in \fref{fig:typicalWF_RDC}.
Due to the high hit rate, several positrons can hit the DS-RDC in one time window; in this example, 
two pulses are visible in the PSs and four in the LYSO crystals.
First, the signal from each PS is analysed to detect hits and to measure their times $t_{\positron,\textrm{RDC}}$. 
The time response of the PSs is fast enough to reconstruct such pile-up hits separately.  
The time resolution is $\sigma_{t_{\positron,\textrm{RDC}}} < \SI{90}{\ps}$ for all bars measured using a $^{90}\mathrm{Sr}$ source in laboratory tests \cite{baldini_2018}.
Then, the signals of the LYSO crystals, whose decay time constant is approximately \SI{50}{\ns}, are analysed for each hit found in the PSs.
The integration window is defined by $t_{\positron,\textrm{RDC}}$, and the 
energy for the hit, $E_{\positron, \textrm{RDC}}$, is reconstructed as the sum of 
$ E_{i,\mathrm{LYSO}}$. 
The energy resolution was estimated to be $\sigma_{E_{\positron, \textrm{RDC}}} = \SI{7.5(3)}{\percent}$ at \SI{1}{\MeV} by fitting the spectrum of self-luminescence and the \SI{1.836}{\MeV} peak of a $^{88}\mathrm{Y}$ source.
%
For \SI{\sim17}{\percent} of the events at $R_{\muon} = \SI{3e7}{\per\second}$,
$E_{\positron, \textrm{RDC}}$
is not reconstructed due to pile-up in LYSO. 
These events are flagged and treated separately in the following analysis.

The RDC is not used as a veto counter but serves as an additional discriminant in the likelihood analysis for the \megp\ search (see \sref{sec:sensitivity}).
The following two observables are used as  discriminating variables between signal and background:
the time difference between DS-RDC and the LXe detector, $t_{\positron,\textrm{RDC}}-t_{\photon,\rm{LXe}}$, and $E_{\positron, \textrm{RDC}}$. 
If multiple hits are reconstructed in one event, the hit with the smallest $|t_{\positron,\textrm{RDC}}-t_{\photon,\rm{LXe}}|$ is selected.
Since the distributions for the RMD observables are correlated with the $\egamma$ 
measured by the LXe detector, the three-dimensional distribution in $(t_{\positron,
\textrm{RDC}}-t_{\photon,\rm{LXe}}, E_{\positron, \textrm{RDC}},\egamma)$ 
for the signal and the background is evaluated and used in the likelihood function. 


\begin{figure}[tbp]
  \centering
  \includegraphics[width=1\columnwidth]{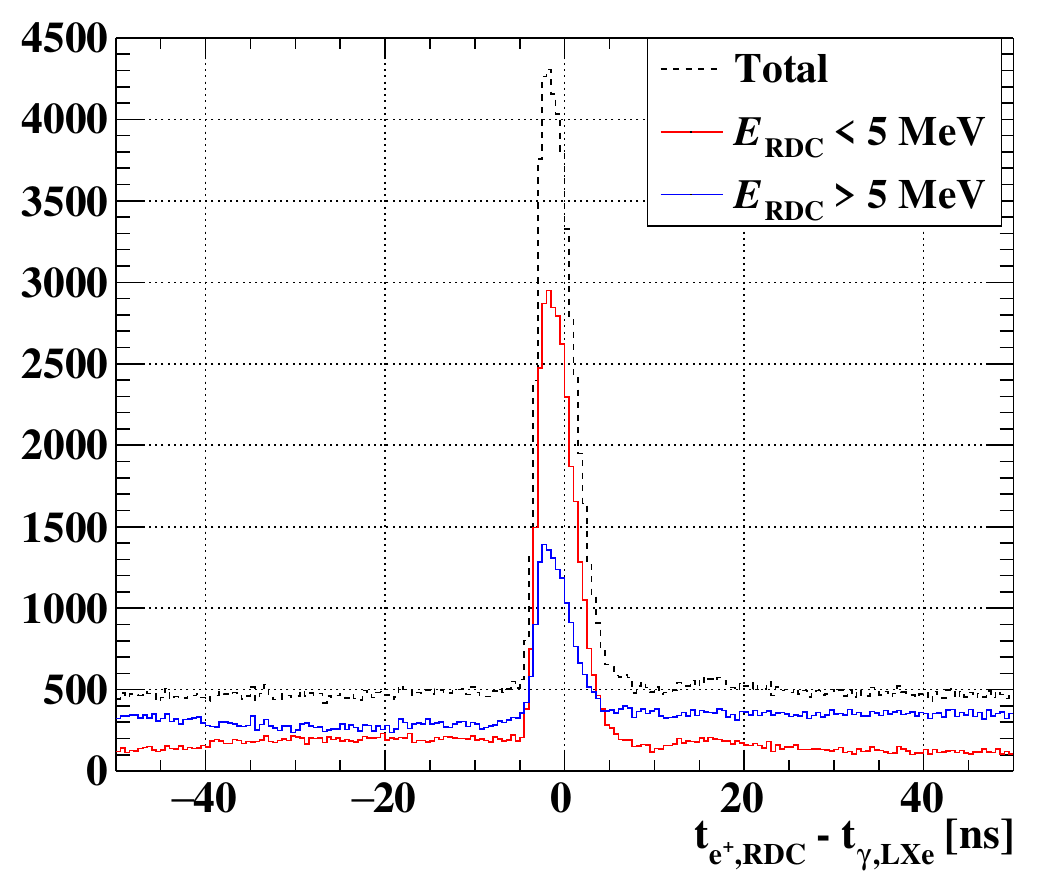}
  \caption{Distributions of time difference between positrons detected in DS-RDC and $\photon$-rays detected in the LXe detector with $\egamma \in [\SI{48}{\MeV},\SI{58}{\MeV}]$.}
  \label{fig:rdc-tdiff}
\end{figure}

\Fref{fig:rdc-tdiff} shows the measured distribution for $t_{\positron,\textrm{RDC}}-t_{\photon,\rm{LXe}}$.
The peak around \SI{0}{\ns} is due to RMD events while the flat distribution below it is due to accidental hits.
The peak timing value was stable within \SI{14}{\ps} throughout the 2021 physics run.

To show that the peak is indeed due to RMD events and to highlight the effect of 
energy measurement, the distributions in different $E_{\positron, \textrm{RDC}}$ regions are also shown.
As expected, the number of RMD events is larger for events with $E_{\positron, \textrm{RDC}}<\SI{5}{\MeV}$.
\textcolor{black}{
While the time distribution of the accidental hits without the $E_{\positron, \textrm{RDC}}$ cut is flat, we see a left to right asymmetry in the plots with $E_{\positron, \textrm{RDC}}$ cut.
This is because the energy reconstruction is biased by a pile-up hit associated with a $\photon$ from RMD; namely hits in the region close to the peak are likely to have a pile-up hit after (before) them, which results in an asymmetric energy bias around the peak region.
In the signal events, such an effect is not expected because RDC hits are only accidental.
Since pile-up hits with the time difference larger than \SI{40}{\ns} do not bias (or have a negligible impact) the energy reconstruction, such region is used to emulate the RDC response in the signal events.
}
The region with $t_{\positron,\textrm{RDC}}-t_{\photon,\mathrm{LXe}}\in [\SI{-8}{\ns}, \SI{8}{\ns}]$ is defined as ``on-timing" and 
\textcolor{black}{
$t_{\positron,\textrm{RDC}}-t_{\photon,\mathrm{LXe}}\in [\SI{-60}{\ns}, \SI{-50}{\ns}]$ 
}
as ``off-timing" events.

\begin{figure}[tbp]
\begin{center}
\includegraphics[width=1\columnwidth]{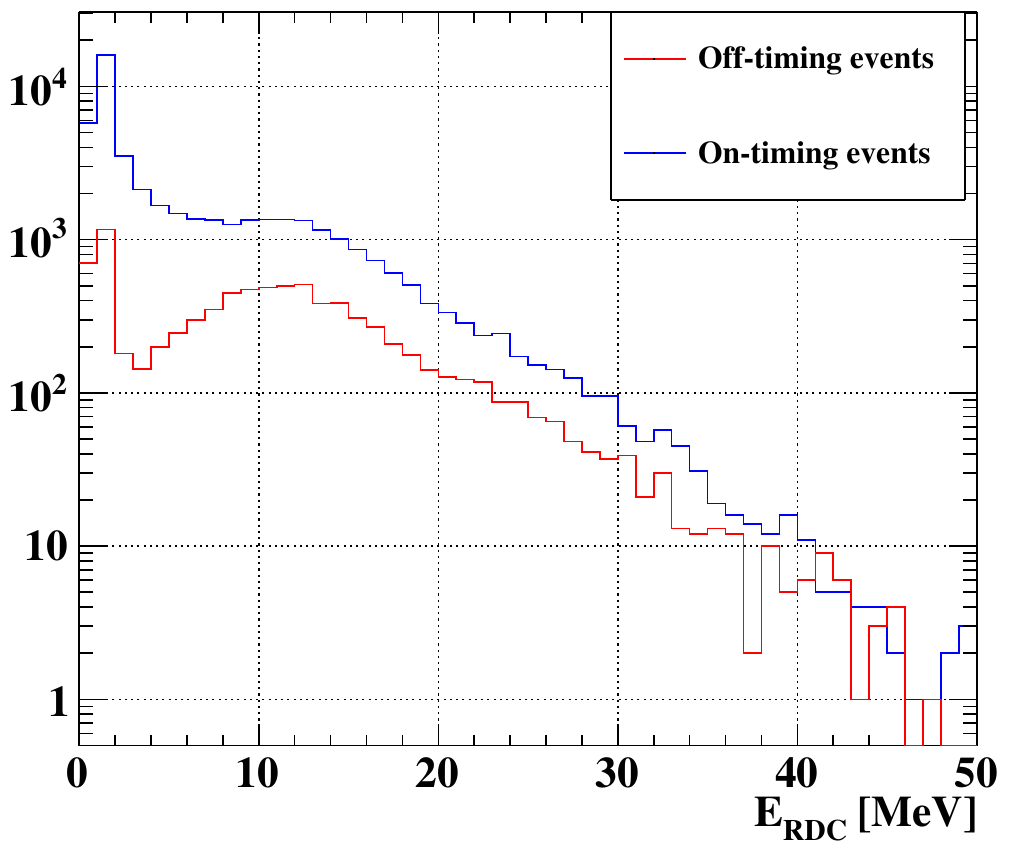}
\caption{Energy spectra in the LYSO crystals for on-timing (red) and off-timing (blue) events.}
\label{fig:E_RDC}
\end{center}
\end{figure}

The $E_{\positron, \textrm{RDC}}$ distributions are shown separately in \fref{fig:E_RDC} for the on-timing and off-timing events. 

Since the probability of getting a DS-RDC hit from the \megp\ decay is very low (\SI{<0.3}{\percent}), the hits for the signal events are dominated by the accidental hits.
Therefore, the distribution for the off-timing events represents the distribution for the \megp\ signal events, while 
the distribution for the accidental background consists of a combination of the on-timing and off-timing distributions.

The fraction of accidental background events tagged by the DS-RDC (tagged RMD fraction) clearly indicates the DS-RDC's performance. 
It is the product of the fraction of the \photon-rays from RMDs to total background \photon-rays, the fraction of positrons emitted downstream,
the fraction of time with DS-RDC at the measurement position in the physics run,
and the detection efficiency of DS-RDS, and is calculated by the ratio of the number of on-timing events, 
subtracting off-timing events, to the total number of events 
in the \megp\ trigger data set.

\begin{figure}[tbp]
\begin{center}
\includegraphics[width=1\columnwidth]{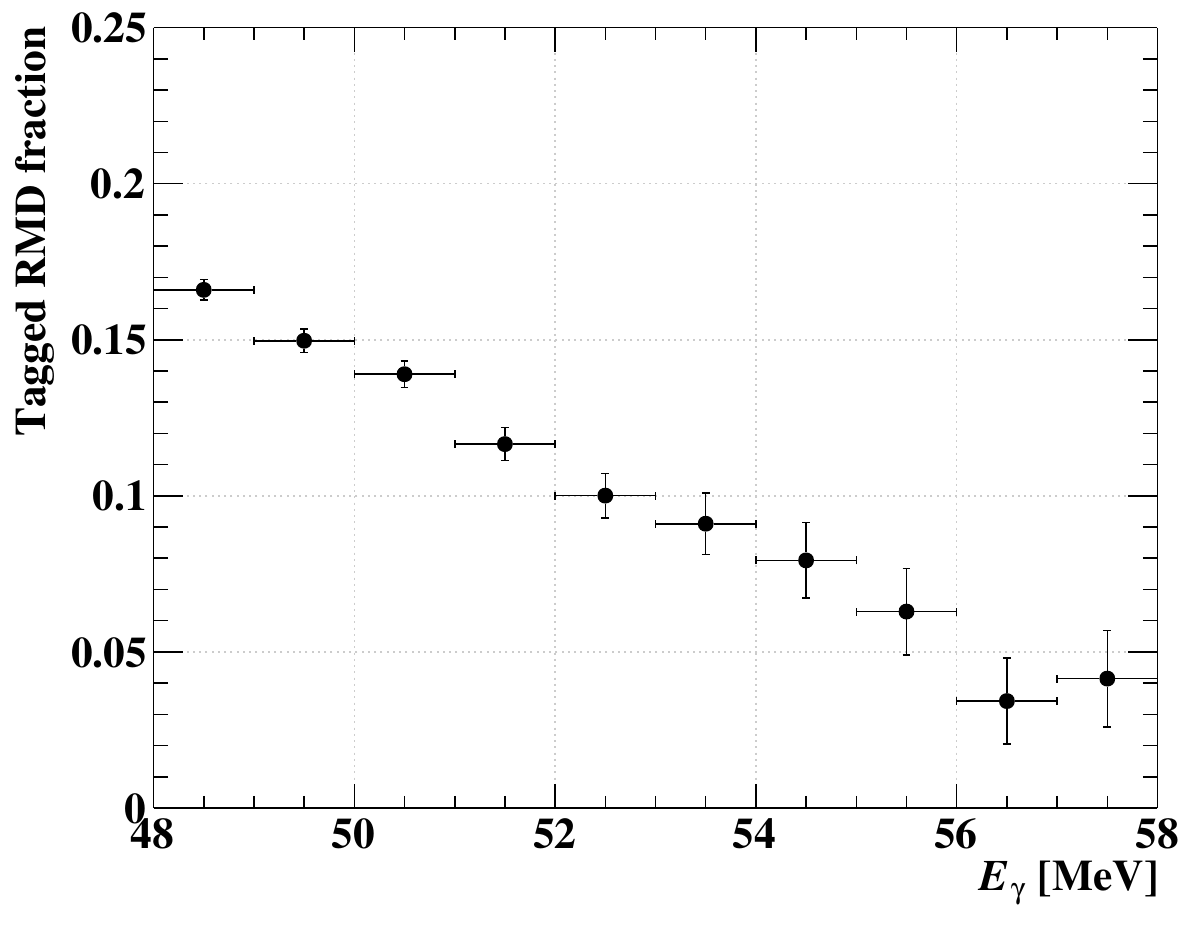}
\caption{Tagged RMD fraction as a function of $\egamma$. See the text for the definition.}
\label{fig:RMDfraction}
\end{center}
\end{figure}

The result for the 2021 data is shown in \fref{fig:RMDfraction} as a function of $\egamma$; 
the average fraction is \SI{14.1(0.2)}{\percent}, while the expected fraction is \SI{20}{\percent}.
The breakdown of this fraction can be found in \tref{tab:rdc_eff}.
The discrepancy between the measured and the expected fraction is listed as an unknown contribution. 
\textcolor{black}{Even though this discrepancy itself does not become a systematic uncertainty in the \megp\ search since the measured fraction is used,
we are trying to understand and improve it for the future analysis.}
One of the possible causes is that more background \photon-rays come from positron annihilation-in-flight events, 
leading to a reduction in the RMD fraction. 
The small fraction of the period in the measurement position is due to problems 
with the interlock system and would be close to 1 in the following years.

\begin{figure}[tbp]
\begin{center}
\includegraphics[width=1\columnwidth]{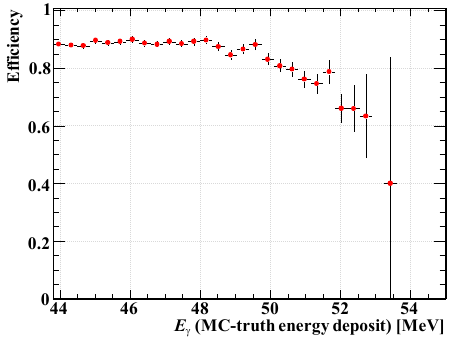}
\caption{Detection efficiency of DS-RDC as a function of $\egamma$ evaluated with a MC simulation. The efficiency is defined as the ratio of the number of RMD events with a hit in DS-RDC to the number of RMD events in which the positrons are emitted downstream \cite{onda_2021}.}
\label{fig:rdc_detection_eff}
\end{center}
\end{figure}

The detection efficiency of DS-RDC versus $\egamma$ derived from MC simulations is shown in \fref{fig:rdc_detection_eff}. 
The efficiency decreases at higher $\egamma$ because the higher $\egamma$ is, the lower the energy of the \positron\ from the RMD is.
At very low energy the $\positron$s may go undetected because they stop in the gas or in the inactive material of DS-RDC, or their energy deposition is too low.
The plateau below \SI{\sim48}{\MeV} is due to the geometric acceptance (\SI{88}{\percent}).

\begin{table}[tb]
\centering
 \caption{Breakdown of the efficiencies for the tagged RMD fraction for $\egamma > \SI{48}{\MeV}$ in 2021.}
 \begin{tabular}{l r} 
 \hline
 RMD fraction & 0.70 \\
 Emission to downstream & 0.48 \\
 Period in measurement position & 0.73 \\
 Detection efficiency & 0.82\\
 Unknown contribution & 0.70\\
 \hline
 \end{tabular}
 \label{tab:rdc_eff}
\end{table}

\begin{figure}[tbp]
\begin{center}
\includegraphics[width=1\columnwidth]{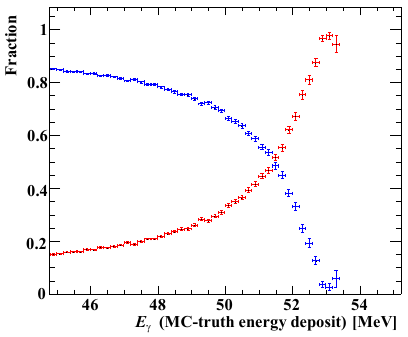}
\caption{Fraction of high-energy \photon-ray sources from a MC simulation: RMD in blue and positron annihilation-in-flight in red \cite{onda_2021}.}
\label{fig:GammaFraction}
\end{center}
\end{figure}

The decreasing trend of the tagged RMD fraction with $\egamma$ is not only due to
the $\egamma$ dependence of the detection efficiency but also to the $\egamma$ dependence of the fraction of
$\photon$-rays originating from RMDs. The $\photon$-rays from positron annihilation-in-flight become dominant at 
higher $\egamma$ as shown in \fref{fig:GammaFraction}. This energy dependence is also the cause of the correlation between $\egamma$ and the DS-RDC observables.

DS-RDC is estimated to improve sensitivity by \SI{7(1)}{\percent}, based on the performance in 2021, under the assumption that the DS-RDC is always at the measurement position during the physics run. 
This value is significantly lower than the \SI{15}{\percent} value given in \cite{baldini_2018}.
The main cause is that the correlation with $\egamma$ was not properly considered.

\subsection{Upstream RDC}
\label{sec:rdcupstream}

Since the upstream RDC will be installed in the path of the muon beam, it must have a low material budget and a high rate capability, as well as good efficiency and timing resolution.
Only the timing can be detected by the US-RDC to meet the requirement of low material budget.
For the US-RDC a resistive plate chamber based on diamond-like carbon (DLC) electrodes is under development \cite{Oya2022, Yamamoto2023}.
The DLC deposited on \SI{50}{\um} thick polyimide foils makes the material budget low enough (\SI{0.1}{\percent} $\mathrm{X_0}$).
A prototype of the detector with a size of \qtyproduct[product-units = bracket-power]{2x2}{\cm} was built and tested with the muon beam.
The results were promising and met the demanding requirements.
Currently, the detector design for a larger detector size 
is being investigated.

\section{Trigger and data acquisition}
\label{sec:rtdaq}

\subsection{Concepts and design}

The experiment trigger and data acquisition system evolved from the experience obtained 
while operating the MEG trigger\cite{trigger2013} and data acquisition system, which 
exploited the DRS4 chip\cite{ritt_2004_nim}.

The MEG~II detector's granularity requires a number of channels approximately four times 
larger than in the MEG detector. The breakdown of channels in the MEG~II experiment is reported in \tref{tab:DAQChannels}. 
Such an increase in number of channels was possible only by merging the previously separated 
trigger and data acquisition branches into the WaveDAQ system~\cite{francesconi2023wavedaq}.

\begin{table}[tb]
\centering
 \caption{Breakdown of detector channels connected to the MEG~II WaveDAQ system}
 \begin{tabular}{l c r r} 
 \hline
 Detector & Channel Type & \# channels & \textcolor{black}{MEG} \\
 \hline
 LXe det. inner face & MPPC & 4092 & \textcolor{black}{216(PMT)}\\ 
 LXe det. & PMT & 668 & \textcolor{black}{630} \\
 pTC & SiPM & 1024 & \textcolor{black}{60}\\
 CDCH & Differential & 2432 & \textcolor{black}{1728}\\
 &  Frontend & \\
 Others & various & 57 & \textcolor{black}{$\approx$ 50} \\
 \hline
 Total & & 8591 & \textcolor{black}{2639}\\
 \hline
 \end{tabular}
 \label{tab:DAQChannels}
\end{table}

 WaveDAQ is a highly integrated custom trigger and 
data acquisition system designed to fulfil the MEG~II requirements in terms of detector 
resolution and background suppression, both at the offline and online stages. The final 
system consists of 35 crates, each containing 
up to 16 WaveDREAM modules. The WaveDREAM is a fully contained 16-channel data 
acquisition platform that employs two DRS4 chips to digitise the analogue signal at sampling speed \textcolor{black}{in the range \SIrange[range-phrase = --,range-units=single]{0.8}{5} GSPS.}

To preserve the timing characteristics of the detector signals, the input channels are 
designed with a programmable gain stage in the range
\numrange[range-phrase = --,range-units=single]{0.25}{100}
 with an analogue bandwidth up to \SI{800}{\mega\hertz} capable 
of providing the bias voltage needed by 
SiPMs.

When an event of interest is identified, the analogue waveform amplitudes are readout by 
an external analogue to digital converter and transmitted to a dedicated Data 
Concentrator Board (DCB) that sends the digitised waveforms 
over the Ethernet network to the readout computer.

The trigger decision is generated by a group of dedicated FPGA Trigger Concentrator Boards (TCBs) 
arranged in a tree layer structure, which must generate the decision in \SI{\sim600}{\nano
\second}, because of constraints on DRS4 buffer length when operating at \SI{1.4}{GSPS}. 

\subsection{Installation and commissioning}

Due to budget constraints, the WaveDREAM system was commissioned using a multi-stage approach \cite{universe7120466}. 
In 2015, we conducted an engineering run in which half of the pTC detector was tested with the PSI muon beam to check the reliability of a WaveDAQ crate. 
From 2016 to 2020, we increased the number of available DAQ crates from four to nine, 
including partial readouts for LXe and CDCH, as well as some of the auxiliary channels.

Furthermore, we installed two specialised WaveDAQ crates in our setup. 
One crate serves as a trigger concentration unit, while the other
crate serves as a central hub for distributing critical system 
control signals, including triggers, clocks, and synchronisation signals. 

Throughout each stage of the commissioning process, we carefully checked and developed the following items:

\begin{itemize}
\item The electronics noise for all detectors and readout schemes, including 
single-ended and low front-end amplification (\numrange[range-phrase = --]{1}{5}) for 
LXe, single-ended and high front-end amplification 
(\numrange[range-phrase = --]{50}{100}) for pTC, and differential for CDCH.
\item The clock distribution and system synchronisation.
\item Basic trigger algorithms for all detectors.
\item The Ethernet readout infrastructure and software readout code.
\end{itemize}

\begin{figure*}[tb]
\centering
  \includegraphics[width=1\textwidth,angle=0] {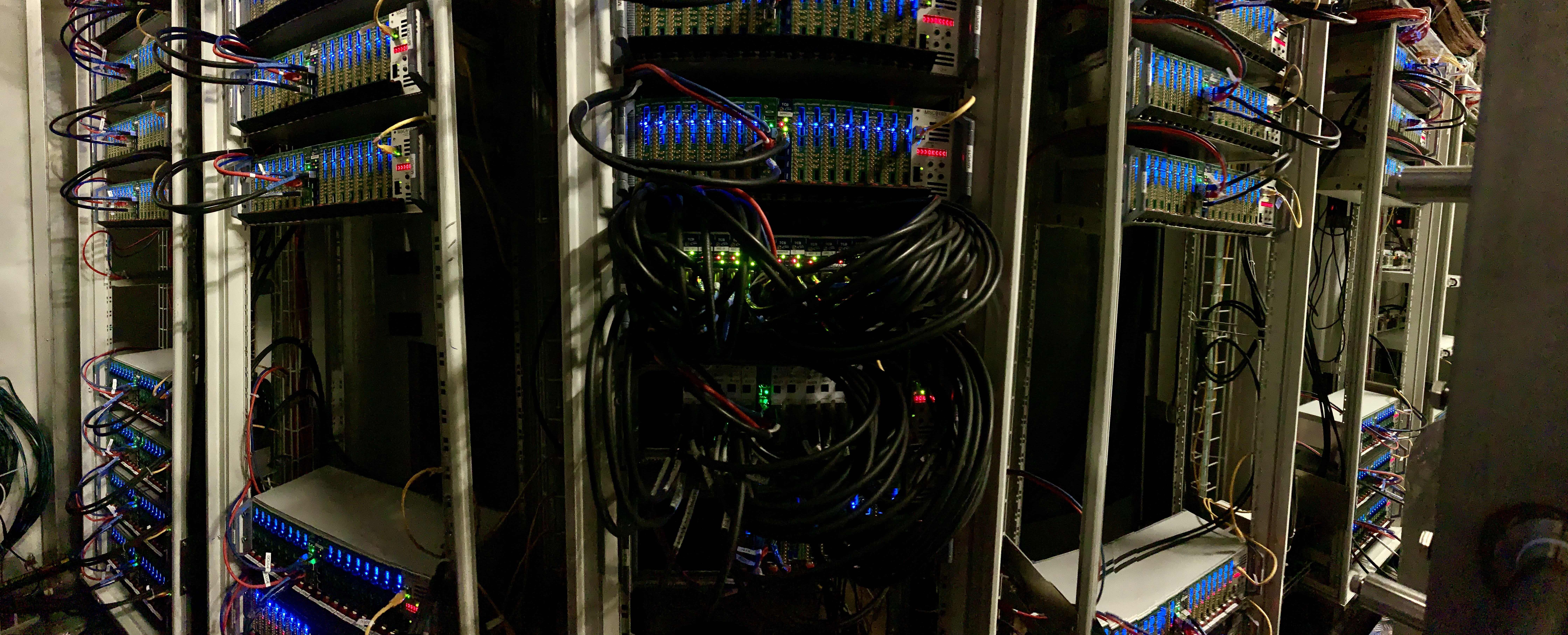}
  \caption{Panoramic view of the full TDAQ system installed in the PiE5 area. The DAQ 
  crates filled with WaveDREAM boards have the blue LED shining, the trigger and clock 
  distribution crates fully cabled are at the centre of the picture.}
 \label{trg:tdaqpic}
\end{figure*}

The full system was installed and commissioned in March 2021, 
ready for the first DAQ campaign, which started in May of the same year. \Fref{trg:tdaqpic} shows a panoramic view of the system.
\subsection{Data acquisition design and performance}
\label{DAQ_design}
The DAQ operates in push-mode, meaning that when it receives a trigger signal, all 
boards prepare the data to be fetched by the DCBs and then delivered to the DAQ server 
via an Ethernet connection. To maximise data transmission rates, the UDP protocol is 
adopted over TCP, since it is not connection-based. However, a drawback of UDP is that 
lost packets are possible, resulting in incomplete event reception and, therefore, DAQ inefficiency.

The typical trigger rate during physics runs ranges from \SIrange[range-units = single]{10}{30}{\hertz}, 
depending on trigger thresholds and beam rate. Given that each waveform is approximately \SI{\sim 1.5}{\kilo\byte}  and the number of WDB boards in the system, a
\SI{10}{\giga\bit\per\second} connection between the WaveDAQ and the DAQ server is necessary.

To accommodate the substantial data rate produced by the system, a private network has 
been installed and managed in terms of Dynamic Host Configuration Protocol (DCHP) and Domain Name System (DNS) servers by the DAQ server, independent of the laboratory network. 
The 35 WaveDAQ crates are distributed among six racks, each of which has a Top-of-Rack 
switch (MikroTik CRS354-48G-4S+2Q+RM~\cite{mikrotik})  to gather the data from all the 
DCBs in the rack and forward them out of the experimental area to the aggregation switch 
(MikroTik CRS326-24S+2Q+RM). Since the DCB link is at \SI{1}{\giga\bit\per\second} and 
the maximum number of DCBs hosted in a rack is eight, we designed the downstream 
connection using two \SI{10}{\giga\bit\per\second} lanes to avoid bottlenecks at this stage.

The aggregation switch collects packets from the six Top-of-Rack switches and provides a 
\SI{10}{\giga\bit\per\second} connection to the DAQ server, which defines the maximum 
DAQ rate to be approximately \SI{50}{\hertz} in line with requirements.

The DAQ server and the DAQ software process up to \num{1e6} packets per second, build 
events, apply DRS chip calibrations, and write events to disk continuously and without loss. 
A multi-threaded software has been designed with four independent processing steps:

\begin{description}
    \item{Collector}: collects all the packets from the kernel and stores them in a local buffer.
    \item{Builder}: fetches fragments of events and merges all the packets belonging to the same event and the same board. A built event is then passed to the next buffer.
    \item{Worker}: applies DRS voltage calibrations to all the waveforms in built events.
    \item{Data handler}: writes the events to disk after applying data reduction algorithms, described in \sref{sec:reduc}.
\end{description}
Each of the steps can be parallelised as needed; in total, we use 32 threads.

\subsubsection{Data reduction}
\label{sec:reduc}

A full MEG~II event is as large as \SI{16}{\mega\byte}, which is a huge value. The event 
size can be substantially reduced without deteriorating the experiment's performance by 
applying data reduction schemes that are tailored to each detector. The methods 
implemented at the data handler stage are:

\begin{enumerate}
    \item Waveform re-binning: merge the waveform bins in groups of $2^n$ ($n$~=~1,2,3,4,5).     
    \item Region of interest (ROI): slice the waveform in a predefined window around the trigger time.
    \item Zero suppression: discard waveform without pulses.
\end{enumerate}

\begin{figure}[tb]
\centering
  \includegraphics[width=1\columnwidth,angle=0] {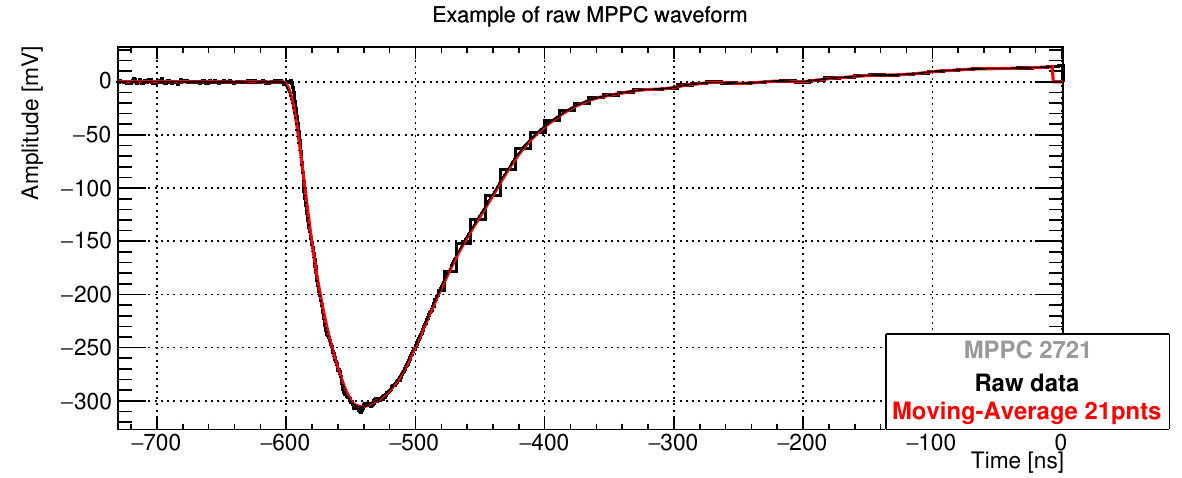}
  \caption{Example of a re-binned LXe waveform, here the re-bin factor is 16. In red the recovered waveform by smoothing the raw data with a moving average.}
 \label{daq:rebin}
\end{figure}

For the LXe detector, re-binning is widely used; waveforms are retained at the nominal 
frequency in a region around the signal edges large enough to preserve their time 
reconstruction, while other bins are re-binned, with a factor ranging from 8 to 32 
dependent on the pulse height. An example of a re-binned waveform is in \fref{daq:rebin}.
Waveforms with very small or no pulses are fully re-binned with a re-bin factor of 8 also in the signal region. 

The pTC detector is highly segmented with a very good signal-to-noise separation, large 
enough to safely apply zero suppression. 
As a result, the size of pTC events is negligible, since only a small fraction of pTC 
counters is hit by a positron in each event. ROI cuts are also applied.

The same approach cannot be used for the CDCH because the signal-to-noise 
separation is not at the same level, so all waveforms are written to disk re-binned by a 
factor of ten, since this has no impact on reconstructed tracks. 
A limited ROI cut of \SI{10}{\percent} of the full range is also applied at the beginning and end of the waveforms.

The overall event size reduction is a factor \num{\approx 10}.

\subsubsection{Performance}

Performance-wise, the infrastructure that was built and commissioned during the early 
stages of data collection guarantees an efficiency of over \SI{99}{\percent} for trigger 
rates up to \SI{35}{\hertz}, corresponding to a traffic rate on the private network of 
about \SI{8}{\giga\bit\per\second}. However, some inefficiency is observed above this threshold.

In the case of detector calibration runs, in which only a subsample of signals are 
required, we can run the system at the current maximum DCB rate of about \SI{52}{\hertz}. 
This is crucial to reduce the detector calibration time to the minimum and increase the sample of physics data. 

Additional improvements are under study to increase the network rate up to \SI{10}{\giga\bit\per\second} 
and the DCB connection speed, which is not at the design value of \SI{1}{\giga\bit\per\second} yet.
  
\subsection{Trigger capabilities}
The WaveDAQ system supports up to 64 independent trigger lines, each with its own prescaling factor, to allow the correct mixing of various conditions in the same data set.
The trigger lines are identified by a number, which is also used as a priority order if 
multiple conditions are matched simultaneously. This is relevant when triggers with 
reduced bias are mixed with the physics trigger for offline studies.
The remaining triggers are dedicated to collecting detector-specific calibration data.

\subsubsection{\megp ~trigger}
\label{sec:meg_trigger}
The \megp\ trigger is based on the simultaneous presence of the following three conditions:
\begin{description}
    \item[\photon-ray energy] The weighted sum of all photosensors in the LXe detector is above the threshold.
    \item[Time coincidence] The time difference $T_{\positron\photon}$ between the group of MPPC closest to the $\photon$-ray conversion and the positron hit on the timing counter is within a programmable time window.
    \item[Direction match] The \photon-ray conversion point in the LXe detector and the impact position of the positron in the pTC are compatible with a two-body muon decay at rest from the target.
\end{description}

Three dedicated triggers, each with one of the conditions relaxed, are recorded during the physics run with a large enough prescaling not to dominate the throughput.

\begin{figure}[tb]
\centering
  \includegraphics[width=1\columnwidth,angle=0] {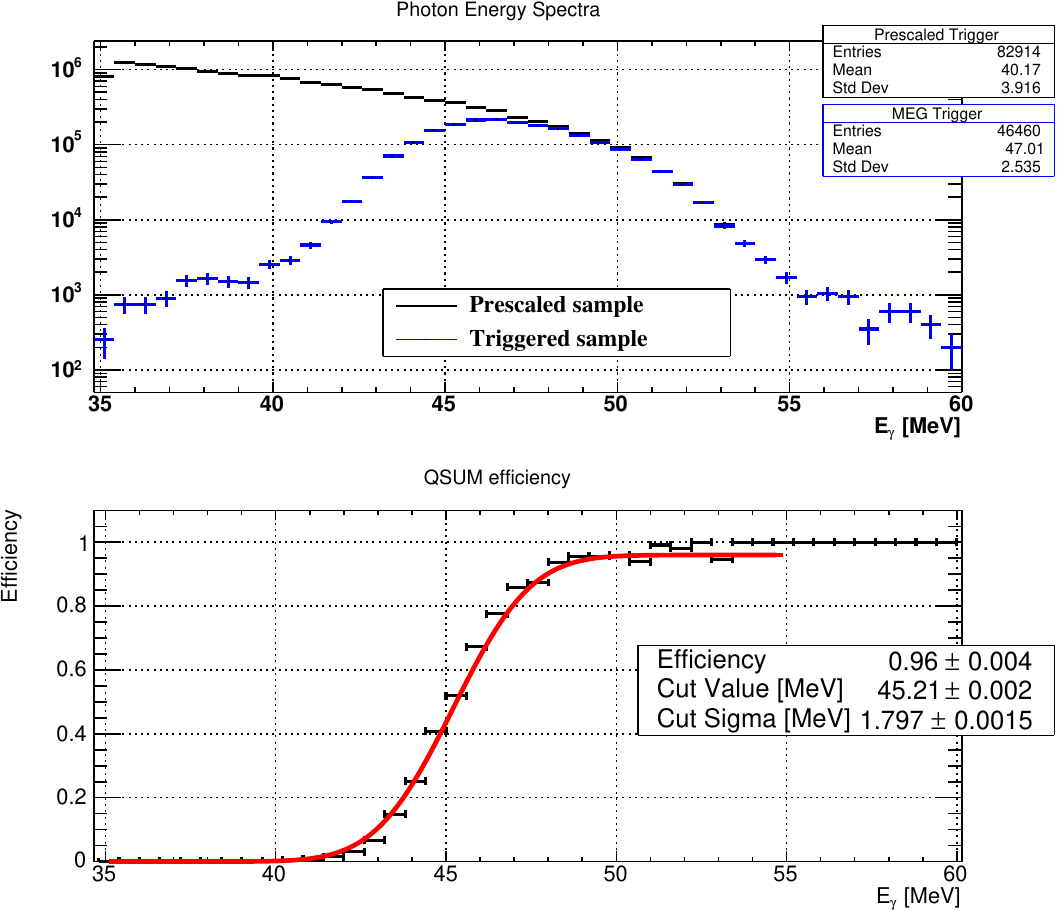}
  \caption{(Top) Offline-computed \photon-ray energy spectra with different thresholds; (bottom) efficiency function obtained from their ratio. The fit yields an online Gaussian resolution of \SI{\sim 4}{\percent} .}
 \label{trigger:egamma}
\end{figure}

By comparing the distribution of offline-reconstructed observables, which are expected 
to have better resolutions than online-computed ones, it is possible to evaluate the 
effect of the online selection: in \fref{trigger:egamma}, the \photon-ray energy spectra 
are reported. To extract the trigger selection resolution, their ratio is fitted with an 
integrated Gaussian function, also known as the Gauss error function. The measured resolution is \SI{\sim 4}{\percent}.

The \photon-ray energy efficiency, $\varepsilon_{\egamma}$, is estimated by multiplying 
the integrated Gaussian function of \fref{trigger:egamma} with the expected response to 
a \SI{52.83}{\MeV} \photon-ray, estimated by scaling the fits of \fref{fig:LXe-55MeVEGamma}.
The current estimate is $\varepsilon_{\egamma}=\SI{96}{\percent}$, limited by the sub optimal response in terms of uniformity over the inner face of the LXe detector. 
The online calibration has been improved in the next runs.

\begin{figure}[tb]
\centering
  \includegraphics[width=1\columnwidth,angle=0] {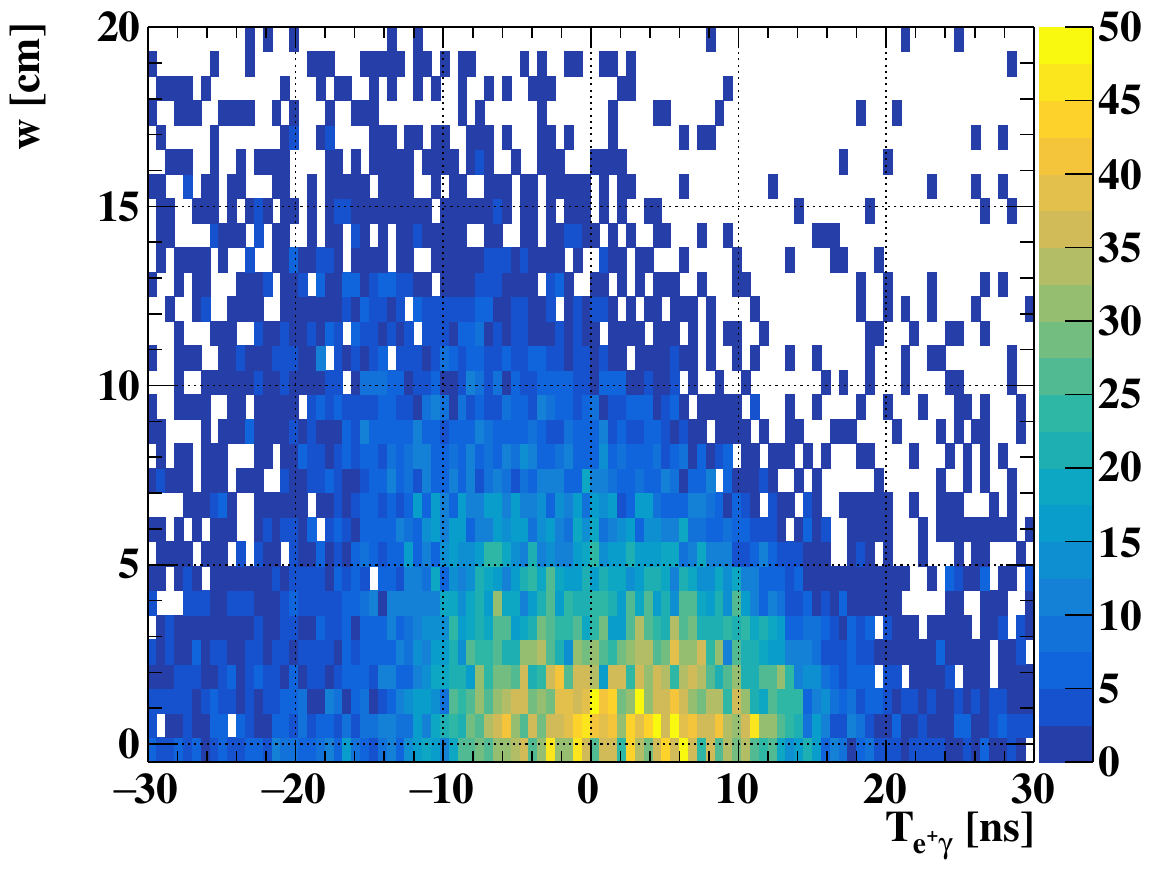}
  \caption{Time distance distribution for \photon-ray--positron pairs $T_{\positron\photon}$ as function of the $\photon$-ray conversion depth $w$ as defined in \ref{sec:intro}.}
 \label{trigger:tegamma}
\end{figure}

\Fref{trigger:tegamma} shows the time selection window $T_{\positron\photon}$, which 
varies depending on the $\photon$-ray conversion depth, denoted as $w$ and defined in~\sref{sec:intro}. 
It is evident that the trigger selection window exhibits a noticeable dependence on this parameter. 
The cause of this behaviour was identified as a significant time walk effect in the time measurement performed by the MPPC detectors. 
Because these detectors have a rise time of a few nanoseconds, the effect is considerably larger than the expectation based on MC studies. 
While we achieve full efficiency for shallow events, the efficiency is significantly reduced when the conversion occurs at $w>$\SI{10}{\cm}. 
We evaluated the efficiency in slices of $w$ and then averaged over its distribution, the resulting value is $\varepsilon_{T_{\positron\photon}} =\SI{94}{\percent}$.
This was improved from the 2022 run, as described in \sref{tdaq2022}.

The efficiency of direction match $\varepsilon_\mathrm{DM}$ can only be measured 
indirectly, using an unbiased set of positron tracks with $\ppositron>\SI{52}{\MeV/
\clight}$, since there is no physics source for back-to-back events.
An artificial set of such events is generated by back-propagating the positron track 
from the target in the LXe detector; the actual positron hit point in the pTC is then paired with the estimated hit point of the $\photon$-ray.
The efficiency is determined by comparing the pTC--LXe hit positions of each event with 
the table in use in the trigger firmware, which is built on MC simulations, to have $\varepsilon_\mathrm{DM}=\SI{88.5}{\percent}$.

The non-optimal efficiency $\varepsilon_\mathrm{DM}$ 
obtained on 2021 data is attributed to a small offset in the beam positioning at the 
COBRA centre and low statistics for double-turn tracks in the MC used to produce the firmware. A new production based on latest results is under study for the incoming runs.
 
The \megp\ trigger efficiency is the product of the efficiencies of the three observables involved:
\begin{equation}
    \varepsilon_\mathrm{TRG} = \varepsilon_{\egamma} \times \varepsilon_{T_{\positron\photon}} \times \varepsilon_\mathrm{DM} \approx \SI{80(1)} {\percent}.
\end{equation}
%
This number was measured at $R_{\muon} = 3\times 10^7 \rm{s^{-1}}$, at higher rates the trigger efficiency is degraded by 2\% at most due to tighter trigger thresholds, in particular the ${T_{\positron\photon}}$ cut.

\paragraph{Improvements since 2022}
\label{tdaq2022}

The $\photon$-ray conversion time algorithm was improved during the 2022 run. Instead of 
using the group of MPPCs closest to the $\photon$-ray conversion point, the signal from 
the first PMT in the LXe detector's back face is used. This modification was made 
because PMTs have a faster response time compared to MPPCs. As a result, we achieved a more precise measurement of $T_{\positron\photon}$, leading, as expected, to 
an improvement in $\varepsilon_{T_{\positron\photon}}$. 

Similarly an improved LXe detector calibration at trigger level returns a more uniform response over the entrance face; therefore $\varepsilon_{\egamma}$ is expected to improve.

With the finalisation of the positron reconstruction algorithm and the availability of a 
large sample of reconstructed positron tracks, we aim to investigate how to extend the 
direction match table to increase $\varepsilon_\mathrm{DM}$.

\subsubsection{Calibration and other triggers}

In experiments such as MEG~II, where a tiny 
signal needs to be detected in a harsh background, a complete and complementary set of 
calibration methods is deployed, most of which are accompanied by dedicated trigger logic. 
The auxiliary crate, a dedicated WaveDAQ unit, collects all the signals from the calibration devices.

Among the various calibrations, the LXe detector calibration is the most challenging. A 
detailed description is available in \sref{sec:XECCalib} and \sref{sec:xec_energy}. About 10 trigger lines are 
dedicated to this task. The energy scale calibration relies on the collection of \photon-rays
of known energies in the detector in the range \SIrange[range-units=single]{9}{130}{\MeV}. 
A \photon-ray is triggered by a threshold on the estimated energy 
deposit, which, in some cases, is in coincidence with an auxiliary device: the neutron 
generator for the \SI{9}{\MeV} events and the BGO calorimeter for the CEX campaign. 

\begin{figure}[tb]
\centering
  \includegraphics[width=1\columnwidth,angle=0] {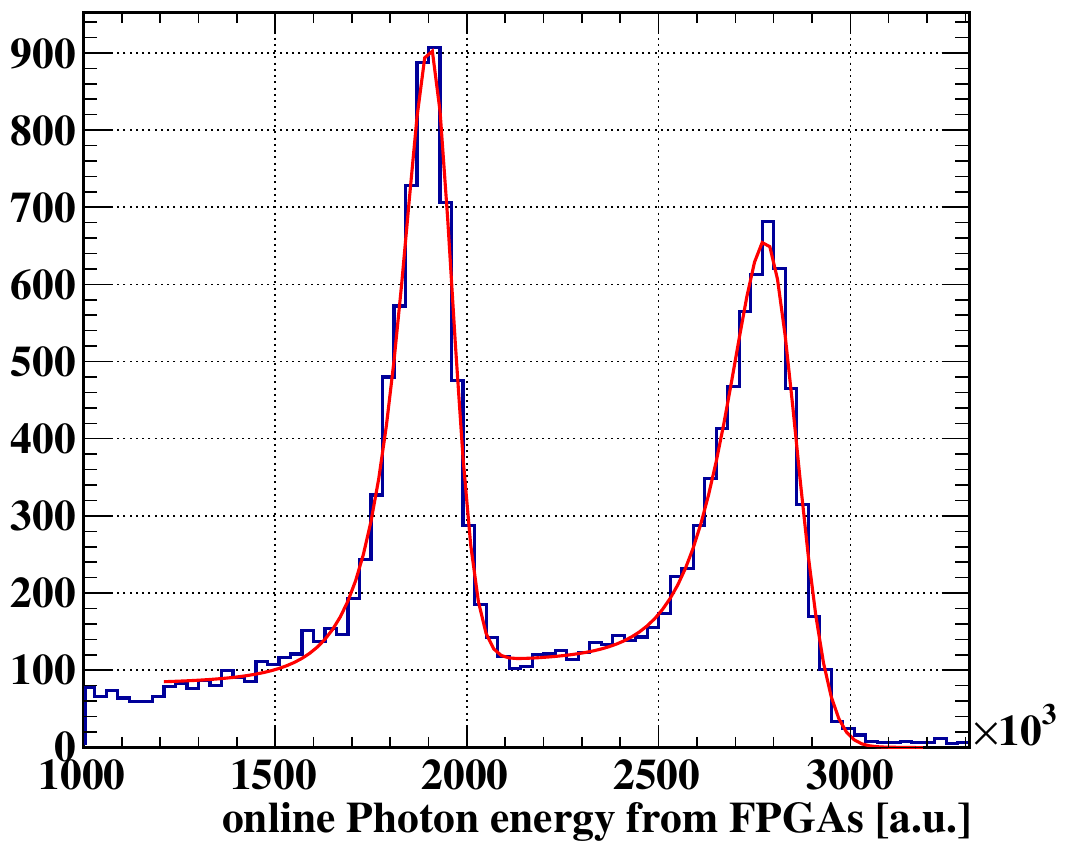}
  \caption{CEX online \photon-ray spectrum as reconstructed by online trigger 
  processing. The two peaks at \qtylist[list-units=single]{55;83}{\MeV} are clearly visible\textcolor{black}{, the CEX configuration is} described in \sref{sec:XECoverview}. Being monochromatic lines, 
  the online resolution here can be measured without relying on offline processing. 
  The function used in the fit is the sum of two response functions as in 
  \eref{eq:LXe-ExpGaus} for the two energy lines plus a flat background. 
  The online resolution at \SI{55}{\MeV} is \SI{3.1}{\percent}.}
 \label{trg:cex}
\end{figure}

\Fref{trg:cex} shows the energy spectrum of the \qtylist[list-units=single]{55;83}{\MeV} 
lines as reconstructed by the trigger logic. The line fit returns the online energy 
resolution at the signal energy.
The sensor gains of the LXe detector are calibrated using a set of LEDs flashing at well-known intensities. 
The LED drivers are connected to the WaveDAQ system and provide TTL signals in time with the LED light emission.

The QEs of the PMTs and the PDEs of the MPPCs are measured by comparing the light measured from the decay of $\alphaup$-particles submerged in the LXe detector with MC simulations. 
These events are selected using a pulse shape discrimination logic, already used in the MEG experiment and described in \cite{Nicolo:2021hzv}. 
A revised version of the trigger logic is implemented in the MEG~II experiment using the signals gathered from all the outer face PMTs. 

The pTC is calibrated daily using the laser-based system to monitor the counter time offsets. 
A dedicated trigger line is used, and the laser pulse generator provides a signal 
synchronous with the laser pulse, which is connected to a dedicated WaveDAQ input channel and used as a trigger. 
The calibration results are shown in \fref{fig:pTCLaser}.


\section{Computing}
\label{sec:computing}

Significant offline computing resources are required to store the data on disk, process them, produce MC simulation samples, and provide the analysis platform to individual users.
The MEG~II offline computing system is based on an on-premise high-performance computing (HPC) cluster, operated by the Science IT Infrastructure and Services Department of PSI.
It consists of login nodes, computing nodes with 320 CPU cores for batch jobs, and storage servers with a capacity of \SI{2}{\peta\byte}.
Two types of storage architectures are used: one based on IBM Spectrum Scale (known as GPFS) with a capacity of $\SI{1.1}{\peta\byte}$, used as main storage for intensive I/O activities; 
the other based on the standard NAS system. 
All nodes and storage are connected via the InfiniBand (EDR \SI{100}{\giga\bit\per\second}) network to ensure high throughput and low latency.
The batch jobs are managed by Slurm \cite{slurm}.
The availability of the system was 98\% (or 99.8\% during the DAQ period) in 2022.
Additional \num{\sim 500} CPU cores are available for offline processing using a general-purpose HPC cluster at PSI.

The data rate during 2022 was \SI{180}{\mega\byte\per\second} at $R_{\muon} = \SI{5e7}{\per\second}$.
The raw data in the MIDAS file format \cite{midas} are transferred from the DAQ machine to the offline cluster over a \SI{10}{\giga\bit\per\second} local area network and then compressed with the bzip2 algorithm to reduce the size by a factor of 3.
The compressed raw data are stored on the disk for offline analysis and, in parallel, sent to a tape-based archive system at
 Swiss National Supercomputing Centre (CSCS) in Lugano over a dedicated $\SI{100}{\giga\bit\per\second}$ network between PSI and CSCS.
The raw data, as well as the processed data sets used for publications of physics results, will be preserved and published there.

 The total raw data in 2022 amounts to \SI{420}{\tera\byte}. After the first reconstruction phase, followed by a pre-selection of events, the raw data are deleted from the disk.
 
 The MEG~II software suite consists of several programs for simulation and analysis:
 a MC program for event generation, particle tracking and detector simulation; a program for simulation of event mixing and readout electronics, a program for reconstruction and a program for statistical analysis. 
The overall structure is similar to that of the MEG experiment \cite{softwaretns}, while the MC program was renewed with Geant4 \cite{geant4, Geant4_3}.
All the programs are based on ROME \cite{rome}, a framework generator, and are written in C++17. 
The file format in ROOT \cite{root} is used throughout for I/O, while the reconstruction program can also read the (compressed) MIDAS files.  

The framework relies heavily on packages that are widely used in the high-energy physics community, such as Geant4, ROOT and GENFIT, while it can also easily integrate external C++ packages. In order to implement deep-learning-based algorithms, the reconstruction program supports the application of models in the ONNX format \cite{onnx} by means of ONNX Runtime C++ API \cite{ort}.
Therefore, users can build and train models in their preferred machine-learning framework and environment while maintaining the single common interface in the reconstruction program. On the other hand, this framework limits the processors available in the inference stage to CPUs. 

The analysis environment is also provided as a container for Docker \cite{docker} and Apptainer \cite{singularity} so that the analysis can be performed seamlessly in other systems, e.g. in a local cluster in other institutes or even on a personal laptop. 

The average time to simulate a signal event is $\SI{37}{\second}$ using a single processor core, with \SI{80}{\percent} of the time spent in tracking scintillation photons in LXe.
The average time to reconstruct an event at $R_{\muon} = \SI{5e7}{\per\second}$ is $\SI{50}{\second}$ with the current reconstruction algorithms; the largest contribution (\SI{64}{\percent}) comes from track-finding in CDCH at high occupancy.
All the physics data in 2022 can be processed in \num{\sim60} days with \num{500} cores in parallel.

\section{Sensitivity}
\label{sec:sensitivity}

A confidence interval for the branching ratio $\mathcal{B}$ of the \megp\ decay is extracted with a likelihood analysis over six or seven variables that have
discriminating power for the signal versus the background: the positron energy 
$\epositron$ and the \photon-ray energy $\egamma$ (both of which have a peak value at \SI{\sim52.83}{\MeV} for signal events); the relative angle $\Thetaegamma$ 
or two independent projections thereof, $\thetaegamma$ and $\phiegamma$,
as defined in~\cite{baldini_2016}; the relative time $\tegamma$ 
(which is expected to have a peak value at zero); two quantities that exploit the RDC: 
the RDC energy deposit $E_\mathrm{\positron,RDC}$ and the relative time $t_{\positron,\textrm{RDC}}-t_{\photon,\mathrm{LXe}}$ (whose 
distributions are different for events caused by RMDs and accidental RDC--LXe coincidences that can also occur for signal events).
The interval (or upper limit if the lower limit includes zero) for $\mathcal{B}$(\megp) with a confidence level (C.L.) of \SI{90}{\percent} 
is extracted using a frequentist approach that also takes systematic uncertainties into account
as described in~\cite{baldini_2016}.

The probability density functions (PDFs) for the accidental background 
used to construct the likelihood function are extracted from the side-bands of an analysis region defined by $\egamma \in [\SI{48}{\MeV},\SI{58}{\MeV}]$ 
and $\tegamma \in [\SI{-500}{\ps}, \SI{500}{\ps}]$. 
The PDFs for the signal and RMD are obtained by convolving the 
expected distributions with resolution functions extracted from the data, with minor MC-based corrections.

Two different analysis strategies are adopted. In one, resolution functions of some of the variables that 
change from event to event are assigned on the 
event-by-event basis (e.g. based on the uncertainties in the positron kinematic variables that can be estimated from the track fit, 
or the reconstructed position of the \photon-ray conversion point in the LXe detector). 
In the other, the events are divided into several categories, depending on the quality of the reconstruction, 
and different PDF sets are extracted for each category. 
The first approach allows maximising statistical sensitivity, 
while the second approach is less prone to systematic uncertainties.

The sensitivity of the experiment reflects the resolution functions and the efficiencies. 
\tref{sensitivity:perf} summarises the main results
discussed in the previous sections, compared to the predictions from~\cite{baldini_2018,MEGII:2021fah}.
Average values are given here, although event-by-event PDFs are used in the likelihood analysis.
If the resolution function is described by a sum of Gaussian PDFs, the width of the principal component (\emph{core resolution}) is given.
For positron observables, the typical core fraction is 90\%. 
For $\egamma$, the main deviation from Gaussianity is the long tail at low energy.
The positron resolutions are evaluated at $R_\mu = \SI{4e7}{\per\second}$, while a \SIrange[range-phrase=--,range-units = single]{5}{7}{\percent} deterioration is observed from $R_\mu = \SI{3e7}{\per\second}$ to \SI{5e7}{\per\second}.

The signal PDFs
take into account the correlations of $\phie$ with $\epositron$ 
and $\thetae$, which translate into correlations of $\phiegamma$ with 
$\epositron$ and $\thetaegamma$. Since the true values of
$\epositron$ and $\thetaegamma$ for the signal events are known, it is 
possible to correct $\phiegamma$ event-by-event (or, in other words, to 
account for the correlation event-by-event in the PDFs as
in~\cite{baldini_2016}). For this reason, the effective statistical error on $\phie$ that determines
the $\phiegamma$ resolution is lower than the global $\phie$ 
resolution. In the table we quote this effective statistical error and 
since it also depends on $\phie$ itself, we quote the value 
at $\phie = 0$ to allow consistent comparison with the numbers 
quoted in previous papers, where the same convention was used.

The $\tegamma$ resolution was evaluated from the peak in the $\tegamma$ distribution due to the coincident RMD events,
with taking into account the $\egamma$- and $\epositron$-dependence of the time resolutions.
The result is consistent with the combination of the resolutions measured for $\photon$-ray and \positron\ individually.

\begin{table}
\caption{ \label{sensitivity:perf}Resolutions (Gaussian $\sigma$) and efficiencies measured at $R_\mu = \SI{4e7}{\per\second}$, compared with the predictions from~\cite{baldini_2018,MEGII:2021fah}.}
\centering
  \begin{minipage}{1\linewidth}
   \renewcommand{\thefootnote}{\alph{footnote})}	
   \renewcommand{\thempfootnote}{\alph{mpfootnote})}	
\centering
\newcommand{\minu}{\hphantom{$-$}}
\newcommand{\cc}[1]{\multicolumn{1}{c}{#1}}
\begin{tabular}{@{}lll}
\hline
  {\bf Resolutions }  & \minu Foreseen & \minu  Achieved  \\ 
\hline\noalign{\smallskip}
$\epositron$ (\unit{keV})  & \minu 100 & \minu 89 \\
$\phie\footnotemark[1],\thetae$ (\unit{mrad})    & \minu 3.7/6.7 & \minu 4.1/7.2 \\
$\ypos,\zpos$ (\unit{mm})   & \minu 0.7/1.6 & \minu 0.74/2.0  \\
$\egamma$(\%)  ($w\SI{<2}{\cm}$)/($w\SI{>2}{\cm}$)  & \minu 1.7/1.7 & \minu 2.0/1.8 \\
$\ugamma, \vgamma, \wgamma$ (\unit{mm})          & \minu 2.4/2.4/5.0 & \minu {2.5/2.5/5.0} \\
$\tegamma$ (\unit{ps})   & \minu 70  & \minu  78 \\
\hline
{\bf  Efficiency (\%)} & &  \\ 
\hline
$\varepsilon_{\photon}$     & \minu  69  & \minu 62 \\
$\varepsilon_{\positron}$   & \minu  65 & \minu 67 \\
$\varepsilon_\mathrm{TRG}$  & $\approx$99  & \minu 80  \\
\hline
\end{tabular}
\footnotetext[1]{At $\phie=0$ with correlation taken into account. See text for the details.}
\end{minipage}
\end{table}

\begin{figure}[tb]
\centering
  \includegraphics[width=0.45\textwidth,angle=0] 
  {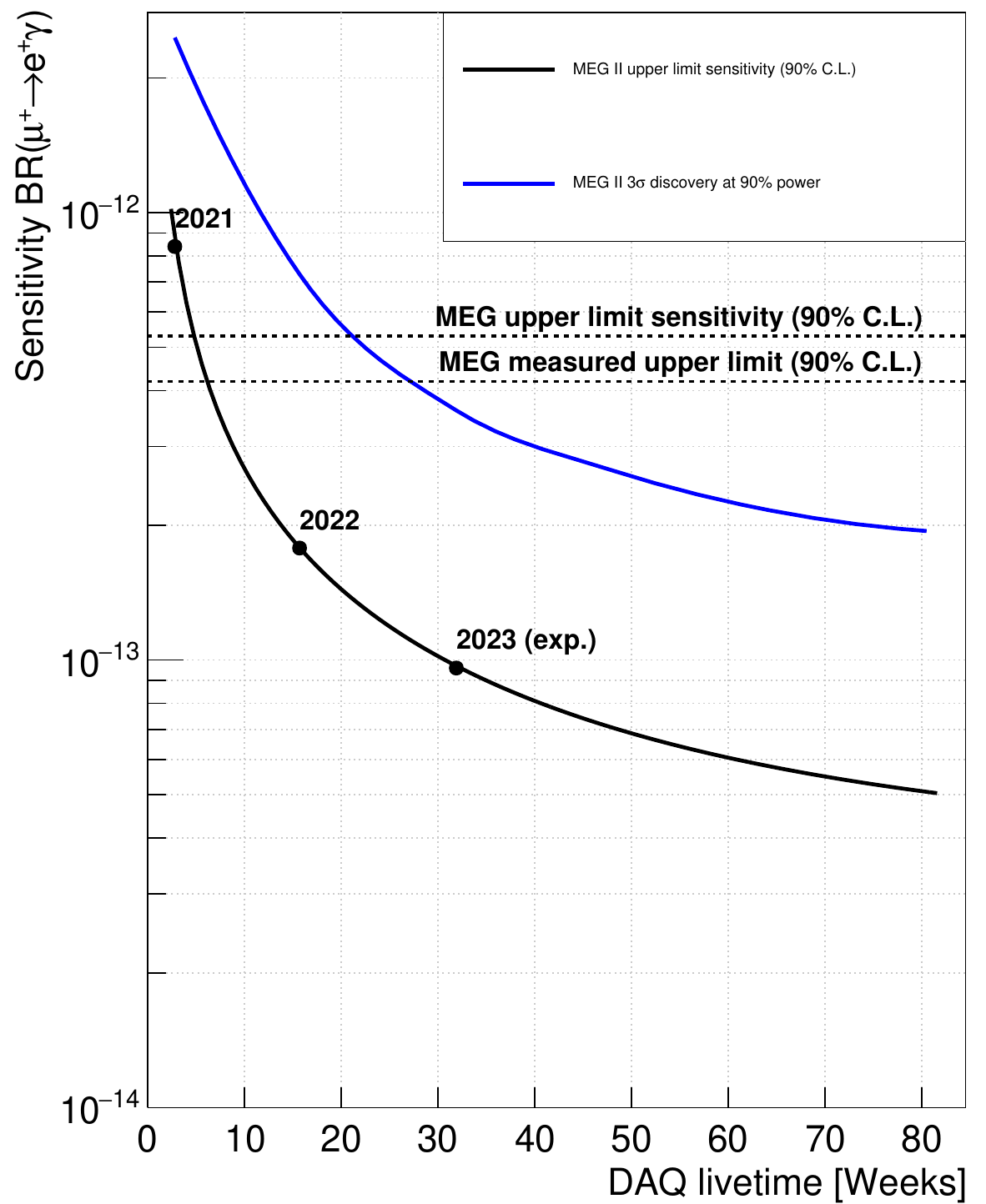} 
  \caption{Expected MEG~II \textcolor{black}{upper limit sensitivity (90\%\ C.L.)} 
  (black curve) and \textcolor{black}{$3 \sigma$} discovery potential 
  (blue curve) versus DAQ time, compared with MEG \textcolor{black}{upper limit 
  sensitivity (90\%\ C.L.)} and \textcolor{black}{measured upper limit (90\%\ C.L.) \cite{baldini_2016}}. 
  The points mark the sensitivity
  for the past years and the estimation for the current one.
  See text for details.
  }
  \label{sensitivity:senfit}
\end{figure}

Simulated pseudo-experiments are used to evaluate the sensitivity of the experiment ${\cal S}_{90}$,
defined as the median value of the distribution of the \SI{90}{\percent} C.L. upper limits 
resulting from the likelihood analysis for a null signal hypothesis.
Simulated pseudo-experiments are also used to evaluate
the $3\sigma$ discovery at \SI{90}{\percent} power.
\Fref{sensitivity:senfit} shows the projected sensitivity and the projected discovery limit versus the DAQ lifetime 
assuming that the resolutions in \tref{sensitivity:perf} remain stable in the coming years and that the analysis remains the same as today. 
Systematic uncertainties, which are expected to give a minor contribution, are not included. 

The prediction reflects the current status and knowledge of the detector, and the quality of the current data analysis. The evolution of the detector behaviour, the continuous improvement of the reconstruction algorithms and the increasingly better understanding of the systematic uncertainties will determine the final sensitivity of the experiment. Nevertheless, we are confident that
the design sensitivity of ${\cal S}_{90} =  \num{6e-14}$ will be achieved by the end of 2026.

\section{Conclusions}
\label{sec:conclusions}

We have presented a detailed overview of the operation of 
all components of the MEG~II detector during the engineering 
runs in the years 2016--2020 and in the physics run in the years 2021--2022.
We found several problems, to name a few: CDCH breakages, faster than expected decrease of MPPC PDEs for LXe and mild
degradation of time resolution of pTC, both due to radiation damage. 
After having accumulated some delay in the schedule, those problems were solved or tackled so as to minimise their 
effects on the sensitivity. Additional planned steps should further reduce these effects.

Since 2021 the MEG~II experiment has regularly operated with performances close to its design values.
On this basis, we estimated the sensitivity to the \megp\ decay versus the DAQ lifetime and calendar year. 
The sensitivity obtained with the 2021--2022 data will be significantly better than the existing MEG limit.

\section*{Acknowledgments}

We are grateful for the support and co-operation provided  by PSI as the host laboratory and to the technical and engineering staff of our institutes.
This work is supported by 
DOE DEFG02-91ER40679 (USA); 
INFN (Italy); 
H2020 Marie Skłodowska-Curie ITN Grant Agreement 858199;
JSPS KAKENHI numbers JP26000004, 20H00154, 21H04991, 21H00065, 22K21350 and JSPS Core-to-Core Program, A. Advanced Research Networks JPJSCCA20180004 (Japan);
Schweizerischer Nationalfonds (SNF) Grants 206021\_177038,
206021\_157742, 200020\_172706, 200020\_162654 and 200021\_137738 (Switzerland); the Leverhulme Trust, LIP-2021-01 (UK).
\bibliographystyle{my}
\bibliography{MEGIIDetector}

\end{document}